\documentclass[twocolumn]{aastex63}

\usepackage{graphicx}
\usepackage[T1]{fontenc}
\usepackage{aecompl}
\usepackage[]{amsmath}

\usepackage{color}

\usepackage{xspace}

\usepackage{soul}

\newcommand{\msun}{\mathrm{M}_\odot}
\graphicspath{{figures/}}

\newcommand{\ud}{\mathrm{d}}

\def\lsim{ \lower .75ex \hbox{$\sim$} \llap{\raise .27ex \hbox{$<$}} }

\usepackage{lineno}

\submitjournal{ApJ}

\shorttitle{DESI satellite}
\shortauthors{Wang et al.}

\graphicspath{{./}{figures/}}

\begin{document}

\title{Luminosity and stellar mass functions of faint photometric satellites around spectroscopic central galaxies from DESI Year-1 Bright Galaxy Survey}

\author[0000-0002-5762-7571]{Wenting Wang}\thanks{wenting.wang@sjtu.edu.cn}
\affiliation{Department of Astronomy, School of Physics and Astronomy, and Shanghai Key Laboratory for Particle Physics and Cosmology, Shanghai Jiao Tong University, Shanghai 200240, People's Republic of China}
\affiliation{State Key Laboratory of Dark Matter Physics, School of Physics and Astronomy,Shanghai Jiao Tong University, Shanghai 200240, China}
\author{Xiaohu Yang}\thanks{xyang@sjtu.edu.cn}
\affiliation{Tsung-Dao Lee Institute, and Key Laboratory for Particle Physics, Astrophysics and Cosmology, Ministry of Education,
Shanghai Jiao Tong University, Shanghai 200240, People's Republic of China}
\affiliation{Department of Astronomy, School of Physics and Astronomy, and Shanghai Key Laboratory for Particle Physics and Cosmology, Shanghai Jiao Tong University, Shanghai 200240, People's Republic of China}
\author{Yipeng Jing}\thanks{ypjing@sjtu.edu.cn}
\affiliation{Tsung-Dao Lee Institute, and Key Laboratory for Particle Physics, Astrophysics and Cosmology, Ministry of Education,
Shanghai Jiao Tong University, Shanghai 200240, People's Republic of China}
\affiliation{Department of Astronomy, School of Physics and Astronomy, and Shanghai Key Laboratory for Particle Physics and Cosmology, Shanghai Jiao Tong University, Shanghai 200240, People's Republic of China}
\author{Ashley J. Ross}
\affiliation{Center for Cosmology and AstroParticle Physics, The Ohio State University, 191 West Woodruff Avenue, Columbus, OH 43210, USA}
\author{Malgorzata Siudek}
\affiliation{Institute of Space Sciences, ICE-CSIC, Campus UAB, Carrer de
Can Magrans s/n, 08913 Bellaterra, Barcelona, Spain}
\affiliation{Instituto Astrofisica de Canarias, Av. Via Lactea s/n, 38205 La
Laguna, Spain}
\author{John Moustakas}
\affiliation{Department of Physics and Astronomy, Siena College, 515 Loudon Road, Loudonville, NY 12110, USA}
\author{Samuel G. Moore}
\affiliation{Institute for Computational Cosmology, Department of Physics, Durham University, South Road, Durham DH1 3LE, UK}
\author{Shaun Cole}
\affiliation{Institute for Computational Cosmology, Department of Physics, Durham University, South Road, Durham DH1 3LE, UK}
\author{Carlos Frenk}
\affiliation{Institute for Computational Cosmology, Department of Physics, Durham University, South Road, Durham DH1 3LE, UK}
\author{Jiaxi Yu}
\affiliation{Kavli IPMU (WPI), UTIAS, The University of Tokyo, Kashiwa, Chiba 277-8583, Japan}
\author{Sergey E. Koposov}
\affiliation{Institute for Astronomy, University of Edinburgh, Royal Observatory, Blackford Hill, Edinburgh EH9 3HJ, UK}
\affiliation{Institute of Astronomy, University of Cambridge, Madingley Road, Cambridge CB3 0HA, UK}
\author{Jiaxin Han}
\affiliation{Department of Astronomy, School of Physics and Astronomy, and Shanghai Key Laboratory for Particle Physics and Cosmology, Shanghai Jiao Tong University, Shanghai 200240, People's Republic of China}
\author{Zhenlin Tan}
\affiliation{Department of Astronomy, School of Physics and Astronomy, and Shanghai Key Laboratory for Particle Physics and Cosmology, Shanghai Jiao Tong University, Shanghai 200240, People's Republic of China}
\author{Kun Xu}
\affiliation{Center for Particle Cosmology, Department of Physics and Astronomy, University of Pennsylvania, Philadelphia, PA 19104, USA}
\author{Yizhou Gu}
\affiliation{Tsung-Dao Lee Institute, and Key Laboratory for Particle Physics, Astrophysics and Cosmology, Ministry of Education,
Shanghai Jiao Tong University, Shanghai 200240, People's Republic of China}
\author{Yirong Wang}
\affiliation{Department of Astronomy, School of Physics and Astronomy, and Shanghai Key Laboratory for Particle Physics and Cosmology, Shanghai Jiao Tong University, Shanghai 200240, People's Republic of China}
\author{Oleg~Y.~Gnedin}
\affiliation{Department of Astronomy, University of Michigan, Ann Arbor, MI 48109, USA }
\author{Jessica Nicole~Aguilar}
\affiliation{Lawrence Berkeley National Laboratory, 1 Cyclotron Road, Berkeley, CA 94720, USA}
\author{Steven~Ahlen}
\affiliation{Physics Dept., Boston University, 590 Commonwealth Avenue, Boston, MA 02215, USA}
\author{Davide~Bianchi}
\affiliation{Dipartimento di Fisica ``Aldo Pontremoli'', Universit\`a degli Studi di Milano, Via Celoria 16, I-20133 Milano, Italy}
\affiliation{INAF-Osservatorio Astronomico di Brera, Via Brera 28, 20122 Milano, Italy}
\author{David~Brooks}
\affiliation{Department of Physics \& Astronomy, University College London, Gower Street, London, WC1E 6BT, UK}
\author{Todd~Claybaugh}
\affiliation{Lawrence Berkeley National Laboratory, 1 Cyclotron Road, Berkeley, CA 94720, USA}
\author{Axel~de la Macorra}
\affiliation{Departamento de F\'{\i}sica, Centro de Investigaci\'{o}n y de Estudios Avanzados del IPN, Av. Polit\'{e}cnico 2508, Col. Sn. Pedro Zacatenco, Del. Gustavo A. Madero, 07010 CDMX, M\'{e}xico}
\affiliation{Departamento de F\'{i}sica, Instituto Nacional de Investigaciones Nucleares, Carreterra M\'{e}xico-Toluca S/N, La Marquesa, Ocoyoacac, Edo. de M\'{e}xico C.~P.~52750, M\'{e}xico}
\affiliation{Departamento de F\'{\i}sica, DCI-Campus Le\'{o}n, Universidad de Guanajuato, Loma del Bosque 103, Le\'{o}n, Guanajuato C.~P.~37150, M\'{e}xico.}
\affiliation{Instituto de F\'{\i}sica, Universidad Nacional Aut\'{o}noma de M\'{e}xico, Circuito de la Investigaci\'{o}n Cient\'{\i}fica, Ciudad Universitaria, Cd. de M\'{e}xico C.~P.~04510, M\'{e}xico}
\author{Arjun~Dey}
\affiliation{NSF NOIRLab, 950 N. Cherry Ave., Tucson, AZ 85719, USA}
\author{Peter~Doel}
\affiliation{Department of Physics \& Astronomy, University College London, Gower Street, London, WC1E 6BT, UK}
\author{Jaime E.~Forero-Romero}
\affiliation{Departamento de F\'isica, Universidad de los Andes, Cra. 1 No. 18A-10, Edificio Ip, CP 111711, Bogot\'a, Colombia}
\affiliation{Observatorio Astron\'omico, Universidad de los Andes, Cra. 1 No. 18A-10, Edificio H, CP 111711 Bogot\'a, Colombia}
\author{Enrique~Gazta\~naga}
\affiliation{Institut d'Estudis Espacials de Catalunya (IEEC), c/ Esteve Terradas 1, Edifici RDIT, Campus PMT-UPC, 08860 Castelldefels, Spain}
\affiliation{Institute of Cosmology and Gravitation, University of Portsmouth, Dennis Sciama Building, Portsmouth, PO1 3FX, UK}
\author{Satya~Gontcho A Gontcho}
\affiliation{Lawrence Berkeley National Laboratory, 1 Cyclotron Road, Berkeley, CA 94720, USA}
\author{Gaston~Gutierrez}
\affiliation{Fermi National Accelerator Laboratory, PO Box 500, Batavia, IL 60510, USA}
\author{Klaus~Honscheid}
\affiliation{The Ohio State University, Columbus, 43210 OH, USA}
\affiliation{Department of Physics, The Ohio State University, 191 West Woodruff Avenue, Columbus, OH 43210, USA}
\affiliation{Center for Cosmology and AstroParticle Physics, The Ohio State University, 191 West Woodruff Avenue, Columbus, OH 43210, USA}
\author{Mustapha~Ishak}
\affiliation{Department of Physics, The University of Texas at Dallas, 800 W. Campbell Rd., Richardson, TX 75080, USA}
\author{Theodore~Kisner}
\affiliation{Lawrence Berkeley National Laboratory, 1 Cyclotron Road, Berkeley, CA 94720, USA}
\author{Martin~Landriau}
\affiliation{Lawrence Berkeley National Laboratory, 1 Cyclotron Road, Berkeley, CA 94720, USA}
\author{Laurent~Le~Guillou}
\affiliation{Sorbonne Universit\'{e}, CNRS/IN2P3, Laboratoire de Physique Nucl\'{e}aire et de Hautes Energies (LPNHE), FR-75005 Paris, France}
\author{Marc~Manera}
\affiliation{Institut de F\'{i}sica d’Altes Energies (IFAE), The Barcelona Institute of Science and Technology, Edifici Cn, Campus UAB, 08193, Bellaterra (Barcelona), Spain}
\affiliation{Departament de F\'{i}sica, Serra H\'{u}nter, Universitat Aut\`{o}noma de Barcelona, 08193 Bellaterra (Barcelona), Spain}
\author{Aaron~Meisner}
\affiliation{NSF NOIRLab, 950 N. Cherry Ave., Tucson, AZ 85719, USA}
\author{Ramon~Miquel}
\affiliation{Institut de F\'{i}sica d’Altes Energies (IFAE), The Barcelona Institute of Science and Technology, Edifici Cn, Campus UAB, 08193, Bellaterra (Barcelona), Spain}
\affiliation{Instituci\'{o} Catalana de Recerca i Estudis Avan\c{c}ats, Passeig de Llu\'{\i}s Companys, 23, 08010 Barcelona, Spain}
\author{Seshadri~Nadathur}
\affiliation{Institute of Cosmology and Gravitation, University of Portsmouth, Dennis Sciama Building, Portsmouth, PO1 3FX, UK}
\author{Claire~Poppett}
\affiliation{University of California, Berkeley, 110 Sproul Hall \#5800 Berkeley, CA 94720, USA }
\affiliation{Space Sciences Laboratory, University of California, Berkeley, 7 Gauss Way, Berkeley, CA 94720, USA}
\affiliation{Lawrence Berkeley National Laboratory, 1 Cyclotron Road, Berkeley, CA 94720, USA}
\author{Francisco~Prada}
\affiliation{Instituto de F\'{i}sica Te\'{o}rica (IFT) UAM/CSIC, Universidad Aut\'{o}noma de Madrid, Cantoblanco, E-28049, Madrid, Spain}
\affiliation{Instituto de Astrof\'{i}sica de Andaluc\'{i}a (CSIC), Glorieta de la Astronom\'{i}a, s/n, E-18008 Granada, Spain}
\affiliation{Instituto de Astrof\'{\i}sica de Canarias, C/ V\'{\i}a L\'{a}ctea, s/n, E-38205 La Laguna, Tenerife, Spain}
\author{Ignasi~P\'erez-R\`afols}
\affiliation{Departament de F\'isica, EEBE, Universitat Polit\`ecnica de Catalunya, c/Eduard Maristany 10, 08930 Barcelona, Spain}
\author{Graziano~Rossi}
\affiliation{Department of Physics and Astronomy, Sejong University, 209 Neungdong-ro, Gwangjin-gu, Seoul 05006, Republic of Korea}
\author{Eusebio~Sanchez}
\affiliation{Barcelona-Madrid RPG - Centro de Investigaciones Energéticas, Medioambientales y Tecnológicas}
\author{David~Schlegel}
\affiliation{Lawrence Berkeley National Laboratory, 1 Cyclotron Road, Berkeley, CA 94720, USA}
\author{Hee-Jong~Seo}
\affiliation{Department of Physics \& Astronomy, Ohio University, 139 University Terrace, Athens, OH 45701, USA}
\author{Joseph Harry~Silber}
\affiliation{Lawrence Berkeley National Laboratory, 1 Cyclotron Road, Berkeley, CA 94720, USA}
\author{David~Sprayberry}
\affiliation{NSF NOIRLab, 950 N. Cherry Ave., Tucson, AZ 85719, USA}
\author{Gregory~Tarl\'{e}}
\affiliation{University of Michigan, 500 S. State Street, Ann Arbor, MI 48109, USA}
\author{Benjamin Alan~Weaver}
\affiliation{NSF NOIRLab, 950 N. Cherry Ave., Tucson, AZ 85719, USA}
\affiliation{Department of Physics and Center for Cosmology and Particle Physics, New York University, New York, NY 10003, USA}
\author{Hu~Zou}
\affiliation{National Astronomical Observatories, Chinese Academy of Sciences, A20 Datun Rd., Chaoyang District, Beijing, 100012, P.R. China}

\begin{abstract}
We measure the luminosity functions (LFs) and stellar mass functions (SMFs) of photometric satellite galaxies around spectroscopically identified isolated central galaxies (ICGs). The photometric satellites are from the DESI Legacy Imaging Surveys (DR9), while the spectroscopic ICGs are selected from the DESI Year-1 BGS sample. We can measure satellite LFs down to $r$-band absolute magnitudes of $M_{r,\mathrm{sat}}\sim-7$, around ICGs as small as $7.1<\log_{10}M_{\ast,\mathrm{ICG}}/\msun<7.8$, with the stellar mass of ICGs measured by the DESI \textsc{Fastspecfit} pipeline. The satellite SMF can be measured down to $\log_{10}M_{\ast,\mathrm{sat}}/\msun\sim 5.5$. Interestingly, we discover that the faint/low-mass end slopes of satellite LFs/SMFs become steeper with the decrease in the stellar masses of host ICGs, with smaller and nearby host ICGs capable of being used to probe their fainter satellites. The steepest slopes can be $-2.298\pm0.656$ and $-$2.888$\pm$0.916 for satellite LF and SMF, respectively. Detailed comparisons are performed between the satellite LFs around ICGs selected from DESI BGS or from the SDSS NYU-VAGC spectroscopic Main galaxies over $7.1<\log_{10}M_{\ast,\mathrm{ICG}}/\msun<11.7$, showing reasonable agreement, but we show that differences between DESI and SDSS stellar masses for ICGs play a role to affect the results. We also compare measurements based on DESI \textsc{Fastspecfit} and \textsc{Cigale} stellar masses used to bin ICGs, with the latter including the modeling of AGN based on WISE photometry, and we find good agreements in the measured satellite LFs by using either of the DESI stellar mass catalogs. 
\end{abstract}

\keywords{large scale structure of Universe; Galaxy dark matter halos (1880)}

\section{Introduction}
\label{sec:intro}

In the hierarchical cosmic structure formation paradigm of our Universe, galaxies form within dark matter halos. Smaller halos form first, which, together with the galaxies they host, merge to form and contribute to the growth of larger halos/galaxies. After falling into host halo systems, these smaller subhalos and galaxies are called subhalos and satellite galaxies, that orbit around the central dominant galaxy and would eventually merge with the central galaxy.

The abundance, spatial distribution and properties of low-mass satellite galaxies in dense galaxy cluster or group environments are of great importance for both galaxy evolution and cosmology studies. Firstly, these low-mass satellites carry important information about how the galaxy cluster/group environments affect their star formation and morphological evolutions. Moreover, the important cosmological implication comes from the fact that cold dark matter (CDM) and warm dark matter (WDM) models predict different amounts of low-mass substructures \citep[e.g.][]{2012MNRAS.420.2318L}, hence the abundance of such low-mass objects can be used to distinguish different dark matter models, though there still exists the difficulty of how to connect luminous low-mass satellite galaxies to dark matter substructures.

The low-mass satellite galaxies, however, often do not have spectroscopic observations due to their faintness, so it is difficult to directly get their precise distance modulus and intrinsic rest-frame properties. To circumvent this issue, many studies have developed the method of counting photometric satellite galaxies around spectroscopically identified central galaxies. In such methodologies, the intrinsic properties of faint photometric companion galaxies can be estimated by assuming these companions are at the same redshifts of their spectroscopic bright central galaxies at first, and then foreground and background companions can be statistically subtracted off by counting companions around random points or using cross correlation techniques. Such efforts can be traced back to as early as \cite{1994MNRAS.269..696L} using observations from the Palomar photometric plates. 

With the developments and progress of later large surveys such as the Sloan Digital sky Survey (SDSS), the method of counting photometric satellites around spectroscopically identified central galaxies has led to many important measurements and conclusions, including pushing the satellite luminosity function (LF) measurement down to fainter magnitudes and inference of the faint end slopes \citep[e.g.][]{2011AJ....142...13L,2011MNRAS.417..370G,2012MNRAS.424.2574W,2012ApJ...760...16J,2016MNRAS.459.3998L,2021MNRAS.500.3776W}, satellite color distribution and star-forming quenching mechanisms \citep[e.g.][]{2012MNRAS.424.2574W}, projected number density profiles \citep[e.g.][]{2011ApJ...734...88W,2012MNRAS.427..428G,2014MNRAS.442.1363W,2021ApJ...919...25W,2023ApJ...947...19A}, redshift evolution \citep[e.g.][]{2013ApJ...772..146N,2014ApJ...792..103K}, abundance and the connection to host halo, central galaxy properties and cosmic environments \citep[e.g.][]{2012MNRAS.424.2574W,2015ApJ...800..112G,2021MNRAS.505.5370T,2021ApJ...919...25W,2023ApJ...947...19A}, measurement of the stellar-halo mass relation and global stellar mass function \citep[e.g.][]{2022ApJ...925...31X,2022ApJ...926..130X,2022ApJ...939..104X,2023ApJ...944..200X}, comparisons with our Milky Way (MW) and Local Group (LG) satellite systems \citep[e.g.][]{2021MNRAS.500.3776W}, angular distribution of satellite galaxies to make intuitions on whether there is the existence of satellite planes to violate the standard theory \citep[e.g.][]{2015MNRAS.449.2576C} and so on. 

Many of the studies mentioned so far perform comparisons with predictions by modern numerical simulations to verify the theory of cosmic structure formation and galaxy evolutions \citep[e.g.][]{2012MNRAS.424.2574W,2013MNRAS.434.1838G,2014MNRAS.442.1363W,2015MNRAS.449.2576C}. In general, good agreements have been reported for comparisons with CDM semi-analytical model predictions or hydro-dynamical simulations. However, when splitting either central or satellite galaxies by color, the associated satellite spatial and property distributions often show more delicate differences between simulation predictions and the real observation \citep[e.g.][]{2013MNRAS.434.1838G,2014MNRAS.442.1363W,2023ApJ...947...19A}. And more recently, \cite{2024NatAs...8..538G} pointed out the excess of co-rotating satellite pairs in numerical simulation than in SDSS. As limited by the mass or magnitude range that can be pushed down so far for satellites, and due to complications brought in of how to associate observable low-mass satellite galaxies to low-mass subhalos, no solid and strong constraints have been so far made to favor or disfavor CDM or WDM models. 

Interests are growing towards searching and studying companion satellites around central galaxies with mass or luminosity comparable to or even lower than the Large Magellanic Cloud \citep[e.g.][]{2013MNRAS.428..573S,2021MNRAS.500.3776W,2021MNRAS.502.1205R}, including some Local Volume searches \citep[e.g.][]{2016ApJ...828L...5C,2020ApJ...891..144C,2020A&A...644A..91M}. Low-mass central galaxies, if spectroscopically identified, can be used to study their even fainter photometric companion satellites down to much lower mass end, where the satellite statistics are still complete above the corresponding photometric survey flux limit, hence pushing the measurement of satellite LFs down to even fainter magnitude ranges for cosmological inferences. 

In a previous study of \cite{2021MNRAS.500.3776W}, we measure the LF of photometric satellites around spectroscopically identified central galaxies from SDSS, and the satellite LF is presented down to $V$-band absolute magnitude of about $-$12 to $-$10, but $-10$ is not the faintest limit, as the main purpose of \cite{2021MNRAS.500.3776W} is to measure satellite LFs in MW-mass systems. 

After SDSS, the Dark Energy Spectroscopic Instrument \citep[DESI;][]{Snowmass2013.Levi,desiScience,desiInstrument,desi-collaboration22a,SurveyOps.Schlafly.2023,DESI2023a.KP1.SV,2023AJ....165...50M}\footnote{\cite{desi-collaboration22a} and \cite{DESI2023a.KP1.SV} are DESI Collaboration Key Papers.} is one of the ongoing and foremost multi-object spectrographs for wide-field surveys. The main science of DESI is to achieve the most precise constraint on the expansion history of the
Universe to date \citep{Snowmass2013.Levi}. The early data release \citep[EDR;][]{DESI2023b.KP1.EDR}\footnote{DESI Collaboration Key Paper} of DESI has been made, and the first data release (DR1;  DESI Collaboration et al. 2025, in prep) will be made very soon this Spring, which provide valuable data products including millions of quasars \citep{2023ApJ...944..107C}, Emission Line Galaxies \citep[ELG;][]{2023AJ....165..126R}, Luminous Red Galaxies \citep[LRG;][]{2023AJ....165...58Z}, bright galaxies \citep[BG;][]{2022APS..APRH13003H,2023AJ....165..253H,2025AJ....169..157J} and MW sources \citep{2023ApJ...947...37C,2024MNRAS.533.1012K}, that has led to key results on two-point correlation function statistics and galaxy clustering measurements \citep{DESI2024.II.KP3,DESI2024.V.KP5}, Baryon Acoustic Oscillation (BAO) measurements and studies from the Lyman alpha forest \citep{2024arXiv240403000D}, cosmological constraints from BAO and galaxy clustering \citep{2024arXiv240403002D,DESI2024.VII.KP7B}\footnote{\cite{DESI2024.II.KP3,DESI2024.V.KP5,2024arXiv240403000D,2024arXiv240403002D,DESI2024.VII.KP7B} are DESI Collaboration Key Papers.}.

The Tier-1 bright galaxy sample of DESI Bright Galaxy Survey \citep[BGS;][]{2022APS..APRH13003H,2023AJ....165..253H} module goes down to Galactic extinction corrected $r$-band flux limit of $r_\mathrm{dustcorr}<19.5$. It promisingly extends the spectroscopic observation of low redshift galaxies by about two magnitudes fainter than SDSS spectroscopic Main galaxies, which is $r<17.7$. Hence at the same distance, DESI would lead to more lower-mass central galaxies to be spectroscopically observed. In this paper, we try to push the satellite measurements down to the faintest magnitudes or mass around lower-mass central galaxies, with the central galaxies selected based on spectroscopically observed DESI bright galaxies and the photometric satellites from the DESI Legacy imaging Survey. We demonstrate the performance based on central hosts selected from DESI Year-1 BGS, with detailed comparisons with SDSS. We show that we can at most push down to $M_{r,\mathrm{sat}}\sim-7$ or $\log_{10}M_{\ast,\mathrm{sat}}/\msun\sim5.5$ in the measured satellite LFs, around central hosts as small as $7.1<\log_{10}M_{\ast,\mathrm{ICG}}/\msun<7.8$. We also discover the general trend that the faint or low-mass end slopes of satellite LFs get steeper with the decrease in the stellar masses of central hosts. 

However, we will also show that satellite counts fainter than $r$-band absolute magnitude of $M_{r,\mathrm{sat}}\sim-10$ are mainly contributed by nearby hosts with redshifts $z<0.01$. At such low redshifts, central galaxies with $\log_{10}M_{\ast,\mathrm{ICG}}/\msun>7.1$ are already mostly observed given the flux limit of $r<17.7$, and thus the fainter flux limit of $r_\mathrm{dustcorr}<19.5$ almost does not include more central galaxies with $\log_{10}M_{\ast,\mathrm{ICG}}/\msun>7.1$. Over smaller stellar mass range of $5.7<\log_{10}M_{\ast,\mathrm{ICG}}/\msun<7.1$ and at redshifts $z<0.01$, the deeper flux limit can increase the sample of central galaxies, but the total sample size is still small. Hence the improvement in the measured satellite LF around ICGs selected from DESI Year-1 BGS with $r_\mathrm{dustcorr}<19.5$ and with $5.7<\log_{10}M_{\ast,\mathrm{ICG}}/\msun<7.1$ is limited, compared with the results based on the shallower flux limit of $r<17.7$.

The layout of this paper is as follows. We introduce DESI BGS, our selections of isolated central galaxies from DESI BGS and SDSS spectroscopic Main galaxies, and photometric satellites in Section~\ref{sec:data}. The methodologies of satellite counting, background subtraction, corrections for fiber assignment incompleteness and correction for incomplete projected area due to survey boundaries and masks are introduced in Section~\ref{sec:methods}. Results are presented in Section~\ref{sec:results}, including detailed comparisons between DESI and SDSS. We conclude in the end (Section~\ref{sec:concl}). The cosmological parameters used in this paper are based on Planck 2018 cosmology \citep{2020A&A...641A...6P} (combining CMB, lensing and BAO), with $\Omega_{\Lambda,0}=0.6889$, $\Omega_{m,0}=0.3111$ and h=0.6766.

\section{Data}
\label{sec:data}
\subsection{DESI Bright Galaxy Survey}

DESI performs wide-field surveys of galaxies and stars within our MW \citep{Snowmass2013.Levi,desiScience,desiInstrument,desi-collaboration22a,SurveyOps.Schlafly.2023}. It features a $3.2^\circ$ diameter field of view at the prime focus of the Mayall 4m telescope at Kitt Peak National Observatory, with 5,000 fibers available. The fibers connect to ten identical 3-arm spectrographs ($B$, $R$ and $Z$-arms from the blue to red ends), altogether span the rest-frame wavelength range of 3,600-9,824~\AA~ with a FWHM resolution of $\sim$1.8~\AA \citep{desiInstrument,silber22a,Corrector.Miller.2023,FiberSystem.Poppett.2024}.

The DESI observations are divided into dark and bright time\footnote{Bright and dark time observations refer to the nights with and without significant moon light contamination.}. Effective exposure times are $\sim1000$~seconds and $\sim180$~seconds for dark and bright time targets, respectively \citep{Spectro.Pipeline.Guy.2023}. While dark time observations are mostly assigned to higher redshift galaxies, the DESI Bright Galaxy Survey (BGS) is performed at the same time with the DESI Milky Way Survey (MWS) at bright time, with sources belonging to BGS assigned higher priorities than MW main survey targets. BGS mainly probes more nearby galaxies at redshifts of $z<0.4$. The readers can refer to \citep{2023AJ....165..253H} for details, and here we only have a brief introduction about the DESI BGS sample. 

DESI BGS is expected to cover an area of $\sim$14,000 square degrees in 4 passes of the sky. The observation divides into two tiers: 1) Tier-1: the primary BGS sample (BGS Bright sample) is an approximately flux limited sample down to Galactic extinction corrected $r$-band apparent magnitude of $r_\mathrm{dustcorr}=19.5$; 2) Tier-2: The BGS faint sample covers $19.5<r_\mathrm{dustcorr}<20.175$, and it also includes color dependent cuts to the fiber magnitudes in the following

\begin{equation} \label{eq:rfib_color_cut}
    r_{\rm fiber} <
    \begin{cases}
        20.75   & \text{if}~(z - W1) - 1.2 (g - r) + 1.2 < 0\\
        21.5    & \text{if}~(z - W1) - 1.2 (g - r) + 1.2 \ge 0, 
    \end{cases}
\end{equation}
where $grz$ and $W1$ are the DESI and WISE photometry. Throughout this paper, when we talk about the Galactic extinction corrected $r$-band apparent magnitudes, we denote it using $r_\mathrm{dustcorr}$. 

The BGS Bright and Faint samples are expected to have target number densities of 864~targets/deg$^2$ and 1,400~targets/deg$^2$, respectively. The faint sample helps to increase the overall BGS target density, allowing small-scale clustering measurements with higher signal-to-noise ratio (S/N). DESI Tier-1 BGS sample goes about two magnitudes fainter and has a median redshift of about twice higher than SDSS spectroscopic Main galaxies\footnote{The median redshift of SDSS Main galaxies is about $z=0.1$.}. Using the DESI BGS sample to construct central galaxies, we are able to measure the LFs of photometric satellites around smaller central galaxies. 

In the next section, we introduce our selection of the isolated central galaxies based on DESI BGS. 

\subsection{Isolated central galaxies in DESI}
\label{sec:desiicg}

\begin{table*}
\caption{The average halo virial radii ($R_{200,\mathrm{mock}}$, second column) for our isolated central galaxies (ICGs) grouped by stellar mass (first column), which are estimated using ICGs selected from a mock galaxy catalog of Guo et al. (2011). In each stellar mass bin, the numbers of ICGs selected under different conditions are provided. The third column shows the total number of ICGs selected from Tier-1 BGS of DESI Year-1 data ($N_\mathrm{ICG,DESI}$), with flux limit of $r_\mathrm{dustcorr}<19.5$. Here $r_\mathrm{dustcorr}$ is Galactic extinction corrected. The numbers are calculated using $1/p_\mathrm{obs}$ as weights to account for fiber incompleteness in DESI (see Section~\ref{sec:fiberincomplete}). The fourth column gives the numbers of SDSS ICGs in these stellar mass bins, after median stellar mass corrections to match the stellar mass of DESI (see Figure~\ref{fig:masscorr}) and with fiber incompleteness corrections (see Section~\ref{sec:fiberincomplete}).  The last column is similar to the third column, but we further includes a flux cut of $r<17.7$ to DESI ICGs, with the other selections identical to the third column, to have a more fair comparison with SDSS.}
\begin{center}
\begin{tabular}{lcccc}\hline\hline
$\log M_{\ast,\mathrm{ICG}}/\msun$ & \multicolumn{1}{c}{$R_{200,\mathrm{mock}}$ [kpc]} & $N_\mathrm{ICG,DESI}$ & $N_\mathrm{ICG,SDSS}$ & $N_\mathrm{ICG,DESI}$ \\
 &  &  &  $M_\ast$ corr (resel$+$rebin) & $(r<17.7)$  \\\hline
11.4-11.7 & 758.65 & 112585  &  19375  & 29968  \\
11.1-11.4  & 459.08 & 468736  &  54718  & 92448  \\
10.8-11.1 & 288.16 &  892661  &  85557  &  130368 \\
10.5-10.8 &  214.80 & 955531   &  82512  &  109634 \\
10.2-10.5 &  173.18  & 723499    &  60217  &  69635 \\
9.9-10.2 & 142.85  &  488155  &  36340   &  41541 \\ 
9.2-9.9 &  114.64 &  586346    &  41229  &  46936 \\
8.5-9.2 & 82.68 & 180217   &  13582  & 15168  \\
7.8-8.5 & 61.42 & 41515   &  3219  &  3394 \\
7.1-7.8 & 61.42 &  7305  & 593   & 768 \\
6.4-7.1 & 61.42 & 1383   &  102  &  211 \\
5.7-6.4 & 61.42 & 228  &  11  & 48 \\
\hline
\label{tbl:r200}
\end{tabular}
\end{center}
\end{table*}

To identify a sample of galaxies that are highly likely the central galaxy of host dark matter halos (purity), we follow the approach of selecting the brightest galaxies within given projected and line-of-sight distances, and we call these galaxies isolated central galaxies (ICGs). In a few previous studies, we have selected ICGs based on the SDSS Main galaxy sample \citep[e.g.][]{2012MNRAS.424.2574W,2014MNRAS.442.1363W,2019MNRAS.487.1580W,2021MNRAS.500.3776W,2021ApJ...919...25W,2023ApJ...947...19A}, and in this paper we apply a similar approach to the DESI Year-1 large scale structure BGS sample \citep{2024arXiv240516593R} after hardware and imaging veto masks \citep{2024arXiv240516593R,2024arXiv241112020D}. To have a clean sample of BGS with robust redshift measurements, we require $\mathrm{ZWARN}=0$, and $\Delta \chi^2$, which is the $\chi^2$ values between the best and next-best-fit redshifts of the pipeline, to be greater than 40. Our parent BGS sample for selection is limited to the DESI Tier-1 BGS sample with $r_\mathrm{dustcorr}<19.5$, which is a nearly flux limited sample \citep{2023AJ....165..253H}. Note, however, in the most dense galaxy cluster or group environments, the completeness can be sometimes as low as $\sim$20\% \citep[e.g.][]{2019MNRAS.484.1285S,2024ApJ...973...82S}.

Stellar masses of DESI BGS are estimated with the DESI \textsc{Fastspecfit} pipeline\footnote{\url{https://fastspecfit.readthedocs.io/en/latest/}} (\cite{2023ascl.soft08005M}; J. Moustakas et al., in prep) assuming a \cite{2003PASP..115..763C} initial mass function, and this would be the default choice of DESI based stellar mass for the majority of results in this paper. Moreover, based on the same initial mass function, DESI Year-1 data also has the \textsc{Cigale} stellar mass measurements with modeling of AGN through the combination of DESI and WISE photometry \citep{2024A&A...691A.308S}, which we will use in a later section of this paper as a comparison to confirm the results.

We adopt the following selection criteria. Galaxies should be the brightest within the projected virial radius, $R_{200}$, of their host dark matter haloes\footnote{$R_{200}$ is defined to be the radius within which the average matter density is 200 times the mean critical density of the universe. The virial radius and velocity here are derived through the abundance matching formula between stellar mass and halo mass \citep{2010MNRAS.404.1111G} for each individual galaxy. In addition, based on mock catalogs it was demonstrated that the choice of three times virial velocity along the line of sight is a safe criterion that identifies all true companion galaxies.} and within three times the virial velocity along the line of sight. To compensate the incompleteness of the DESI BGS sample and avoid the situation that when a true companion is actually brighter but the galaxy mistakenly pass our selection because this companion does not have spectroscopic observations, we further discard galaxies that have a photometric companion satisfying the magnitude requirement, whose spectroscopic redshift information is not available but its photometric redshift (photoz), $z_\mathrm{phot}$, satisfies the condition of $|z_\mathrm{phot}-z_\mathrm{cen,spec}|<2.5\max(\sigma_p,0.05)$. Here $z_\mathrm{cen,spec}$ is the spectroscopic redshift of the spectroscopic central galaxy, and $\sigma_p$ is the 1-$\sigma$ error of the photoz. $z_\mathrm{phot}$ and $\sigma_p$ are both taken from the DESI Legacy imaging Survey (see Section~\ref{sec:photosample} below). This criterion follows \cite{2011MNRAS.417..370G}.

As tested with a mock galaxy catalog based on the semi-analytical model of \cite{2010MNRAS.404.1111G}, the completeness of ICGs among all true halo central galaxies is about 90\%. The purity is above 82\%, which reaches $>$90\% at $\log_{10}M_{\ast,\mathrm{ICG}}/\msun>11.5$. The readers can check more details in  \cite{2019MNRAS.487.1580W}. However, the mock galaxy catalog does not incorporate the effect of incomplete spectroscopic redshifts, and the readers may wonder whether the chosen criterion for photoz above is appropriate, given the large relative photoz uncertainties at low redshifts. To test this, we have tried to reject the central ICG, as long as it has a brighter companion galaxy projected within the virial radius, and does not have spectroscopic redshift measurement. The number of selected ICGs decreases at most by $\sim$20\% in the most massive bin, and for most the other mass bins the number of ICGs only decreases by $<\sim$10\%, indicating our results are unlikely affected by the photoz selection significantly.

In this study, we will calculate the LF for satellites projected within the halo virial radius of ICGs. To ensure consistencies with our previous studies \citep{2019MNRAS.487.1580W,2021MNRAS.500.3776W,2021ApJ...919...25W,2023ApJ...947...19A}, here we choose the same $R_{200}$ for all ICGs in the same bin of stellar mass. Historically, this $R_{200}$ for each stellar mass bin was calculated based on ICGs selected from the mock galaxy catalog of \cite{2011MNRAS.413..101G}, and we maintain the choice in this paper. Note, however, the $R_{200}$ here is different from the $R_{200}$ we adopted above when selecting ICGs, that vary for each individual galaxy and was calculated from the abundance matching formula of \cite{2010MNRAS.404.1111G}. So hereafter, we choose to denote this $R_{200}$, within which we calculate satellite number counts, explicitly as $R_{200,\mathrm{mock}}$. 

The values of $R_{200,\mathrm{mock}}$ in these different bins\footnote{The bins are defined in log stellar mass following \cite{2012MNRAS.424.2574W}, with the sizes tuned to ensure enough number of ICGs in bins at both the most and least massive ends.} are provided in Table~\ref{tbl:r200}. For the three least massive bins, we simply fix their values of $R_{200,\mathrm{mock}}$ to the bin of $7.8<\log_{10}M_{\ast,\mathrm{ICG}}/\msun<8.5$, as the decrease in host halo mass is very slow with the decrease in stellar mass at the low mass end \citep[e.g.][]{2010MNRAS.404.1111G,2021ApJ...919...25W}. Moreover, in a previous study of \cite{2021MNRAS.500.3776W}, we have pointed out the existence of fake sources due to mistakenly deblended parts of the central galaxy, especially some star-forming regions along the spiral arms of late type galaxies, can contaminate the measured satellite LF. We avoid such deblending mistakes by applying inner radius cuts in projection, and the choice is $>$30~kpc for $\log_{10}M_{\ast,\mathrm{ICG}}/\msun>10.2$ and $>$10~kpc for $\log_{10}M_{\ast,\mathrm{ICG}}/\msun\leq 10.2$.

Table~\ref{tbl:r200} also provides the total number of ICGs in each bin under different conditions. The third column shows the total number of ICGs selected from Tier-1 BGS in DESI Year-1 observation with $r_\mathrm{dustcorr}<19.5$ ($N_\mathrm{ICG, DESI}$). For DESI, the ICG numbers are weighted sum of $1/p_\mathrm{obs}$, to account for fiber incompleteness in DESI BGS. The readers can check Section~\ref{sec:fiberincomplete} below about how is $1/p_\mathrm{obs}$ defined. We will back discussing the other columns slightly later.

\subsection{Isolated central galaxies in SDSS}
\label{sec:sdssicg}

In this paper we will perform detailed comparisons of satellite LF measurements around ICGs selected from both DESI BGS and from SDSS spectroscopic Main galaxies, and in this subsection we introduce how ICGs are selected from SDSS.

The parent sample used for selection is the NYU Value Added Galaxy Catalogue 
\citep[NYU-VAGC;][]{2005AJ....129.2562B}, which is based on the spectroscopic Main galaxy sample from the seventh data release of the Sloan Digital Sky Survey \citep[SDSS/DR7;][]{2009ApJS..182..543A}. The sample is flux 
limited down to an apparent magnitude of $\sim$17.7 in SDSS $r$-band, with most of the objects below redshift $z=0.25$. Stellar masses in NYU-VAGC were estimated from the K-corrected galaxy colours by fitting the stellar population synthesis model \citep{2007AJ....133..734B} assuming 
a \cite{2003PASP..115..763C} initial mass function.

In the previous study of \cite{2021MNRAS.500.3776W}, we have selected ICGs from SDSS, with almost the same isolation criteria adopted in this paper, except for the compensating selection criteria adopted to photometric companions when their spectroscopic redshift measurements are not available due to fiber collisions. For such cases, \cite{2021MNRAS.500.3776W} adopted the photoz probability distribution from \cite{2009MNRAS.396.2379C} for ICG selections, whereas the selection criteria we adopted above do not require a full photoz probatility distribution, but instead relies on the 1-$\sigma$ error, $\sigma_p$, of photoz measurements ($|z_\mathrm{phot}-z_\mathrm{cen,spec}|<2.5\max(\sigma_p,0.05)$, see Section~\ref{sec:desiicg} above). 

Thus to ensure fair comparisons between SDSS and DESI, we now force the compensating selection to be exactly the same between DESI and SDSS. Instead of using the SDSS photoz catalogs, we adopt the same photoz catalog from the DESI Legacy Survey (see Section~\ref{sec:photosample} below), as when we select ICGs from DESI BGS. We select ICGs from SDSS spectroscopic Main galaxies using exactly the same criteria as in Section~\ref{sec:desiicg}.

\subsection{Photometric satellites}
\label{sec:photosample}

Our satellite galaxies are counted using those photometric companions around ICGs selected in the previous subsection. The photometric companions are from the DESI Legacy imaging Survey, which image the sky in three optical bands ($g$, $r$ and $z$), comprising 14,000 deg$^2$ of sky area, bounded by $-18\deg<\mathrm{Dec.}<84\deg$ in celestial coordinates and $|b|>18\deg$ in Galactic coordinates \citep{2019AJ....157..168D}. The surveys are comprised of 3 imaging projects, the Beijing-Arizona Sky 
Survey \citep[BASS;][]{2017PASP..129f4101Z}, the Mayall $z$-band Legacy Survey (MzLS) and the Dark Energy Camera Legacy Survey (DECaLS). The survey footprints are observed at least once, while most fields are observed twice or more times. In fact, the depth varies over the sky. 

Data used in this study are downloaded from the ninth data release (DR9) of the DESI Legacy surveys (Schlegel et al., in prep). All data from the Legacy Surveys 
are first processed through the NOAO Community Pipelines \citep{2019AJ....157..168D}. The photometric source product is constructed by \textsc{tractor}\footnote{\url{https://github.com/dstndstn/tractor}}. \textsc{tractor} \citep{2016ascl.soft04008L} 
generates an inference-based model of the sky which best fits the real data. The sources are detected using the sky subtracted and PSF convolved stacked images with a threshold of $6\sigma$. For each detected source, \textsc{tractor} models its pipeline-reduced images from different exposures and in multiple bands simultaneously. This is achieved by fitting parametric profiles including a delta function (for 
point sources), a de Vaucouleurs law, an exponential model or de Vaucouleurs plus exponential to each image simultaneously. The model is assumed to be the same for all images and is convolved with the corresponding PSF in different exposures and bands before fitting to each image. \textsc{tractor} also outputs 
the quantity which can be used to distinguish extended sources (galaxies) from point sources (stars).

Starting from the DESI Legacy Survey database sweep files, we remove all sources with TYPE classified as ``PSF'' 
, and require BITMASK not containing any of the following: BRIGHT, SATUR\_G (saturated), SATUR\_R, ALLMASK\_G\footnote{ALLMASK\_X denotes a source that touches a pixel with problems in all of a set of overlapping X-band images. Explicitly, such pixels include BADPIX, SATUR (saturated) , INTERP (interpolated), CR (hit by cosmic ray) or EDGE (edge pixels).}, ALLMASK\_R. In \cite{2021MNRAS.500.3776W}, we discussed that the average number counts of sources keep rising to $r\sim23$. However, given the variation of the depth over the sky, we only use the regions with depth deeper than 23 in $r$. This is achieved by at first selecting bricks with at least three exposures in both $g$ and $r$-bands, and we also require the $r$-band GALdepth 
of the bricks to be deeper than 23. We only include galaxies within these selected bricks. Finally, when using the Legacy survey photometric sources to calculate companion LFs, we further make the flux limit 0.5 magnitude brighter, i.e., we adopt a flux cut of $r<22.5$ to define a safe flux limited photometric sample throughout our analysis in this study. 

\begin{figure} 
\includegraphics[width=0.49\textwidth]{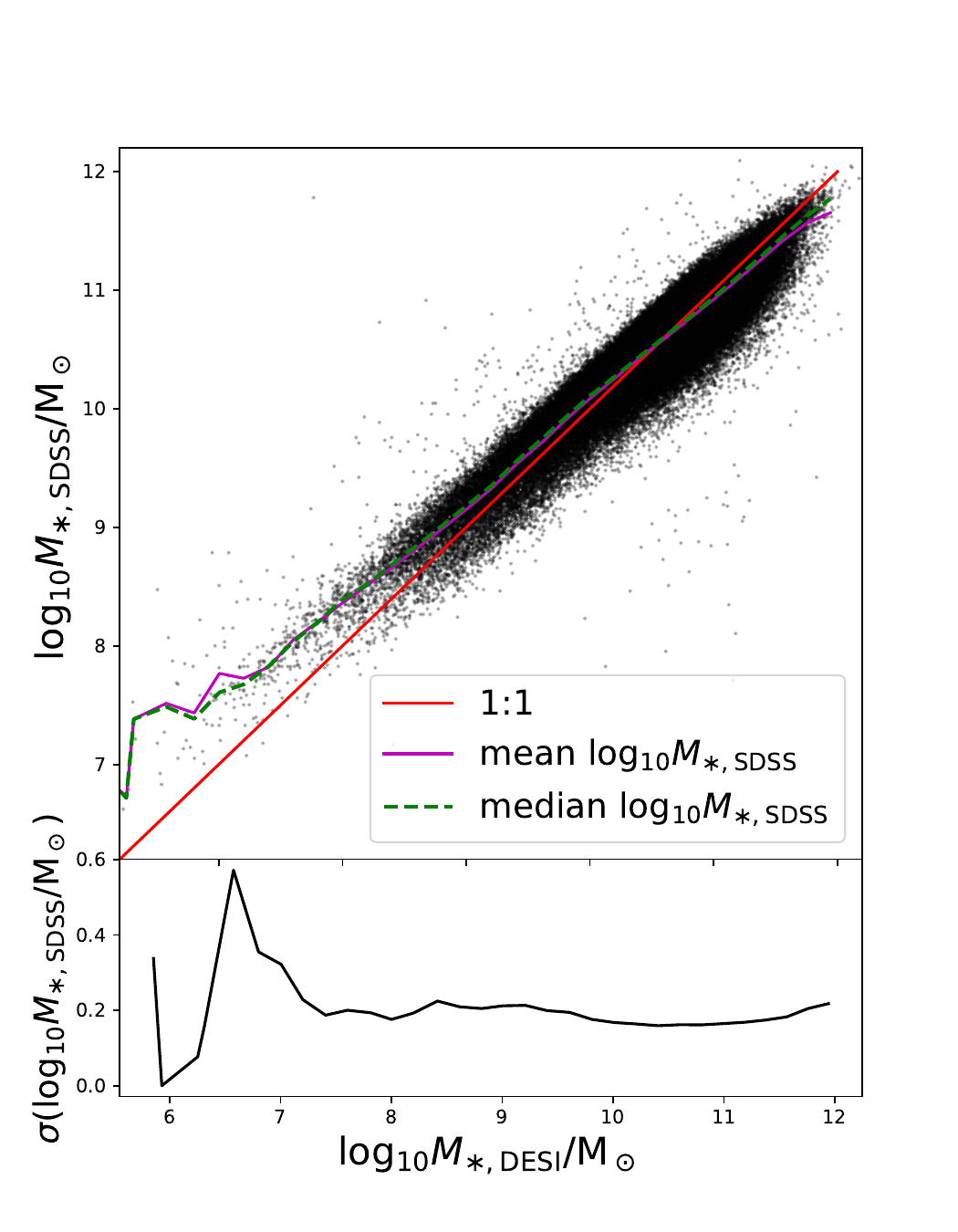}%
\caption{{\bf Top:} DESI \textsc{Fastspecfit} stellar masses versus SDSS NYU-VAGC stellar masses. Black dots show a sample of DESI Year-1 BGS matched to SDSS. The red solid line marks $y=x$ to guide the eye, while the magenta solid and green dashed curves mark the mean and median SDSS stellar mass at fixed DESI stellar mass. {\bf Bottom:} The scatter in log SDSS stellar mass, as a function of DESI \textsc{Fastspecfit} stellar mass.}
\label{fig:masscorr}
\end{figure}

\begin{figure*}
\includegraphics[width=0.99\textwidth]{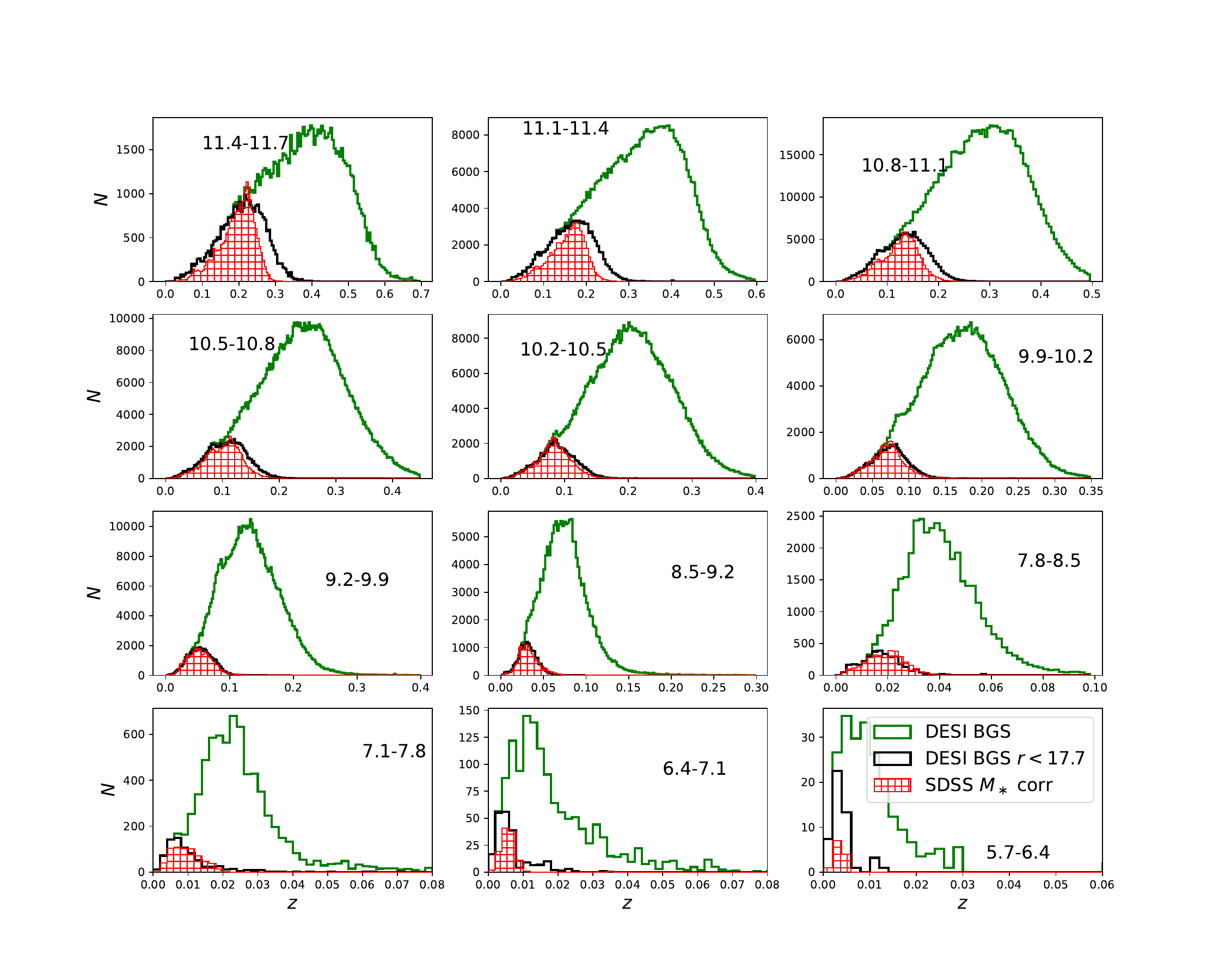}%
\caption{Redshift distribution for ICGs selected from DESI Year-1 data (green plain histogram) with $r_\mathrm{dustcorr}<19.5$ (extinction corrected according to DESI criteria) and SDSS spectroscopic Main galaxies (red hatched histogram) with $r<17.7$ (no extinction correction according to SDSS NYU-VAGC documents), in 12 stellar mass bins, as indicated by the text in each panel ($\log_{10}M_{\ast,\mathrm{ICG}}/\msun$). Here SDSS ICGs have been corrected for the median bias of stellar mass from DESI (see Figure~\ref{fig:masscorr}). Black histograms are based on a subset of DESI BGS but with flux cut of $r<17.7$ (no extinction correction), which is the flux limit for SDSS Main galaxies. Here the green and black plain histograms are all plotted with $1/p_\mathrm{obs}$ as weights, to account for fiber incompleteness in DESI (see Section~\ref{sec:fiberincomplete}). Red hatched histograms are based on the fiber collision corrected catalog from NYU-VAGC.}
\label{fig:zhistall}
\end{figure*}

\section{Methodology}
\label{sec:methods}

\subsection{Satellite counting and background subtraction}

Our method of counting photometric satellites around spectroscopically identified ICGs and computing the intrinsic luminosities of satellites follows \cite{2012MNRAS.424.2574W}. For each ICG in a given stellar mass bin, we at first count all its photometric companions down to the flux limit of $r=22.5$ and projected within the halo virial radius, $R_{200,\mathrm{mock}}$ (see Table~\ref{tbl:r200}). The physical scale is calculated based on the spectroscopic redshift and angular diameter distance of the ICG. However, without redshift information and accurate distance measurements for photometric companions, the companion counts can only be counted as a function of apparent magnitude and observed-frame color, and the total counts not only include true satellites, but also contamination by fore/background sources. We will discuss the subtraction of foreground and background sources shortly.

We derive the intrinsic luminosities and rest-frame colors using the following method. For each companion, we employ the empirical K correction of \cite{2010PASP..122.1258W} to estimate its rest frame color using the observed color and also assuming that the companion is at the same redshift as the ICG. The distance modulus is also calculated from the redshift of the ICG, in combination with the K correction, to infer the intrinsic luminosity or absolute magnitude. This is a reasonable approximation, because physically associated satellite galaxies are expected to share very similar redshifts as the ICG. 

In this paper, we mainly present measurements of satellite LFs, but we will also show our measurements of satellite SMFs. To obtain the stellar mass of satellites, \cite{2012MNRAS.424.2574W} derived a relation between the $r$-band stellar mass-to-light ratio and the $g-r$ galaxy color from SDSS spectroscopic Main galaxies. In this paper, we update the relation using DESI Year-1 BGS as $M_\ast/L_r=2.0012\times^{0.1}(g-r)-0.3133$. Here the upper index of 0.1 means all galaxies are K-corrected to $z=0.1$. Note that for all $r$-band absolute magnitudes presented in this paper, the K correction is always done to $z=0.1$, but we do not show the upper index of 0.1 to make the symbols short and easy to read.

To ensure the completeness of satellite number counts in different luminosity bins, for each ICG, we convert the photometric flux limit ($r<22.5$, see Section~\ref{sec:photosample}) to a K-corrected absolute magnitude, $M_{r,{\rm lim}}$, using the redshift of the ICG and a color chosen to be on the red envelope of the intrinsic color distribution for galaxies at that redshift. Here, $M_{r,{\rm lim}}$ can also be converted to a limit in stellar mass, $M_{\ast,{\rm lim}}$ based on the same color on the red envelope. Then for a given luminosity or stellar mass bin, ICGs are allowed to contribute to the final averaged companion counts only if $M_{r,{\rm lim}}$ (or $M_{\ast,{\rm lim}}$) is fainter (smaller) than the fainter (smaller) bin boundary. As a result, the number of actual ICGs contributing to different luminosity or stellar-mass bins for satellites can vary. The fainter or smaller the bin, the less number of ICGs can contribute to the satellite counts, and their redshifts are lower. In the end, the total companion counts are divided by the total number of ICGs which actually contribute to the satellite counts in each bin. This provides the complete and average companion LF or SMF per ICG. 

To subtract fore/background contamination, we use a sample of random points, which are assigned the same redshift and stellar mass distributions as true central primaries, but their coordinates have been randomized within the survey footprint. The averaged companion counts per random point are calculated in exactly the same way as around real ICGs above, and then be subtracted off from the averaged counts around real primaries, to obtain the background-subtracted averaged satellite counts per host ICG.  

A red end cut of $^{0.1}(g-r)<0.065\log_{10}M_\ast/\msun+0.35$ is applied to the photometric companions to reduce the number of background sources which are too red to be at the same redshift of the ICG, and hence increase the signal-to-noise (S/N). The color cut is drawn from the color distribution of SDSS spectroscopic Main galaxies in \cite{2012MNRAS.424.2574W}. Again we emphasize that the upper index of 0.1 means all galaxies are K-corrected to $z=0.1$, and for all $r$-band absolute magnitudes in this paper, the K-correction is always done to $z=0.1$.

\subsection{Correction for fiber incompleteness}
\label{sec:fiberincomplete}

Though the DESI Tier-1 BGS sample is a nearly flux limited sample, its completeness fraction in fact varies significantly with the density of the local environment. In dense galaxy cluster or group environments, the worse completeness fraction can be as low as $\sim$20\% \citep[e.g.][]{2019MNRAS.484.1285S,2024ApJ...973...82S,2024arXiv241112025B}. In order to account for the incompleteness of the BGS sample due to fiber assignments, we weight each ICG by the inverse of $p_\mathrm{obs}$ (the PROB\_OBS column) from the DESI Year-1 large scale structure catalog \citep{2024arXiv240516593R}. The final averaged satellite LF per host ICG is in fact the weighted average, with $1/p_\mathrm{obs}$ as weights. Here $p_\mathrm{obs}$ is defined as $p_\mathrm{obs}=N_\mathrm{assign}/129$, with $N_\mathrm{assign}$ the number of DESI mock fiber assignment realizations in which the target is assigned. The readers can check more details in \cite{2025JCAP...01..127L}. There are a total of 128 mock DESI fiber assignment simulations generated with different initial random seeds, and this is why the value in the denominator is 129=128+1. Adopting such a weighting strategy will increase the contribution by satellite counts around ICGs with lower fiber assignment completeness fractions.

For SDSS spectroscopic Main galaxies, we use the fiber collision corrected files provided by the NYU-VAGC website \footnote{\url{http://sdss.physics.nyu.edu/vagc/}} to account for possible effects due to fiber assignment incompleteness. The collision correction is done simply by assigning the galaxy that does not have a spectroscopic redshift observation due to fiber collisions the redshift of the nearest object in angular separation. 

\subsection{Correcting incomplete projected area}

Due to the survey boundary and photometric masks, satellite counts in a projected circular region around each ICG may be masked or outside the survey footprint. We should estimate the completeness of the projected area
around primaries. This is achieved by using the DESI Legacy survey photometric random samples provided in the database. We apply exactly the same selection and masks to random points. The completeness of the projected area is estimated as 

\begin{equation}
    f_\mathrm{complete}=\frac{\mathrm{number\ of\ actual\ random\ points}}{\mathrm{area}\times \mathrm{(surface\ density\ of\ random\ points)}}.
\end{equation}

Our actual companion counts around both real and random primaries are divided by $f_\mathrm{complete}$ for incompleteness corrections.

\begin{figure}
\includegraphics[width=0.49\textwidth]{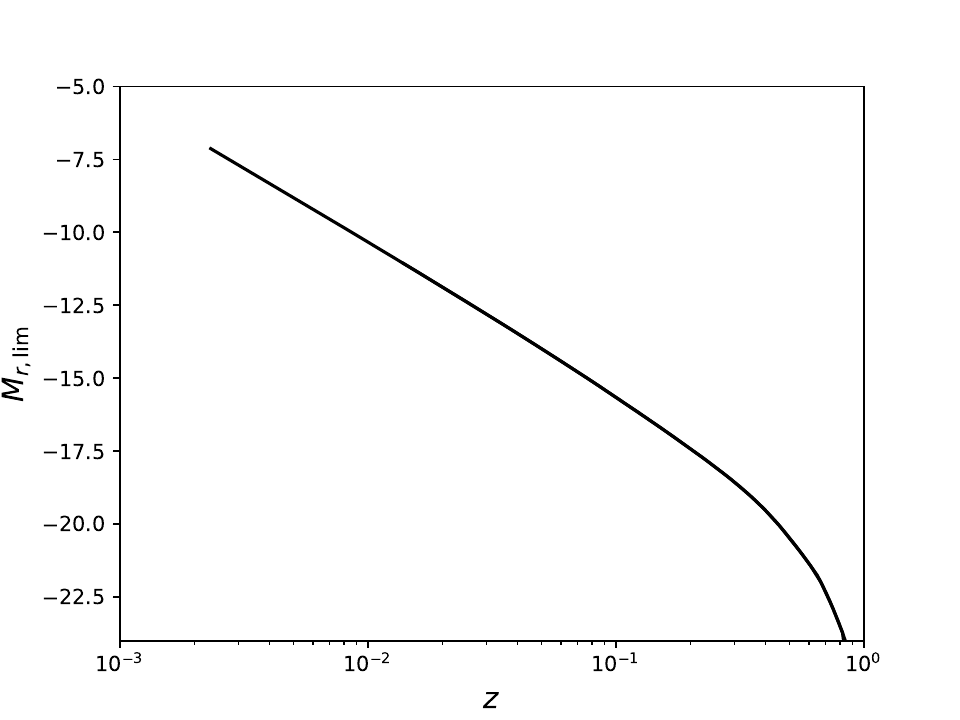}%
\caption{The limiting $r$-band absolute magnitudes ($M_{r,\mathrm{lim}}$) at different redshifts, brighter than which the satellite counts around ICGs at that redshift is still complete above the flux limit of $r=22.5$ chosen for photometric sources from the DESI Legacy imaging Survey. }
\label{fig:zlimit}
\end{figure}

\section{Results}
\label{sec:results}

\subsection{Redshift distribution, limiting magnitude and expectations}
\label{sec:expect}

Before presenting the satellite LF and SMF measurements, we first compare ICGs selected from SDSS spectroscopic Main galaxies and from DESI Tier-1 BGS sample in the Year-1 observation, and use the  comparison to discuss the expected improvements in measured satellite LFs and SMFs, around the deeper DESI BGS sample. 

Firstly, Figure~\ref{fig:masscorr} shows a comparison between the SDSS and DESI \textsc{Fastspecfit} stellar mass measurements, for a subsample of DESI BGS matched to SDSS. In general, the black dots go through the red solid diagonal line. The bottom panel of Figure~\ref{fig:masscorr} shows the scatter in SDSS stellar mass at fixed DESI stellar mass. The scatters are mostly $\sim$0.2~dex. The magenta solid and green dashed curves mark the mean and median of log SDSS stellar mass, at fixed stellar mass in DESI. As we can see, the mean and median values indicate lower SDSS stellar masses at the most massive end and higher SDSS stellar masses at lower masses, as compared with DESI \textsc{Fastspecfit} stellar masses. Though here we compare DESI \textsc{Fastspecfit} and SDSS stellar masses, similar signs and amounts of median deviations are found, if we compare SDSS stellar masses to DESI \textsc{Cigale} stellar masses. 

We do not know exactly which stellar mass is closer to the ground truth, but we take the median bias and correct for the bias of SDSS stellar mass, to force them agree with DESI \textsc{Fastspecfit} stellar mass on average\footnote{Our selection of ICGs depend on stellar mass to infer the virial radius (see Section~\ref{sec:data}). We have tried to either reselect or not reselect SDSS ICGs after the median stellar mass correction, and find with the reselection, the total numbers of SDSS ICG become slightly smaller, but the redshift distributions and measured satellite LFs/SMFs are very similar.}. This is for fair comparisons between DESI and SDSS. After the correction, SDSS ICGs are binned into 12 different stellar mass bins, with their redshift distributions shown by the red hatched histograms in different panels of Figure~\ref{fig:zhistall}. Here the redshift distributions are all plotted after fiber incompleteness corrections. In the fourth column of Table~\ref{tbl:r200}, we provide the number for ICGs selected from SDSS Main galaxies after this stellar mass correction. Due to some modification in selection criteria (see Section~\ref{sec:sdssicg}) and the stellar mass correction, the numbers differ from \cite{2021MNRAS.500.3776W}.

The redshift distributions for ICGs selected from DESI Tier-1 BGS sample in the Year-1 observation is shown by the green plain histograms in Figure~\ref{fig:zhistall}. Here to plot the histograms, we adopt $1/p_\mathrm{obs}$ as weights (see Section~\ref{sec:fiberincomplete}). With a deeper flux limit of $r_\mathrm{dustcorr}<19.5$, the green plain histograms of DESI Year-1 BGS (Tier-1) can extend to much higher redshifts. In particular, when the numbers of SDSS ICGs (red hatched histograms) already start to drop with the increase in redshifts, showing significant incompleteness, the numbers of DESI ICGs still keep rising with the increase in redshifts, which start to drop at much higher redshifts. 

We also choose a subset of ICGs from DESI BGS with the same flux limit as SDSS, i.e., $r<17.7$, which is shown as the black plain histogram. With the same flux limit, the SDSS and DESI histograms are more similar\footnote{The total footprints of SDSS Spectroscopic Main galaxy from NYU-VAGC and of DESI Year-1 BGS are 7,818~deg$^2$ \citep{2005AJ....129.2562B} and 7,473~deg$^2$ \citep{2024arXiv241112020D}. Assuming the same number density, we expect the SDSS and DESI ICG histograms to be close in amplitudes.}. Note our corrections to the SDSS stellar mass (Figure~\ref{fig:masscorr}) has redistributed SDSS ICGs in a few intermediate mass bins to the most and least massive ends, bringing in much better agreement between the DESI and SDSS ICG redshift distributions under the same flux limit of $r<17.7$. Without the median stellar correction, the black plain and red hatched histograms show more significant differences in the few most and least massive panels.

However, in the two least massive panels, the black plain histogram of DESI is still higher than the red hatched histogram of SDSS after the median stellar mass correction and with the same flux limit. This is at least partially related to the accuracy in our median stellar mass correction, as the number of low-mass galaxies adopted to deduce the median bias is very limited in the two least massive panels. The median correction shown in Figure~\ref{fig:masscorr} is very noisy at the low-mass end. Moreover, the red hatched histograms for SDSS in the three most massive panels of Figure~\ref{fig:zhistall} are still narrower than the black plain histograms for DESI. This can be due to the remaining scatter between the DESI and SDSS stellar masses, which is impossible to be easily corrected.

With Figure~\ref{fig:zhistall} and Table~\ref{tbl:r200} showing the redshift distribution and number of ICGs in different stellar mass bins, now we move on to talk about the completeness of satellite counts, and what is the expected improvement of measuring satellite LFs around ICGs selected from DESI BGS, compared with SDSS. Figure~\ref{fig:zlimit} shows the limiting absolute magnitude, $M_{r,\mathrm{lim}}$, above which the photometric satellite counts are complete above the flux limit of our photometric sample, and as a function of redshift. Here $M_{r,\mathrm{lim}}$ is estimated with $r$-band flux limit of $r=22.5$, and the readers can check Section~\ref{sec:methods} for details about how we calculate $M_{r,\mathrm{lim}}$. With the decrease in redshifts, $M_{r,\mathrm{lim}}$ gets fainter and fainter. This means that satellite LFs at fainter magnitudes are only contributed by more nearby ICGs, because satellites around more distant ICGs become incomplete. In particular, if we want to push fainter than $M_{r,\mathrm{sat}}=-10$ for satellite LF measurement, only ICGs with redshifts lower than $z=0.01$ can contribute complete satellite counts.

Despite the fact that DESI bright galaxies can extend to much higher redshifts with a deeper flux limit of $r_\mathrm{dustcorr}<19.5$ in Figure~\ref{fig:zhistall}, as compared with SDSS Main galaxies with $r<17.7$, stellar mass bins more massive than $\log_{10}M_{\ast,\mathrm{ICG}}/\msun=7.8$ in Figure~\ref{fig:zhistall} show similar number distributions of DESI bright galaxies and SDSS Main galaxies at $z<0.01$. This is because for galaxies more massive than $\log_{10}M_{\ast,\mathrm{ICG}}/\msun=7.8$, they are already well observed at $z<0.01$ with the shallower flux limit of $r<17.7$. Only in the two least massive bins of $6.4<\log_{10}M_{\ast,\mathrm{ICG}}/\msun<7.1$ and $5.7<\log_{10}M_{\ast,\mathrm{ICG}}/\msun<6.4$, there can be more galaxies from DESI BGS with the deeper flux limit of $r_\mathrm{dustcorr}<19.5$ than the flux limit of $r<17.7$ and at $z\sim0.01$. For the bin of $7.1<\log_{10}M_{\ast,\mathrm{ICG}}/\msun<7.8$, the green plain histogram is slightly higher than the black histograms at $z\sim0.01$ in Figure~\ref{fig:zhistall}, but the difference is very small. Thus, if we want to push fainter than $M_{r,\mathrm{sat}}=-10$ with ICGs selected from DESI Tier-1 BGS sample in the Year-1 observation, and achieve better measurements based on the flux limit of $r_\mathrm{masscorr}<19.5$ than $r<17.7$, from which ICGs are selected, only the measurements around ICGs in the two least massive stellar mass bins may be helpful.

\begin{figure*}
\includegraphics[width=0.99\textwidth]{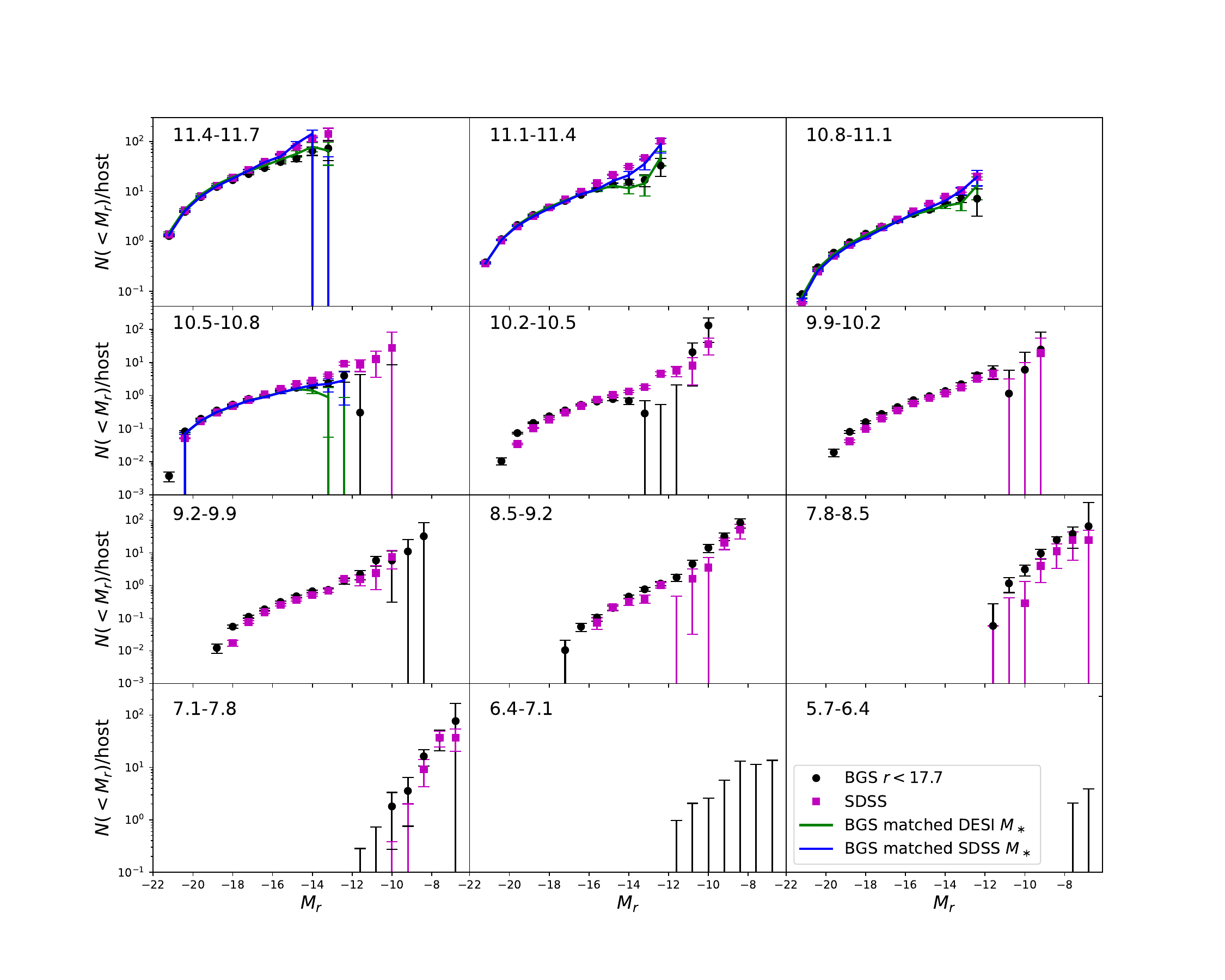}%
\caption{Cumulative satellite LFs around ICGs in a few different stellar mass bins (the text in each panel gives the $\log_{10} M_{\ast,\mathrm{ICG}}\msun$ stellar mass range for ICGs). For ICGs selected from DESI BGS, we apply median corrections to the DESI stellar mass (Figure~\ref{fig:masscorr}), to bring the DESI stellar mass to be consistent with SDSS on average. In addition, we apply a flux cut of $r<17.7$ to ICGs from DESI. Black dots and magenta squares with errorbars are based on DESI and SDSS ICGs. Errorbars are calculated from the 1-$\sigma$ scatter among 100 bootstrapped subsamples of the ICGs. Green and magenta curves in the few most massive panels are based on a matched subsample between DESI and SDSS ICGs, but using DESI and SDSS stellar masses, respectively. The cumulative LF sometimes decreases at some point. This is due to statistical fluctuations. After the background subtraction, sometimes the differential LF can have negative measurements.}
\label{fig:LFcompare}
\end{figure*}

\begin{figure*}
\includegraphics[width=0.99\textwidth]{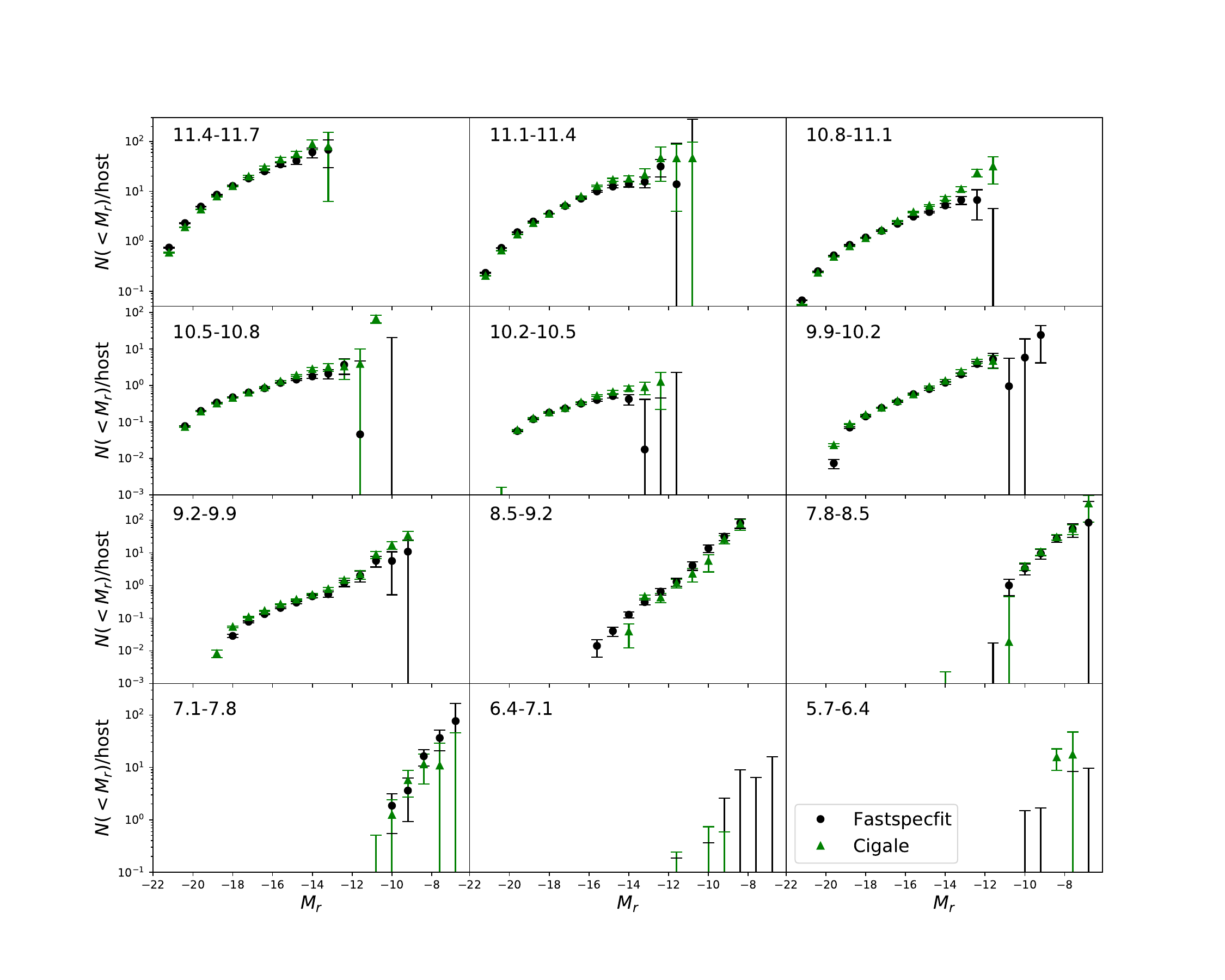}%
\caption{Cumulative satellite LFs around ICGs in 12 different stellar mass bins. Each panel refers to a given stellar mass bin of ICGs, with the text in each panel gives the $\log_{10} M_{\ast,\mathrm{ICG}}/\msun$ stellar mass range for ICGs. We adopt either DESI Year-1 \textsc{Fastspecfit} (black dots) or \textsc{Cigale} (green triangles) stellar mass measurements to bin ICGs into different stellar mass bins. This plot is based on photometric companion counts around ICGs selected from DESI Year-1 BGS with extinction corrected $r$-band flux limit of $r_\mathrm{dustcorr}<19.5$. Errorbars are calculated from the 1-$\sigma$ scatter among 100 bootstrapped subsamples of the ICGs.}
\label{fig:LF}
\end{figure*}

\begin{figure}
\includegraphics[width=0.49\textwidth]{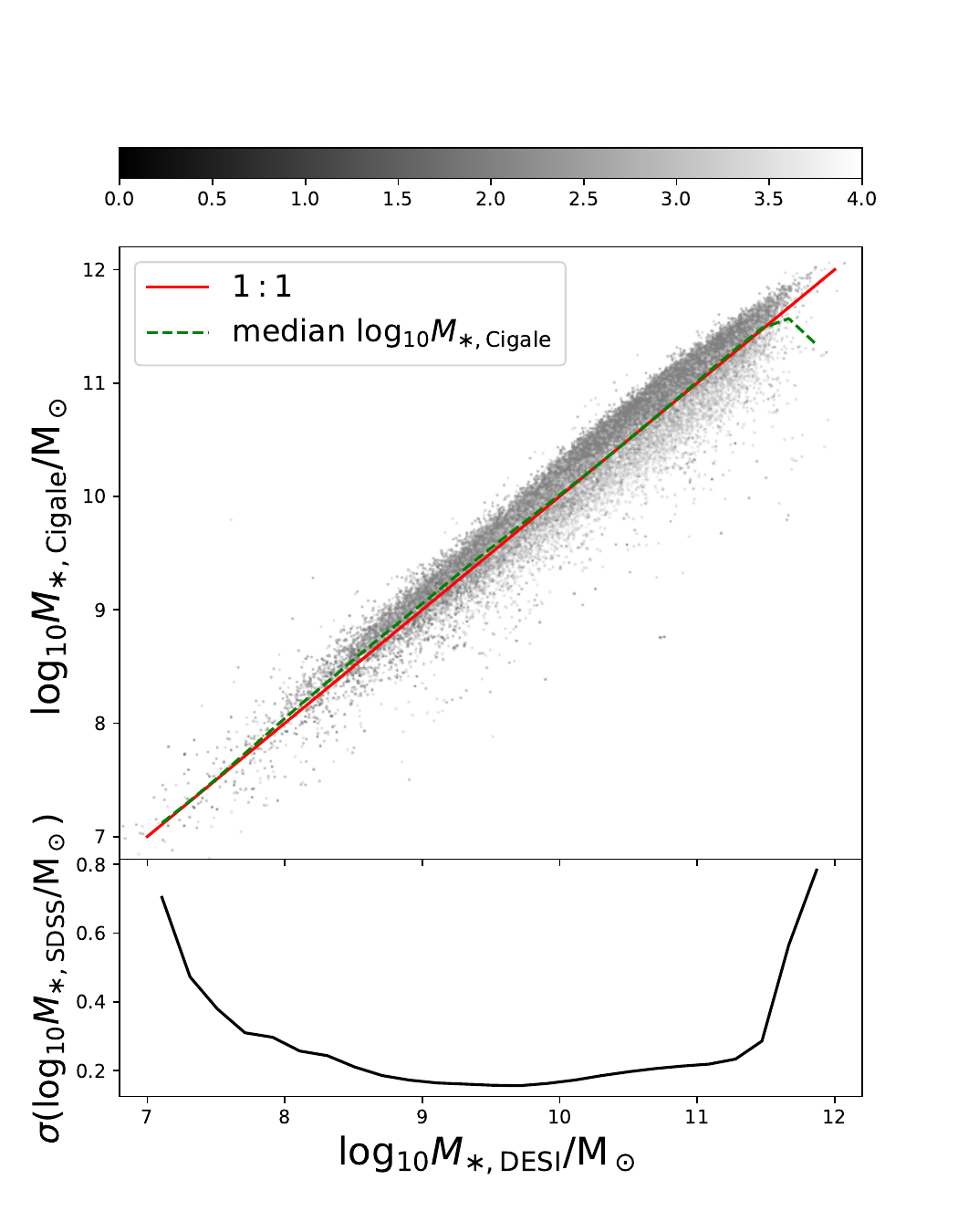}%
\caption{{\bf Top:} A comparison between DESI \textsc{cigale} ($y$-axis) and \textsc{Fastspecfit} ($x$-axis) stellar mass measurements. The red solid line marks $y=x$ to guide the eye. The green dashed curve marks the median of \textsc{cigale} stellar mass at fixed \textsc{Fastspecfit} stellar mass. The scatter points are based on a 1/100 random subsample, and are coded by in gray scale by the FLAGINFRARED column of the \textsc{Cigale} stellar mass catalog, which indicates the number of WISE W1-4 bands with S/N$\geq3$ (see the colorbar on top). {\bf Bottom:} The scatter in log \textsc{Cigale} stellar mass, as a function of DESI \textsc{Fastspecfit} stellar mass.}
\label{fig:fastspeccigalemass}
\end{figure}

\subsection{Comparison between satellite luminosity functions around ICGs from DESI and SDSS}
\label{sec:sdssdesicmp}

We first compare the satellite LF measurements, around ICGs selected from SDSS spectroscopic Main galaxies and from DESI Year-1 BGS. This not only helps to validate the robustness of our method, but also provides valuable information about the difference and similarity between the two independent spectroscopic galaxy surveys with large and comparable footprints. Note the photometric galaxies used for satellite counts are always from the DESI Legacy imaging Survey (DR9). So only the spectroscopic ICG sample differs between SDSS and DESI. 

The satellite LFs around ICGs in 12 different stellar mass bins of ICGs are shown in Figure~\ref{fig:LFcompare}. First of all, we would like to note that around more massive ICGs, their satellite LF stops at brighter magnitudes than those around smaller ICGs. This is because fainter satellites are complete at lower redshifts given the photometric flux limit (see Figure~\ref{fig:zlimit}), i.e., only nearby ICGs can contribute fainter satellite counts. At the more nearby Universe, the cosmic volume is small. The number density of the most massive galaxies in our Universe is much lower, and thus there are very limited numbers of massive ICGs that can contribute satellite counts in the local Universe, making the measurements stop at brighter magnitudes for satellites. On the other hand, smaller ICGs are more abundant, so their satellite LFs can be measured down to fainter magnitudes. This is mainly a selection effect given the survey flux limit.

The black dots with errorbars in each panel of Figure~\ref{fig:LFcompare} show the measurement around ICGs from DESI Year-1 BGS, but with a flux limit of $r<17.7$ to be consistent with the SDSS flux limit. The magenta squares with errorbars show the measurements around ICGs from SDSS. Here the stellar masses of SDSS ICGs have been corrected for the median bias from DESI. We have used the new stellar mass for SDSS ICG selection and then bin SDSS ICGs into different panels. 

There is no clear signal around ICGs smaller than $10^{7.1}\msun$, as the numbers of ICGs there are quite low. For the other panels, the black dots and magenta squares are mostly consistent with each other. In the two most massive panels with $\log_{10}M_{\ast,\mathrm{ICG}}/\msun>11.1$, however, the magenta squares are higher than black dots at fainter magnitudes. The difference is not large, but is significant compared with the small errorbars there. The same discrepancy also exists at the faint end in the third, fourth and fifth most massive panels covering $10.2<\log_{10}M_{\ast,\mathrm{ICG}}/\msun<11.1$, though less prominent. 

The discrepancy between DESI and SDSS in the few most massive panels may be due to the scatter in the stellar mass measurements. Despite the fact that we have corrected the median discrepancy in stellar mass between DESI and SDSS, the scatter remains. To demonstrate this, we match ICGs selected from DESI BGS to ICGs selected from SDSS Main galaxies, and we only use the matched sample to calculate the satellite LFs. Based on the matched sample, we repeat our calculations by using the stellar mass either from DESI or from SDSS, to bin ICGs into different stellar mass bins. The results are shown in the four most massive panels as green and blue curves, for DESI and SDSS stellar masses, respectively. Here the SDSS stellar masses are median corrected to DESI stellar masses. It is clear to see that the green curves (DESI stellar mass) are close to the magenta squares, which are based on the full sample of ICGs in DESI. On the other hand, the blue curves (SDSS stellar mass) are close to the black dots, which are based on the full sample of ICGs in SDSS. Since we only use the matched sample of ICGs, the only difference comes from the stellar mass, based on which ICGs are binned into different panels. This shows that the discrepancy between the satellite LFs around ICGs selected from DESI and SDSS in the few most massive panels is likely due to the difference in stellar mass.

We only show the blue and green curves based on the DESI-SDSS cross matched sample of ICGs in the four most massive panels of Figure~\ref{fig:LFcompare}, but not in the other panels, as the agreement between DESI and SDSS in most of the other panels are mostly good. We do not see prominent differences when different stellar masses are used based on the matched sample in other less massive panels. The effect due to the difference between DESI and SDSS stellar masses seem to mainly reveal in the two most massive panels. This is likely because the change in host halo mass and satellite abundance with the change in stellar mass is most rapid at the massive end \citep[e.g.][]{2010MNRAS.404.1111G,2021ApJ...919...25W}, so making the satellite LF measurements more sensitive to the difference in stellar mass. 

\begin{table*}
\caption{The number of ICGs selected from DESI Tier-1 BGS sample in the Year-1 observation and in three least massive stellar mass bins (first column). We fix the redshift range to be $z<0.01$ for all columns. The second and third columns are based on \textsc{Fastspecfit} stellar masses, with two different flux limits ($r_\mathrm{dustcorr}<19.5$ and $r<17.7$), while the fourth and fifth columns are based on \textsc{Cigale} stellar masses with the two different flux limits. No weights are applied to these numbers.
}
\begin{center}
\begin{tabular}{lcccc}\hline\hline
$\log M_{\ast,\mathrm{ICG}}/\msun$ & $N_\mathrm{ICG,DESI}(r_\mathrm{dustcorr}<19.5)$ & $N_\mathrm{ICG,DESI}(r<17.7)$ & $N_\mathrm{ICG,DESI}(r_\mathrm{dustcorr}<19.5)$ & $N_\mathrm{ICG,DESI}(r<17.7)$ \\
 & \textsc{Fastspecfit} & \textsc{Fastspecfit} & \textsc{Cigale} & \textsc{Cigale} \\\hline
7.1-7.8 & 385 & 329 & 422 & 334 \\
6.4-7.1 & 284 & 116 & 287 & 101 \\
5.7-6.4 & 91 & 23 & 89 & 14 \\ 
\hline
\label{tbl:num}
\end{tabular}
\end{center}
\end{table*}

\subsection{Satellite luminosity function based on DESI Year-1 data}

So far we have performed detailed comparisons between the satellite LF measurements, around ICGs selected from DESI and SDSS. The satellite LFs are mostly consistent. Starting from this subsection, we present satellite LF measurements based on ICGs selected from DESI BGS only. And instead of adopting a flux cut of $r<17.7$ as in the previous subsection to compare with SDSS, now all ICGs above the DESI Tier-1 BGS flux limit are used ($r_\mathrm{dustcorr}<19.5$). 

Black dots with errorbars in Figure~\ref{fig:LF} show the measured satellite LFs in 12 different stellar mass bins of ICGs, with the stellar mass measurements based on \textsc{Fastspecfit}, which is the same choice for measurements based on ICGs from DESI BGS in all previous plots. The $\log_{10} M_{\ast,\mathrm{ICG}}/\msun$ stellar mass range is indicated by the text in each panel. We will discuss the green triangles slightly later. The measurement can be achieved around ICGs with $\log_{10}M_{\ast,\mathrm{ICG}}/\msun>7.1$, reaching $M_{r,\mathrm{sat}}\sim-7$ at the faint end of satellite LF.

With Figures~\ref{fig:zhistall} and \ref{fig:zlimit} above, we have also discussed that only in the two least massive stellar mass bins ($6.4<\log_{10}M_{\ast,\mathrm{ICG}}/\msun<7.1$ and $5.7<\log_{10}M_{\ast,\mathrm{ICG}}/\msun<6.4$), there are somewhat more ICGs with $r_\mathrm{dustcorr}<19.5$ than those with $r<17.7$ at $z<\sim0.01$ in Figure~\ref{fig:zhistall}, so the deeper flux limit of DESI Tier-1 BGS may further help to improve the measurement there compared with the shallower flux limit of $r<17.7$. However, there is still no prominent improvement in the two least massive bin now with ICGs selected from DESI Tier-1 BGS sample in the Year-1 observation. We provide in the second and third columns of Table~\ref{tbl:num} the numbers of ICGs in the three least massive stellar mass bins of ICGs, with flux limits of $r_\mathrm{dustcorr}<19.5$ and $r<17.7$. With the deeper flux limit of $r_\mathrm{dustcorr}<19.5$ than $r<17.7$, the numbers of ICGs in the two least massive bins indeed increase, but the total number is still limited (284 and 91 at $6.4<\log_{10}M_{\ast,\mathrm{ICG}}/\msun<7.1$ and $5.7<\log_{10}M_{\ast,\mathrm{ICG}}/\msun<6.4$, respectively).

Moreover, the number of satellite galaxies of such low-mass ICGs is low, whereas for these very nearby low-mass ICGs, their fainter photometric companions are more significantly contaminated by background sources\footnote{Fainter apparent magnitudes or flux limits would result in the inclusion of significantly more faint background sources (see Section~\ref{sec:methods} for details).}. We have checked that for the bin of $7.1<\log_{10}M_{\ast,\mathrm{ICG}}/\msun<7.8$, the signal is only about 3-5\% of the background level. For the two least massive bins of $6.4<\log_{10}M_{\ast,\mathrm{ICG}}/\msun<7.1$ and $5.7<\log_{10}M_{\ast,\mathrm{ICG}}/\msun<6.4$, the background level is expected to be even more significantly larger than the signal. As a result, with limited number of ICGs, it is hard for us to extract the signals given such a large background level in the two lowest mass bins of ICGs.

The black dots in the third least massive panel ($7.1<\log_{10}M_{\ast,\mathrm{ICG}}/\msun<7.8$) look similar between Figure~\ref{fig:LFcompare} ($r<17.7$) and the current Figure~\ref{fig:LF} ($r_\mathrm{dustcorr}<19.5$), despite the difference in flux limits. This is not surprising, as we have already discussed with Figure~\ref{fig:zlimit} that to push fainter than $M_{r,\mathrm{sat}}=-10$, the satellite counts are complete only around ICGs with redshifts $z<0.01$. And for the bin of $7.1<\log_{10}M_{\ast,\mathrm{ICG}}/\msun<7.8$, the redshift distributions of ICGs with the two different flux limits of $r<17.7$ and $r_\mathrm{dustcorr}<19.5$ are similar at $z<0.01$.

\begin{figure*}
\includegraphics[width=0.99\textwidth]{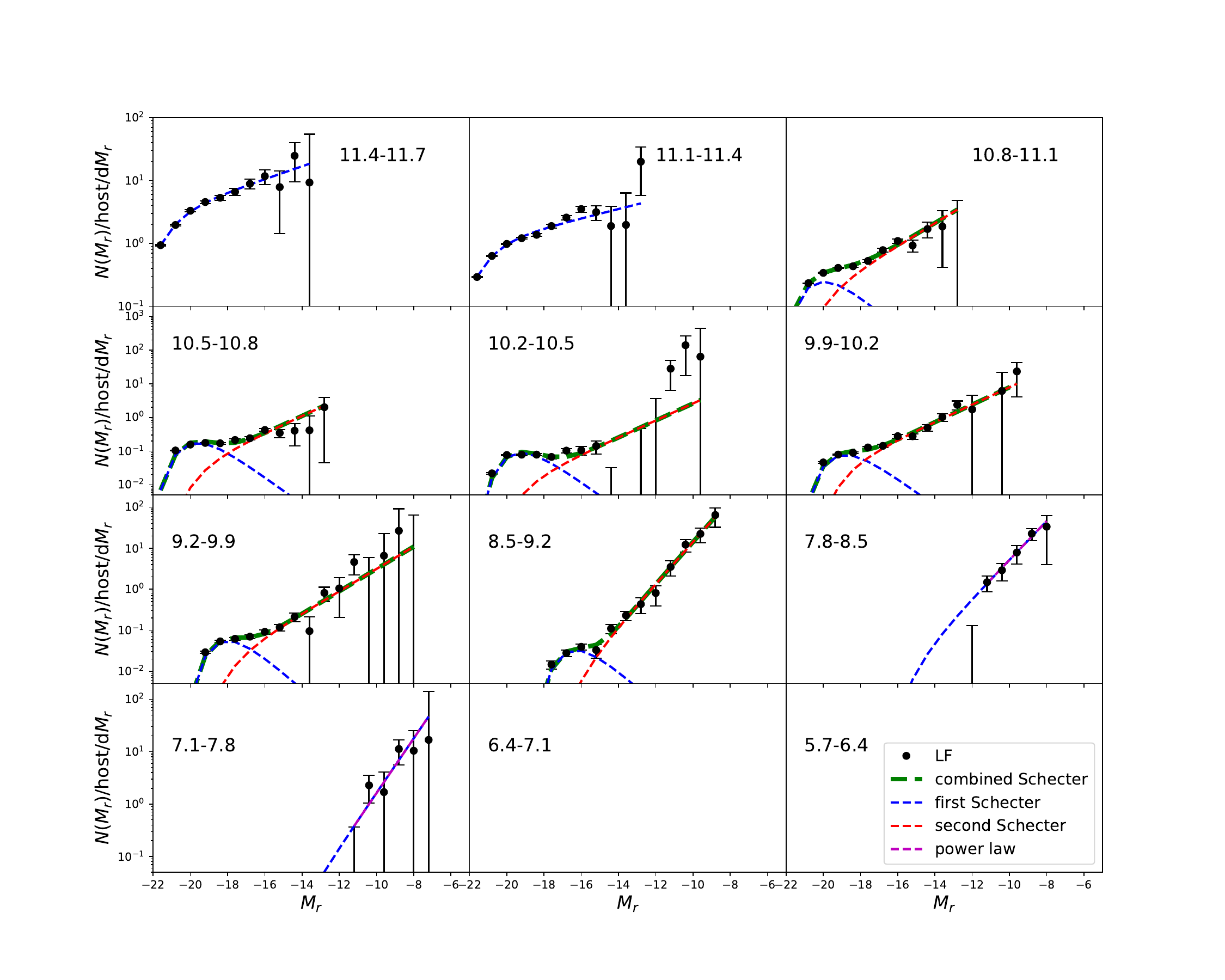}%
\caption{Black dots with errorbars are differential satellite LFs around ICGs selected from DESI Tier-1 BGS sample in the Year-1 observation. The text in each panel gives the $\log_{10} M_{\ast,\mathrm{ICG}}\msun$ stellar mass range for ICGs. We adopt single or double Schecter functions to fit the measured satellite LFs in each panel. When only single Schecter function is used, the best-fit is shown by the blue dashed curve. When we adopt double Schecter function, the first and second components are shown by the blue and red dashed curves, with the green curve showing the total best-fit model by combining the two components. The magenta dashed lines in the two least massive panels of $7.8<\log_{10}M_{\ast,\mathrm{ICG}}/\msun<8.5$ and $7.1<\log_{10}M_{\ast,\mathrm{ICG}}/\msun<7.8$ are single power law fits, which are almost identical to the faint end slopes of the single Schecter functions. }
\label{fig:schecterfit}
\end{figure*}

\begin{figure}
\includegraphics[width=0.49\textwidth]{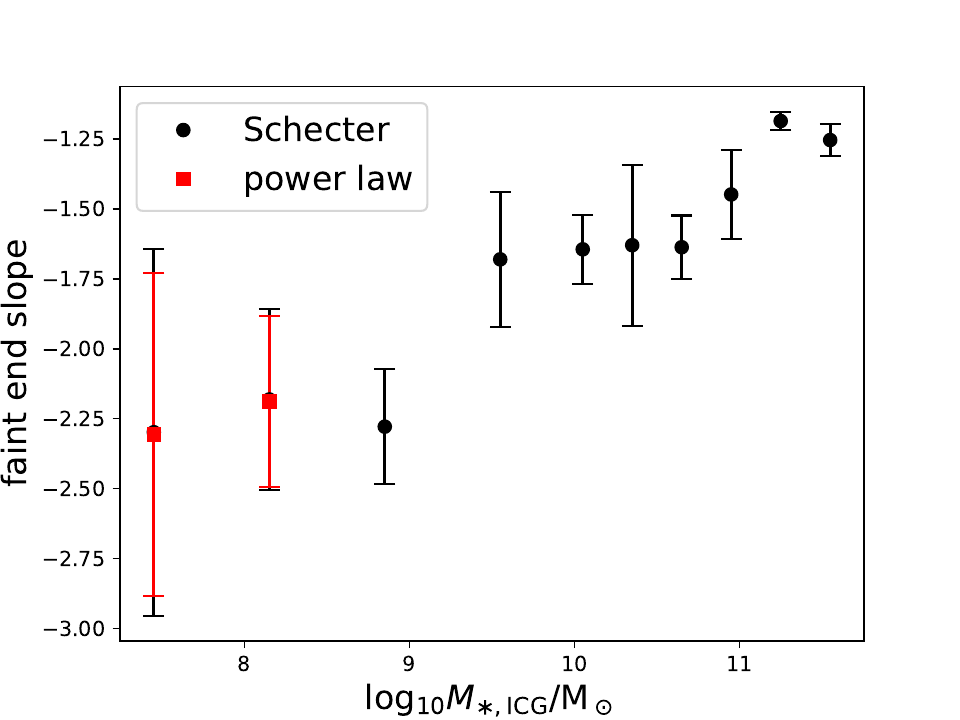}%
\caption{The best-fit faint end slope of satellite LFs versus the middle point ($\log_{10}M_{\ast,\mathrm{ICG}}/\msun$) for the stellar mass bin of ICGs. Black dots with errorbars are the best-fit slopes ($\beta$) from single or double Schecter functions. Two red squares show best fits assuming single power law functional form of $\log_{10}\Phi(M_r)=-0.4(\beta-1)M_r+C$.}
\label{fig:slope}
\end{figure}

\begin{table*}
\caption{The best-fit double Schecter function parameters to satellite LFs around ICGs in 10 stellar mass bins.}
\begin{center}
\begin{tabular}{lccccc}\hline\hline

$\log M_{\ast,\mathrm{ICG}}/\msun$ & $M_{r,0}$ & $\Phi_{\ast,1}$ & $\alpha$ & $\Phi_{\ast,2}$  & $\beta$  \\ \hline
11.4-11.7 & $-$21.48$\pm$0.15 & 3.171$\pm$0.530 & $-$1.254$\pm$0.058 & -   &  -  \\
11.1-11.4  &  $-$21.41$\pm$0.08  & 1.073$\pm$0.093 & -1.186$\pm$0.033 & -  & -  \\
10.8-11.1 &  $-$20.51$\pm$0.16  & 0.669$\pm$0.078 & $-$0.381$\pm$0.299 & 0.154$\pm$0.113 & $-$1.448$\pm$0.160  \\
10.5-10.8 &  $-$19.65$\pm$0.03  & 0.525$\pm$0.030 & 0.000$\pm$0.020 & 0.044$\pm$0.015 & $-$1.637$\pm$0.113 \\
10.2-10.5 &  $-$19.25$\pm$0.06  &  0.264$\pm$0.024  & 0.000$\pm$0.049  & 0.013$\pm$0.011 & $-$1.630$\pm$0.288 \\
9.9-10.2 &  $-$18.85$\pm$0.08  & 0.230$\pm$0.029 & 0.000$\pm$0.123  & 0.044$\pm$0.015 & $-$1.645$\pm$0.123\\
9.2-9.9  &  $-$18.02$\pm$0.12  & 0.162$\pm$0.027 & 0.000$\pm$0.312 & 0.022$\pm$0.014 & $-$1.681$\pm$0.242 \\
8.5-9.2 & $-$16.32 $\pm$0.32  & 0.096$\pm$0.036 & 0.000$\pm$1.878  & 0.009$\pm$0.008 & $-$2.270$\pm$0.207 \\
7.8-8.5 & $-$15.09 $\pm$8.94  & 0.068$\pm$1.183 & $-$2.182$\pm$0.323 & -   & - \\
7.1-7.8 & $-$15.01 $\pm$5.70  & 0.004$\pm$0.027 & $-$2.298$\pm$0.656 & -   & - \\
\hline
\label{tbl:LFpara}
\end{tabular}
\end{center}
\end{table*}

While all the DESI results presented so far above are based on \textsc{Fastspecfit} stellar mass to select ICGs and bin ICGs into different stellar mass bins. We also try a different stellar mass measurement in DESI for comparison, the \textsc{Cigale} stellar mass catalog \citep{2024A&A...691A.308S}. \textsc{Cigale} takes into account the modeling of AGN through the combination of DESI and WISE photometry. We show in Figure~\ref{fig:fastspeccigalemass} a comparison between \textsc{Fastspecfit} and \textsc{Cigale} stellar mass. As we can see, the scattered points go well through the red solid diagonal line, with a bit more points falling below the line, indicating reasonable agreement between the two different stellar mass measurements in DESI. The bottom panel shows the scatter of about 0.2~dex at most places. The scatter increases at the most and least massive ends, which is mainly because the data points more scattered below the diagonal line. The green solid curve in the top plot marks the median of \textsc{cigale} stellar mass at fixed \textsc{Fastspecfit} stellar mass, which is almost unbiased. We code the points in gray scale by the FLAGINFRARED column from the \textsc{Cigale} catalog, which is the WISE photometry quality, defined as the number of WISE W1-4 bands that have S/N$\geq$3. As is clearly shown, the deviation of \textsc{Cigale} stellar mass from \textsc{Fastspecfit} stellar mass demonstrates prominent dependence on FLAGINFRARED. This is consistent with Figure~C2 of \cite{2024A&A...691A.308S}. Because one of the main differences of \textsc{Cigale} from \textsc{Fastspecfit} is the modeling of AGN and the inclusion of WISE photometry, it is not surprising to see such a correlation between the stellar mass difference and FLAGINFRARED. When there is better WISE photometry (number of WISE bands of 4), most of the points falling below the diagonal line, i.e., the \textsc{Cigale} stellar masses are more under estimated due to the inclusion of more WISE bands. 

Green triangles with errorbars in Figure~\ref{fig:LF} show the satellite LF measurements around ICGs binned according to their \textsc{Cigale} stellar mass. Here we choose not to do any corrections to either \textsc{Fastspecfit} or \textsc{Cigale} stellar masses, as we cannot tell which one is closer to the ground truth. Encouragingly, there are very good agreements between the black and green symbols or curves, indicating robustness in our satellite LF measurements, despite the uncertainties in the stellar mass measurements. 

For the bins over $5.7<\log_{10}M_{\ast,\mathrm{ICG}}/\msun<7.8$, the numbers of ICGs binned according to \textsc{Cigale} stellar masses are similar (see  the fourth and fifth columns of Table~\ref{tbl:num}). We still have no robust measurements in the two least massive panels with \textsc{Cigale}. Notably, our criteria adopted to select ICGs (see Section~\ref{sec:data}) depend on stellar mass to infer the virial radius and velocity, and thus the change in stellar mass not only affects how we bin ICGs into different stellar masses, but also affect the selection of ICGs. However, we do have tested by just using ICGs selected above with \textsc{Fastspecfit} stellar mass, and bin them according to \textsc{Cigale} stellar mass. There is no prominent change in our measurements.

\subsection{The faint end slopes of satellite LF}

According to Figure~\ref{fig:LF}, it seems there is a trend for the faint slopes of satellite LFs to become steeper with the decrease in the stellar mass range of host ICGs. To more qualitatively evaluate the trend, we perform single or double Schecter function fits to the differential satellite LFs shown in Figure~\ref{fig:schecterfit}, with the best fits demonstrated by the dashed curves. The double Schecter function takes the following form:

\begin{equation}
    \Phi(L)\ud L= \left\{\Phi_{\ast,1} \left[ \frac{L}{L_0} \right]^\alpha +  \Phi_{\ast,2}\left[ \frac{L}{L_0} \right]^\beta \right\}\exp{\left(-\frac{L}{L_0}\right)}\ud L,
    \label{eqn:LF}
\end{equation}
and because the relation between luminosity and absolute magnitude is $\frac{L}{L_0}=10^{-0.4(M-M_0)}$, 
Equation~\ref{eqn:LF} can be expressed in terms of $r$-band absolute magnitude $M_r$ as
\begin{align}
    \Phi(M_r)\ud M_r &= 0.4\ln{10} \times \nonumber \\ 
    &[\Phi_{\ast,1} 10^{-0.4(M_r-M_{r,0})(\alpha+1)}\nonumber\\
    &+\Phi_{\ast,2} 10^{-0.4(M_r-M_{r,0})(\beta+1)} ]\nonumber  \\
    &\times \exp{\left[-10^{-0.4(M_r-M_{r,0})}\right]} \ud M_r. 
    \label{eqn:schecterm}
\end{align}

In Equation~\ref{eqn:schecterm} above, $M_{r,0}$ is the characteristic magnitude. $\Phi_{\ast,1}$ and $\Phi_{\ast,2}$ are the normalizations for the two components, and $\alpha$ and $\beta$ are the faint end slopes of the two components. 

When we adopt the single Schecter function for fitting, the corresponding functional form is equivalent to simply force $\Phi_{\ast,2}$ to zero. Explicitly, we adopt single Schecter function for the two most massive panels ($11.4<\log_{10}M_{\ast,\mathrm{ICG}}/\msun<11.7$ and $11.1<\log_{10}M_{\ast,\mathrm{ICG}}/\msun<11.4$), and the two panels of $7.8<\log_{10}M_{\ast,\mathrm{ICG}}/\msun<8.5$ and $7.1<\log_{10}M_{\ast,\mathrm{ICG}}/\msun<7.8$. For the two most massive bins, we have tested that using single or double Schecter functions do not bring significant differences at all, with the secondary components significantly lower in amplitudes. For the other two less massive panels, there are much less data points, so impossible to constrain all six free parameters. At the faint end, the behaviour is well described by a single power law. The inclusion of a second component does not improve the fit. In the other panels, double Schecter functions are adopted, with the second fainter component demonstrated by the red dashed curves in corresponding panels of Figure~\ref{fig:schecterfit}. The best-fit model parameters are provided in Table~\ref{tbl:LFpara}.

In addition to using Schecter functions, we notice that the number of data points for the few least massive panels is comparable to the number of free paremeters in the Schecter function, so in order to ensure the robustness in our estimates of the faint end slopes, we also try to fit the data points in the panels of $7.8<\log_{10}M_{\ast,\mathrm{ICG}}/\msun<8.5$ and $7.1<\log_{10}M_{\ast,\mathrm{ICG}}/\msun<7.8$ using a single power law functional form of $\log_{10}\Phi(M_r)=-0.4(\beta-1)M_r+C$. The best fits are shown by the magenta dashed curves, which are almost identical to the faint end slopes by the best-fit Schecter functions. The best-fit parameters are $C=5.47\pm1.20$ and $\beta=-2.189\pm0.306$ for $7.8<\log_{10}M_{\ast,\mathrm{ICG}}/\msun<8.5$, and $C=5.44\pm2.12$ and $\beta=-2.307\pm0.577$ for $7.1<\log_{10}M_{\ast,\mathrm{ICG}}/\msun<7.8$.

We further show in Figure~\ref{fig:slope} the faint end slopes of satellite LFs versus the middle point of the log stellar mass ranges for ICGs. The faint end slopes are taken as $\beta$ from the slope of the second component if double Schecter function is adopted for fitting, while are taken as $\alpha$ when single Schecter function is adopted. For the two least massive points, we also show the faint end slopes of the best-fit single power law function as red squares. The best fits achieved by using Schecter functions or by using single power law functions give almost identical faint end slopes, with the Schecter function fitting gives slightly larger errorbars. Interestingly, we clearly see the trend that the faint end slopes of the satellite LFs become steeper with the decrease in the stellar mass of host ICGs. 

\subsection{Satellite stellar mass function and the low-mass end slope}

\begin{figure*}
\includegraphics[width=0.99\textwidth]{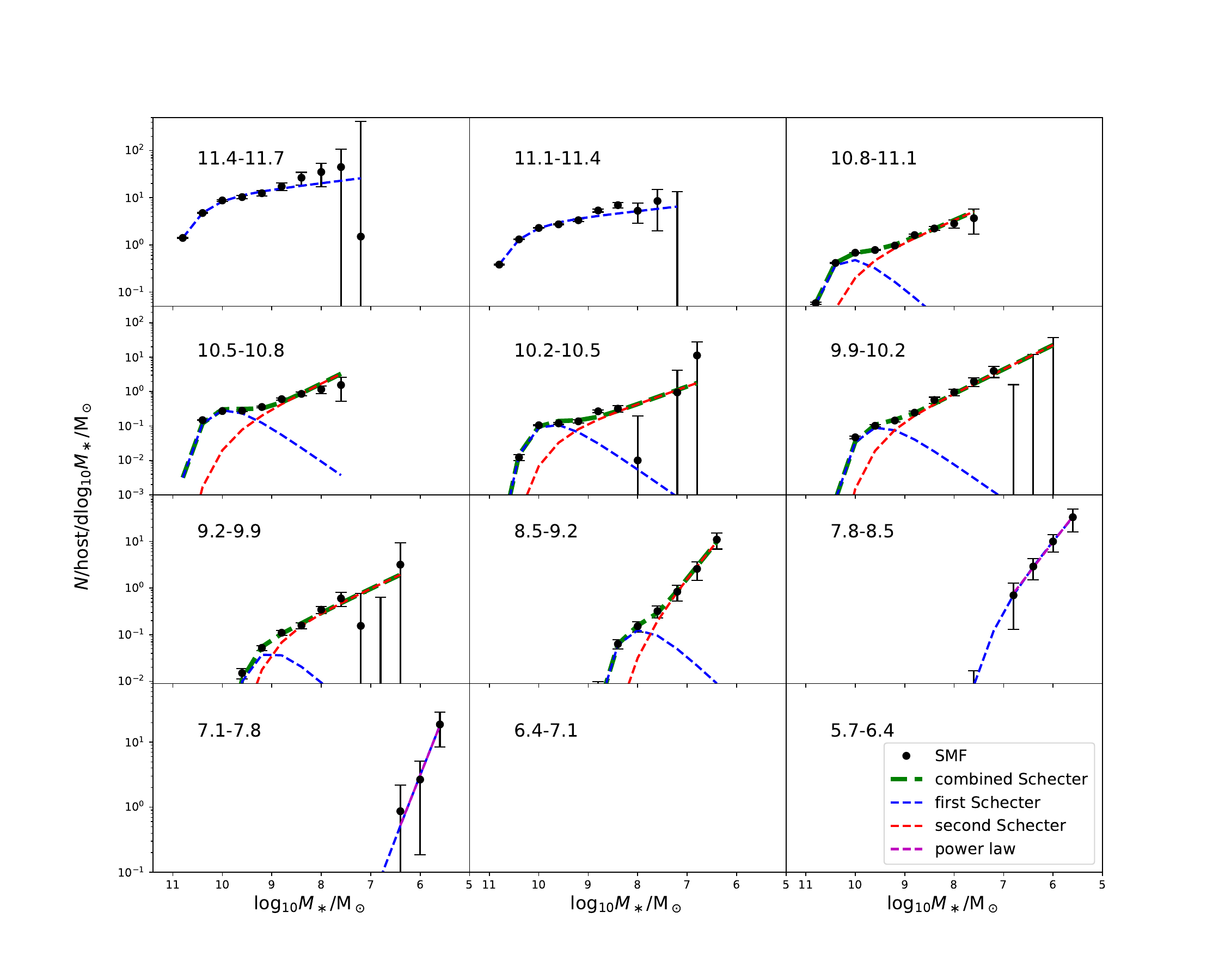}%
\caption{Black dots with errorbars are differential satellite SMFs around ICGs selected from DESI Tier-1 BGS sample in the Year-1 observation. The text in each panel gives the $\log_{10} M_{\ast,\mathrm{ICG}}\msun$ stellar mass range for ICGs. We adopt single or double Schecter functions to fit the measured satellite LFs in each panel. When only single Schecter function is used, the best-fit is shown by the blue dashed curve. When we adopt double Schecter function, the first and second components are shown by the blue and red dashed curves, with the green curve showing the total best-fit model by combining the two components. The magenta dashed lines in the two least massive panels of $7.8<\log_{10}M_{\ast,\mathrm{ICG}}/\msun<8.5$ and $7.1<\log_{10}M_{\ast,\mathrm{ICG}}/\msun<7.8$ are single power law fits, which give almost identical faint end slope as that of the single Schecter function.}
\label{fig:schecterfitMF}
\end{figure*}

\begin{figure}
\includegraphics[width=0.49\textwidth]{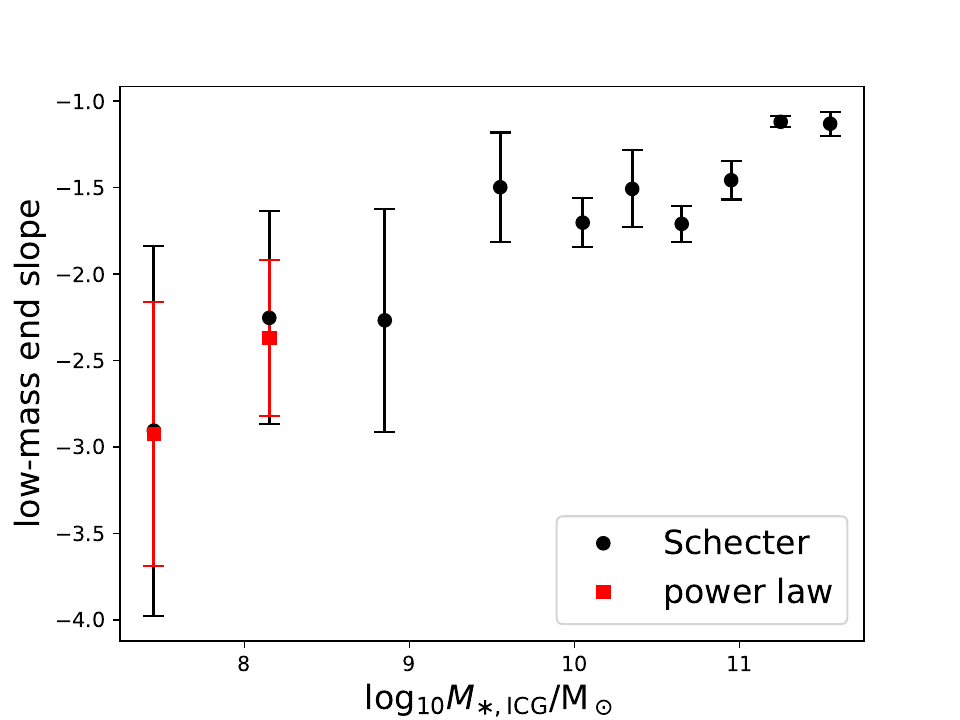}%
\caption{Similar to Figure~\ref{fig:slope}, but shows the low-mass end slopes of satellite SMFs. Black dots with errorbars are the best-fit slopes ($\beta$) from single or double Schecter functions. Two red squares show best fits assuming single power law functional form of $\log_{10}\Phi(\log_{10}M_\ast)=(\beta-1)\log_{10}M_\ast+C$.}
\label{fig:slopeMF}
\end{figure}

\begin{table*}
\caption{The best-fit double Schecter function parameters to satellite SMFs around ICGs in 10 stellar mass bins. }
\begin{center}
\begin{tabular}{lccccc}\hline\hline

$\log M_{\ast,\mathrm{ICG}}/\msun$ & $\log_{10}M_{\ast,0}$ & $\Phi_{\ast,1}$ & $\alpha$ & $\Phi_{\ast,2}$  & $\beta$  \\ \hline
11.4-11.7 & 10.54$\pm$0.04  &  4.133$\pm$0.587  & $-$1.130$\pm$0.070 &  -  &  -  \\
11.1-11.4  &  10.54$\pm$0.02  &  1.129$\pm$0.076   & $-$1.119$\pm$0.032 & -  & -  \\
10.8-11.1 &  10.14$\pm$0.04  &   0.574$\pm$0.061   &   $-$0.092$\pm$0.234   &  0.153$\pm$0.063  &  $-$1.457$\pm$0.112 \\
10.5-10.8 &  9.92$\pm$0.01    &    0.343$\pm$0.021   &    $-$0.000$\pm$0.020    &    0.032$\pm$0.010   &    $-$1.710$\pm$0.104 \\
10.2-10.5 &  9.74$\pm$0.04   &     0.132$\pm$0.023   &    $-$0.000$\pm$0.153   &     0.025$\pm$0.014   &    $-$1.507$\pm$0.222 \\
9.9-10.2 &  9.50$\pm$0.06   &     0.108$\pm$0.026   &    $-$0.000$\pm$0.248   &     0.034$\pm$0.013  &   $-$1.702$\pm$0.140  \\
9.2-9.9  &  9.03$\pm$0.16   &     0.048$\pm$0.041  &     $-$0.000$\pm$1.923   &     0.041$\pm$0.026  &   $-$1.497$\pm$0.317  \\
8.5-9.2 &   8.05$\pm$1.07   &     0.147$\pm$0.193  &   $-$0.383$\pm$1.191   &     0.027$\pm$0.306   &    $-$2.351$\pm$1.028   \\
7.8-8.5 &  7.20$\pm$1.57   &     0.147$\pm$1.421   &   $-$2.253$\pm$0.617   &     -   &    -  \\
7.1-7.8 &   7.40$\pm$2.17    &    0.003$\pm$0.135   &    $-$2.888$\pm$0.916   &     -  &      -  \\  
\hline
\label{tbl:MFpara}
\end{tabular}
\end{center}
\end{table*}

In this last subsection, we present our measurements of the SMFs of satellite galaxies. The readers can check Section~\ref{sec:methods} about details on how we estimate the stellar  masses of satellite galaxies. 

Figure~\ref{fig:schecterfitMF} is similar to Figure~\ref{fig:schecterfit}, but the black dots with errorbars show the differential SMFs of satellites. We can measure the SMF of satellites down to $\log_{10}M_{\ast,\mathrm{sat}}/\msun\sim5.5$ around ICGs with $7.8<\log_{10}M_{\ast,\mathrm{ICG}}/\msun<8.5$ and $7.1<\log_{10}M_{\ast,\mathrm{ICG}}/\msun<7.8$. 

We again perform single or double Schecter function fitting to the measurements, based on Equation~\ref{eqn:schectermass} below
\begin{align}
    \Phi(\log_{10}M_\ast)\ud \log_{10}M_\ast &= \ln{10} \nonumber \\ 
    &\hspace{-2em}\times [\Phi_{\ast,1} 10^{(\log_{10}M_\ast-\log_{10}M_{\ast,0})(\alpha+1)} \nonumber \\
    &+\Phi_{\ast,2} 10^{(\log_{10}M_\ast-\log_{10}M_{\ast,0})(\beta+1)} ]  \nonumber \\
    &\times \exp{[-10^{(\log_{10}M_\ast-\log_{10}M_{\ast,0})}]} \nonumber \\
    &\ud \log_{10}M_\ast. 
    \label{eqn:schectermass}
\end{align}

Similar to Figure~\ref{fig:schecterfit}, single Schecter function is adopted for the two most massive bins and bins of $7.8<\log_{10}M_{\ast,\mathrm{ICG}}/\msun<8.5$ and $7.1<\log_{10}M_{\ast,\mathrm{ICG}}/\msun<7.8$. We provide the best-fitting parameters in Table~\ref{tbl:MFpara}. We also try to fit the data points in panels of $7.8<\log_{10}M_{\ast,\mathrm{ICG}}/\msun<8.5$ and $7.1<\log_{10}M_{\ast,\mathrm{ICG}}/\msun<7.8$ using a single power law functional form of $\log_{10}\Phi(\log_{10}M_\ast)=(\beta-1)\log_{10}M_\ast+C$. The best fits are shown by the magenta dashed curves in the corresponding panels, which again give almost identical faint end slopes as those by the best-fit Schecter functions. The best-fit parameters are $C=9.21\pm2.75$ and $\beta=-2.370\pm0.451$ for $7.8<\log_{10}M_{\ast,\mathrm{ICG}}/\msun<8.5$, and $C=12.03\pm4.37$ and $\beta=-2.934\pm0.765$ for $7.1<\log_{10}M_{\ast,\mathrm{ICG}}/\msun<7.8$.

We plot the low-mass end slopes versus the middle point of the log stellar mass of host ICGs in Figure~\ref{fig:slopeMF}, and we still see the trend that the faint end slopes of satellite SMFs get steeper with the decrease in the stellar mass of host ICGs. 

\section{Discussions}

As we have mentioned at the beginning of Section~\ref{sec:sdssdesicmp}, due to the selection effect given the photometric flux limit of counting satellites (see Section~\ref{sec:photosample}), we can measure satellite LF/SMF down to fainter/smaller ranges only around smaller ICGs. Thus the fact that the best-fit faint end or low-mass end slopes of satellite LF or SMF get steeper with the decrease in the stellar mass of host ICGs, is also telling us that there is an upturn for fainter/smaller satellite counts.

Whether there is an upturn in the measured galaxy LF/SMF, and whether the upturn is in agreement with $\Lambda$CDM predictions, have long been under debates \citep[e.g.][]{2005ApJ...631..208B,2012MNRAS.420.1239L}, with some studies in the past claiming the existence of an upturn or change in slope for the LF of field galaxies or galaxies in dense cluster environments \citep[e.g.][]{1994MNRAS.268..393D,1995ApJ...450..534D,2005A&A...433..415P,2005nfcd.conf..346P,2007ApJ...671.1471B,2007ApJ...666..846J,2007AJ....133..177M,2010ApJ...721L..14B,2011MNRAS.413.1333W,2014MNRAS.444L..34A,2015A&A...581A..11M,2016MNRAS.459.3998L}, and some studies disagree on this \citep[e.g.][]{2008AJ....135.1837R,2009AJ....137.3091H}. Hot debates also rely on observations in our MW and the Local Group \citep[e.g.][]{1999ApJ...522...82K,1999ApJ...524L..19M}, showing that the observed number of low-mass satellites is much smaller than the predicted number of low-mass subhalos in $\Lambda$CDM, though nowadays more and more low-mass satellites have been detected around our MW, which significantly weakens the tension.

If there exists the upturn at the faint end of galaxy LFs, this would bring better agreement with $\Lambda$CDM predictions. On the other hand, and to address the issue of less low-mass satellites observed in our MW than CDM theory, some studies invoke warm dark matter model, which predicts much less surviving small substructures 
\citep[e.g.][]{2014MNRAS.439..300L}, or try to explain through baryonic physics under the CDM framework \cite[e.g.][]{2000ApJ...539..517B,2002MNRAS.333..156B}, such as photoionization process during reionization and supernova 
feedback, which inhibit the star formation in small haloes, \citep[e.g.][]{2000ApJ...539..517B,2002MNRAS.333..156B,
2002MNRAS.333..177B,2002ApJ...572L..23S}, predicting that a significant number of small subhalos do not 
host a galaxy. Our measurements of the faint end and low-mass end slopes thus offer valuable information to be compared with the theory, and help to address critical questions about the nature of dark matter, the reionization process, and also the formation of dwarf galaxies.

In the following, we further discuss two more recent studies on galaxy LF and SMF, that are also based on DESI Year-1 data and have pointed out the existence of low-mass end upturns: 

Based on a similar method of cross correlating Year-1 DESI bright galaxies and DESI Legacy Survey photometric galaxies, a recent study of \cite{Xu2025} has managed to measure the global SMFs of galaxies down to very small masses, which is $10^{5.3}$ for blue galaxies and $10^{6.3}$ for red galaxies. The mass range that can be achieved by \cite{Xu2025} is comparable to ours. The main difference of the science between the current study and \cite{Xu2025} is that: 1) \cite{Xu2025} calculate the global SMF, instead of for satellites; 2) To have enough signal, we do not divide our galaxies by color. 

Interestingly, \cite{Xu2025} find that the low-mass end slopes in their measurements are $-$1.54 and $-$2.5 for blue and red galaxies. There is a strong steepening in the low-mass end slope for red galaxies smaller than $10^{8.5}\msun$. For our satellite SMFs, the steepest slope is $-$2.888$\pm$0.916, which is consistent with the slope for red galaxies of \cite{Xu2025} within the errorbars. 

The steepening of the faint end slopes of galaxy conditional luminosity functions (CLFs), conditional stellar mass functions (CSMFs)\footnote{The CLF or CSMF describes the average number of galaxies as a function of galaxy luminosity or stellar mass in the dark matter halo of a given mass.} and global LFs and SMFs have also been identified in a recent study of \cite{2024ApJ...971..119W}. \cite{2024ApJ...971..119W} adopted galaxy group catalogs constructed from DESI Year-1 and SV observations \citep{2021ApJ...909..143Y} for their measurements. Upturns in the CLF or CSMF, with slopes of about $\beta\sim-1.85$, and upturns in the global LF or SMF, with slopes of about $\beta\sim-1.93$, is reported for stellar mass smaller than $\sim10^9\mathrm{h}^{-2}\msun$ or luminosity fainter than $\sim10^9\mathrm{h}^{-2}\mathrm{L_\odot}$. 

The faint end or low-mass end slopes of satellite LFs, as we have discussed above, are mainly contributed by counts around more nearby ICGs (see Section~\ref{sec:expect} and Figure~\ref{fig:zlimit}). However, it is known that the more local Universe is an under-dense region compared with the more distant Universe, called local void \citep[e.g.][]{1988ngc..book.....T,2010Natur.465..565P,2014MNRAS.441..933X,2019ApJ...872..180C,2021MNRAS.500.3776W}. Based on the constrained ELUCID simulation that resembles the real Universe \citep{2014ApJ...794...94W,2016ApJ...831..164W,2018ApJ...860...30Y,2022ApJ...936...11L,2024ApJ...966..236L}, it has been shown by \cite{2019ApJ...872..180C} that the global SMF of galaxies measured at $z<\sim0.03$ can have significantly underestimated low-mass end slopes. The measured slopes of the global LF and SMF by \cite{2024ApJ...971..119W} as mentioned above has in fact been corrected for the bias of the Local Volume from the more distant Universe. 

We did not include such a correction in our analysis, because the correction by \cite{2019ApJ...872..180C} stops at $z<\sim0.125$, that is affected by the resolution limit of the ELUCID simulation, though \cite{2024ApJ...971..119W} still incorporated this correction down to smaller mass ranges with extrapolations. Moreover, the corrections by \cite{2019ApJ...872..180C} are for global SMFs, and it is unclear whether similar amount of corrections are supposed to be made to satellite LFs and SMFs. Thus it is hard to judge whether such corrections have to be included for our measurements. So we leave our results without such corrections.

\section{Conclusions}
\label{sec:concl}

In this study, we select isolated central galaxies (ICGs) from the DESI Year-1 BGS sample. We limit our selections to the Tier-1 BGS sample, which is nearly volume limited above the flux limit of $r_\mathrm{dustcorr}<19.5$ after Galactic extinction correction. Photometric companions from the DESI Legacy imaging Survey with flux limit of $r<22.5$ are then counted around the selected ICGs in different stellar mass bins, following the method of \cite{2012MNRAS.424.2574W} to get LF and SMF of satellite galaxies and subtract background source contaminations. 

We first perform detailed comparisons with photometric satellite counts around ICGs selected from SDSS spectroscopic Main galaxies. For SDSS, the photometric companions are still from the DESI Legacy imaging Survey with the same flux limit of $r<22.5$. We achieve reasonable agreements between SDSS and DESI for ICGs with $7.1<\log_{10}M_{\ast,\mathrm{ICG}}/\msun<11.7$. For the most massive bins with $\log_{10}M_{\ast,\mathrm{ICG}}/\msun>11.1$, there are some disagreements between SDSS and DESI at fainter magnitudes of satellites, and the discrepancy is likely due to the difference in SDSS and DESI stellar mass measurements.

With ICGs selected from DESI BGS, we can measure their satellite luminosity functions (LFs) around ICGs spanning a wide range in stellar mass of $7.1<\log_{10}M_{\ast,\mathrm{ICG}}/\msun<11.7$. To push down to satellite LFs fainter than $r$-band absolute magnitudes of $M_{r,\mathrm{sat}}\sim-10$, only nearby galaxy systems with redshifts lower than $z=0.01$ can provide complete satellite counts above the photometric flux limit. And we can push down to $M_{r,\mathrm{sat}}\sim-7$ around ICGs in the stellar mass range of $7.1<\log_{10}M_{\ast,\mathrm{ICG}}/\msun<7.8$.  The DESI \textsc{Fastspecfit} and \textsc{Cigale} stellar masses lead to consistent results over the full stellar mass range of ICGs probed. 

We also measure the SMFs of satellite galaxies, and we can push down to about $10^{5.5}\msun$ in the measured mass functions around ICGs in the stellar mass range of $7.1<\log_{10}M_{\ast,\mathrm{ICG}}/\msun<7.8$.

We adopt single or double Schecter functions to fit the measured satellite LFs or SMFs around ICGs spanning different stellar mass ranges. Interestingly, we have discovered that the faint end or low-mass end slopes of the measured satellite LFs or SMFs clearly decrease with the decrease in stellar masses of the host ICGs, with smaller and nearby host ICGs capable of being used to probe fainter satellites. The steepest slopes can be $-2.298\pm0.656$ and $-$2.888$\pm$0.916 for satellite LF and SMF, respectively. The slopes are all shallower than $-2$ for satellites around ICGs more massive than $10^9\msun$.

\acknowledgments

This work is supported by NSFC (12273021,12022307,\\
12133006) and the National Key R\&D Program of
China (2023YFA1605600, 2023YFA1605601). We thank the sponsorship from Yangyang Development Fund. 
The computations of this work are carried on the Gravity supercomputer at the Department of Astronomy, Shanghai Jiao Tong University and on the National Energy Research Scientific Computing Center (NERSC). 

M.S. acknowledges support by the State Research Agency of the Spanish Ministry of Science and Innovation under the grants `Galaxy Evolution with Artificial Intelligence' (PGC2018-100852-A-I00) and 'BASALT' (PID2021-126838NB-I00) and the Polish National Agency for Academic Exchange (Bekker grant BPN/BEK/2021/1/00298/DEC/1). This work was partially supported by the European Union's Horizon 2020 Research and Innovation program under the Maria Sklodowska-Curie grant agreement (No. 754510).

J.M. gratefully acknowledges funding support from the U.S. Department of Energy, Office of Science, Office of High Energy Physics under Award Number DE-SC0020086.

S.K. acknowledges support from the Science \& Technology Facilities Council (STFC) grant ST/Y001001/1.

W.W. is grateful for useful discussions and helps by Yangyao Chen.

We are grateful for the efforts spent on checking and coordinating our paper by the DESI publication handler, David Parkinson, and the DESI GQP working group chair, Rita Tojeiro. 

This paper is based upon work supported by the U.S. Department of Energy (DOE), Office of Science, Office of High-Energy Physics, under Contract No. DE–AC02–05CH11231, and by the National Energy Research Scientific Computing Center, a DOE Office of Science User Facility under the same contract. Additional support for DESI was provided by the U.S. National Science Foundation (NSF), Division of Astronomical Sciences under Contract No. AST-0950945 to the NSF’s National Optical-Infrared Astronomy Research Laboratory; the Science and Technology Facilities Council of the United Kingdom; the Gordon and Betty Moore Foundation; the Heising-Simons Foundation; the French Alternative Energies and Atomic Energy Commission (CEA); the National Council of Humanities, Science and Technology of Mexico (CONAHCYT); the Ministry of Science, Innovation and Universities of Spain (MICIU/AEI/10.13039/501100011033), and by the DESI Member Institutions: \url{https://www.desi.lbl.gov/collaborating-institutions}. Any opinions, findings, and conclusions or recommendations expressed in this material are those of the author(s) and do not necessarily reflect the views of the U. S. National Science Foundation, the U. S. Department of Energy, or any of the listed funding agencies.

The authors are honored to be permitted to conduct scientific research on Iolkam Du’ag (Kitt Peak), a mountain with particular significance to the Tohono O’odham Nation.

For the purpose of open access, the author has applied a Creative
Commons Attribution (CC BY) licence to any Author Accepted
Manuscript version arising from this submission.

The measurements presented in this paper can be accessed at \url{https://zenodo.org/records/15221360}, which contains data points for figures presented in this work.

\bibliography{paper}{}

\begin{thebibliography}{}
\expandafter\ifx\csname natexlab\endcsname\relax\def\natexlab#1{#1}\fi
\providecommand{\url}[1]{\href{#1}{#1}}
\providecommand{\dodoi}[1]{doi:~\href{http://doi.org/#1}{\nolinkurl{#1}}}
\providecommand{\doeprint}[1]{\href{http://ascl.net/#1}{\nolinkurl{http://ascl.net/#1}}}
\providecommand{\doarXiv}[1]{\href{https://arxiv.org/abs/#1}{\nolinkurl{https://arxiv.org/abs/#1}}}

\bibitem[{{Abazajian} {et~al.}(2009){Abazajian}, {Adelman-McCarthy}, {Ag{\"u}eros}, {Allam}, {Allende Prieto}, {An}, {Anderson}, {Anderson}, {Annis}, {Bahcall}, \& et~al.}]{2009ApJS..182..543A}
{Abazajian}, K.~N., {Adelman-McCarthy}, J.~K., {Ag{\"u}eros}, M.~A., {et~al.} 2009, \apjs, 182, 543, \dodoi{10.1088/0067-0049/182/2/543}

\bibitem[{{Agulli} {et~al.}(2014){Agulli}, {Aguerri}, {Sanchez-Janssen}, {Barrena}, {Diaferio}, {Serra}, \& {Mendez-Abreu}}]{2014MNRAS.444L..34A}
{Agulli}, I., {Aguerri}, J.~A.~L., {Sanchez-Janssen}, R., {et~al.} 2014, \mnras, 444, L34, \dodoi{10.1093/mnrasl/slu108}

\bibitem[{{Alonso} {et~al.}(2023){Alonso}, {Wang}, {Zhang}, {Li}, {Shao}, {Guo}, {He}, {Hao}, \& {Shi}}]{2023ApJ...947...19A}
{Alonso}, P., {Wang}, W., {Zhang}, J., {et~al.} 2023, \apj, 947, 19, \dodoi{10.3847/1538-4357/acbf4a}

\bibitem[{{Ba{\~n}ados} {et~al.}(2010){Ba{\~n}ados}, {Hung}, {De Propris}, \& {West}}]{2010ApJ...721L..14B}
{Ba{\~n}ados}, E., {Hung}, L.-W., {De Propris}, R., \& {West}, M.~J. 2010, \apjl, 721, L14, \dodoi{10.1088/2041-8205/721/1/L14}

\bibitem[{{Barkhouse} {et~al.}(2007){Barkhouse}, {Yee}, \& {L{\'o}pez-Cruz}}]{2007ApJ...671.1471B}
{Barkhouse}, W.~A., {Yee}, H.~K.~C., \& {L{\'o}pez-Cruz}, O. 2007, \apj, 671, 1471, \dodoi{10.1086/523257}

\bibitem[{{Benson} {et~al.}(2002{\natexlab{a}}){Benson}, {Frenk}, {Lacey}, {Baugh}, \& {Cole}}]{2002MNRAS.333..177B}
{Benson}, A.~J., {Frenk}, C.~S., {Lacey}, C.~G., {Baugh}, C.~M., \& {Cole}, S. 2002{\natexlab{a}}, \mnras, 333, 177, \dodoi{10.1046/j.1365-8711.2002.05388.x}

\bibitem[{{Benson} {et~al.}(2002{\natexlab{b}}){Benson}, {Lacey}, {Baugh}, {Cole}, \& {Frenk}}]{2002MNRAS.333..156B}
{Benson}, A.~J., {Lacey}, C.~G., {Baugh}, C.~M., {Cole}, S., \& {Frenk}, C.~S. 2002{\natexlab{b}}, \mnras, 333, 156, \dodoi{10.1046/j.1365-8711.2002.05387.x}

\bibitem[{{Bianchi} {et~al.}(2024){Bianchi}, {Hanif}, {Carnero Rosell}, {Lasker}, {Ross}, {Pinon}, {de Mattia}, {White}, {Ahlen}, {Bailey}, {Brooks}, {Burtin}, {Chaussidon}, {Claybaugh}, {Cole}, {de la Macorra}, {Ferraro}, {Font-Ribera}, {Forero-Romero}, {Gazta{\~n}aga}, {Gontcho}, {Gutierrez}, {Guy}, {Hahn}, {Honscheid}, {Howlett}, {Juneau}, {Kirkby}, {Kisner}, {Kremin}, {Landriau}, {Le Guillou}, {Levi}, {McDonald}, {Meisner}, {Miquel}, {Moustakas}, {Palanque-Delabrouille}, {Percival}, {Prada}, {P{\'e}rez-R{\`a}fols}, {Raichoor}, {Rossi}, {Sanchez}, {Schlegel}, {Schubnell}, {Sharples}, {Silber}, {Sprayberry}, {Tarl{\'e}}, {Vargas-Maga{\~n}a}, {Weaver}, {Zarrouk}, {Zhou}, \& {Zou}}]{2024arXiv241112025B}
{Bianchi}, D., {Hanif}, M.~M.~S., {Carnero Rosell}, A., {et~al.} 2024, arXiv e-prints, arXiv:2411.12025, \dodoi{10.48550/arXiv.2411.12025}

\bibitem[{{Blanton} {et~al.}(2005{\natexlab{a}}){Blanton}, {Lupton}, {Schlegel}, {Strauss}, {Brinkmann}, {Fukugita}, \& {Loveday}}]{2005ApJ...631..208B}
{Blanton}, M.~R., {Lupton}, R.~H., {Schlegel}, D.~J., {et~al.} 2005{\natexlab{a}}, \apj, 631, 208, \dodoi{10.1086/431416}

\bibitem[{{Blanton} \& {Roweis}(2007)}]{2007AJ....133..734B}
{Blanton}, M.~R., \& {Roweis}, S. 2007, \aj, 133, 734, \dodoi{10.1086/510127}

\bibitem[{{Blanton} {et~al.}(2005{\natexlab{b}}){Blanton}, {Schlegel}, {Strauss}, {Brinkmann}, {Finkbeiner}, {Fukugita}, {Gunn}, {Hogg}, {Ivezi{\'c}}, {Knapp}, {Lupton}, {Munn}, {Schneider}, {Tegmark}, \& {Zehavi}}]{2005AJ....129.2562B}
{Blanton}, M.~R., {Schlegel}, D.~J., {Strauss}, M.~A., {et~al.} 2005{\natexlab{b}}, \aj, 129, 2562, \dodoi{10.1086/429803}

\bibitem[{{Bullock} {et~al.}(2000){Bullock}, {Kravtsov}, \& {Weinberg}}]{2000ApJ...539..517B}
{Bullock}, J.~S., {Kravtsov}, A.~V., \& {Weinberg}, D.~H. 2000, \apj, 539, 517, \dodoi{10.1086/309279}

\bibitem[{{Carlin} {et~al.}(2016){Carlin}, {Sand}, {Price}, {Willman}, {Karunakaran}, {Spekkens}, {Bell}, {Brodie}, {Crnojevi{\'c}}, {Forbes}, {Hargis}, {Kirby}, {Lupton}, {Peter}, {Romanowsky}, \& {Strader}}]{2016ApJ...828L...5C}
{Carlin}, J.~L., {Sand}, D.~J., {Price}, P., {et~al.} 2016, \apjl, 828, L5, \dodoi{10.3847/2041-8205/828/1/L5}

\bibitem[{{Carlsten} {et~al.}(2020){Carlsten}, {Greco}, {Beaton}, \& {Greene}}]{2020ApJ...891..144C}
{Carlsten}, S.~G., {Greco}, J.~P., {Beaton}, R.~L., \& {Greene}, J.~E. 2020, \apj, 891, 144, \dodoi{10.3847/1538-4357/ab7758}

\bibitem[{{Cautun} {et~al.}(2015){Cautun}, {Wang}, {Frenk}, \& {Sawala}}]{2015MNRAS.449.2576C}
{Cautun}, M., {Wang}, W., {Frenk}, C.~S., \& {Sawala}, T. 2015, \mnras, 449, 2576, \dodoi{10.1093/mnras/stv490}

\bibitem[{{Chabrier}(2003)}]{2003PASP..115..763C}
{Chabrier}, G. 2003, \pasp, 115, 763, \dodoi{10.1086/376392}

\bibitem[{{Chaussidon} {et~al.}(2023){Chaussidon}, {Y{\`e}che}, {Palanque-Delabrouille}, {Alexander}, {Yang}, {Ahlen}, {Bailey}, {Brooks}, {Cai}, {Chabanier}, {Davis}, {Dawson}, {de laMacorra}, {Dey}, {Dey}, {Eftekharzadeh}, {Eisenstein}, {Fanning}, {Font-Ribera}, {Gazta{\~n}aga}, {A Gontcho}, {Gonzalez-Morales}, {Guy}, {Herrera-Alcantar}, {Honscheid}, {Ishak}, {Jiang}, {Juneau}, {Kehoe}, {Kisner}, {Kov{\'a}cs}, {Kremin}, {Lan}, {Landriau}, {Le Guillou}, {Levi}, {Magneville}, {Martini}, {Meisner}, {Moustakas}, {Mu{\~n}oz-Guti{\'e}rrez}, {Myers}, {Newman}, {Nie}, {Percival}, {Poppett}, {Prada}, {Raichoor}, {Ravoux}, {Ross}, {Schlafly}, {Schlegel}, {Tan}, {Tarl{\'e}}, {Zhou}, {Zhou}, \& {Zou}}]{2023ApJ...944..107C}
{Chaussidon}, E., {Y{\`e}che}, C., {Palanque-Delabrouille}, N., {et~al.} 2023, \apj, 944, 107, \dodoi{10.3847/1538-4357/acb3c2}

\bibitem[{{Chen} {et~al.}(2019){Chen}, {Mo}, {Li}, {Wang}, {Yang}, {Zhou}, \& {Zhang}}]{2019ApJ...872..180C}
{Chen}, Y., {Mo}, H.~J., {Li}, C., {et~al.} 2019, \apj, 872, 180, \dodoi{10.3847/1538-4357/ab0208}

\bibitem[{{Cooper} {et~al.}(2023){Cooper}, {Koposov}, {Allende Prieto}, {Manser}, {Kizhuprakkat}, {Myers}, {Dey}, {G{\"a}nsicke}, {Li}, {Rockosi}, {Valluri}, {Najita}, {Deason}, {Raichoor}, {Wang}, {Ting}, {Kim}, {Carrillo}, {Wang}, {Beraldo e Silva}, {Han}, {Ding}, {S{\'a}nchez-Conde}, {Aguilar}, {Ahlen}, {Bailey}, {Belokurov}, {Brooks}, {Cunha}, {Dawson}, {de la Macorra}, {Doel}, {Eisenstein}, {Fagrelius}, {Fanning}, {Font-Ribera}, {Forero-Romero}, {Gazta{\~n}aga}, {Gontcho a Gontcho}, {Guy}, {Honscheid}, {Kehoe}, {Kisner}, {Kremin}, {Landriau}, {Levi}, {Martini}, {Meisner}, {Miquel}, {Moustakas}, {Nie}, {Palanque-Delabrouille}, {Percival}, {Poppett}, {Prada}, {Rehemtulla}, {Schlafly}, {Schlegel}, {Schubnell}, {Sharples}, {Tarl{\'e}}, {Wechsler}, {Weinberg}, {Zhou}, \& {Zou}}]{2023ApJ...947...37C}
{Cooper}, A.~P., {Koposov}, S.~E., {Allende Prieto}, C., {et~al.} 2023, \apj, 947, 37, \dodoi{10.3847/1538-4357/acb3c0}

\bibitem[{{Cunha} {et~al.}(2009){Cunha}, {Lima}, {Oyaizu}, {Frieman}, \& {Lin}}]{2009MNRAS.396.2379C}
{Cunha}, C.~E., {Lima}, M., {Oyaizu}, H., {Frieman}, J., \& {Lin}, H. 2009, \mnras, 396, 2379, \dodoi{10.1111/j.1365-2966.2009.14908.x}

\bibitem[{{de Propris} {et~al.}(1995){de Propris}, {Pritchet}, {Harris}, \& {McClure}}]{1995ApJ...450..534D}
{de Propris}, R., {Pritchet}, C.~J., {Harris}, W.~E., \& {McClure}, R.~D. 1995, \apj, 450, 534, \dodoi{10.1086/176163}

\bibitem[{{DESI Collaboration} {et~al.}(2016{\natexlab{a}}){DESI Collaboration}, {Aghamousa}, {Aguilar}, {Ahlen}, {Alam}, {Allen}, {Allende Prieto}, {Annis}, {Bailey}, {Balland}, \& et~al.}]{desiScience}
{DESI Collaboration}, {Aghamousa}, A., {Aguilar}, J., {et~al.} 2016{\natexlab{a}}, ArXiv e-prints.
\newblock \doarXiv{1611.00036}

\bibitem[{{DESI Collaboration} {et~al.}(2016{\natexlab{b}}){DESI Collaboration}, {Aghamousa}, {Aguilar}, {Ahlen}, {Alam}, {Allen}, {Allende Prieto}, {Annis}, {Bailey}, {Balland}, \& et~al.}]{desiInstrument}
---. 2016{\natexlab{b}}, ArXiv e-prints.
\newblock \doarXiv{1611.00037}

\bibitem[{{DESI Collaboration} {et~al.}(2022){DESI Collaboration}, {Abareshi}, {Aguilar}, {Ahlen}, {Alam}, {Alexander}, {Alfarsy}, {Allen}, {Allende Prieto}, {Alves}, {Ameel}, {Armengaud}, {Asorey}, {Aviles}, {Bailey}, {Balaguera-Antol{\'\i}nez}, {Ballester}, {Baltay}, {Bault}, {Beltran}, {Benavides}, {BenZvi}, {Berti}, {Besuner}, {Beutler}, {Bianchi}, {Blake}, {Blanc}, {Blum}, {Bolton}, {Bose}, {Bramall}, {Brieden}, {Brodzeller}, {Brooks}, {Brownewell}, {Buckley-Geer}, {Cahn}, {Cai}, {Canning}, {Carnero Rosell}, {Carton}, {Casas}, {Castander}, {Cervantes-Cota}, {Chabanier}, {Chaussidon}, {Chuang}, {Circosta}, {Cole}, {Cooper}, {da Costa}, {Cousinou}, {Cuceu}, {Davis}, {Dawson}, {de la Cruz-Noriega}, {de la Macorra}, {de Mattia}, {Della Costa}, {Demmer}, {Derwent}, {Dey}, {Dey}, {Dhungana}, {Ding}, {Dobson}, {Doel}, {Donald-McCann}, {Donaldson}, {Douglass}, {Duan}, {Dunlop}, {Edelstein}, {Eftekharzadeh}, {Eisenstein}, {Enriquez-Vargas}, {Escoffier}, {Evatt}, {Fagrelius}, {Fan}, {Fanning}, {Fawcett},
  {Ferraro}, {Ereza}, {Flaugher}, {Font-Ribera}, {Forero-Romero}, {Frenk}, {Fromenteau}, {G{\"a}nsicke}, {Garcia-Quintero}, {Garrison}, {Gazta{\~n}aga}, {Gerardi}, {Gil-Mar{\'\i}n}, {Gontcho}, {Gonzalez-Morales}, {Gonzalez-de-Rivera}, {Gonzalez-Perez}, {Gordon}, {Graur}, {Green}, {Grove}, {Gruen}, {Gutierrez}, {Guy}, {Hahn}, {Harris}, {Herrera}, {Herrera-Alcantar}, {Honscheid}, {Howlett}, {Huterer}, {Ir{\v{s}}i{\v{c}}}, {Ishak}, {Jelinsky}, {Jiang}, {Jimenez}, {Jing}, {Joyce}, {Jullo}, {Juneau}, {Kara{\c{c}}ayl{\i}}, {Karamanis}, {Karcher}, {Karim}, {Kehoe}, {Kent}, {Kirkby}, {Kisner}, {Kitaura}, {Koposov}, {Kov{\'a}cs}, {Kremin}, {Krolewski}, {L'Huillier}, {Lahav}, {Lambert}, {Lamman}, {Lan}, {Landriau}, {Lane}, {Lang}, {Lange}, {Lasker}, {Le Guillou}, {Leauthaud}, {Le Van Suu}, {Levi}, {Li}, {Magneville}, {Manera}, {Manser}, {Marshall}, {McCollam}, {McDonald}, {Meisner}, {Mezcua}, {Miller}, {Miquel}, {Montero-Camacho}, {Moon}, {Martini}, {Meneses-Rizo}, {Moustakas}, {Mueller}, {Mu{\~n}oz-Guti{\'e}rrez},
  {Myers}, {Nadathur}, {Najita}, {Napolitano}, {Neilsen}, {Newman}, {Nie}, {Ning}, {Niz}, {Norberg}, {Noriega}, {O'Brien}, {Obuljen}, {Palanque-Delabrouille}, {Palmese}, {Zhiwei}, {Pappalardo}, {Peng}, {Percival}, {Perruchot}, {Pogge}, {Poppett}, {Porredon}, {Prada}, {Prochaska}, {Pucha}, {P{\'e}rez-Fern{\'a}ndez}, {P{\'e}rez-R{\'a}fols}, {Rabinowitz}, {Raichoor}, {Ramirez-Solano}, {Ram{\'\i}rez-P{\'e}rez}, {Ravoux}, {Reil}, {Rezaie}, {Rocher}, {Rockosi}, {Roe}, {Roodman}, {Ross}, {Rossi}, {Ruggeri}, {Ruhlmann-Kleider}, {Sabiu}, {Safonova}, {Said}, {Saintonge}, {Salas Catonga}, {Samushia}, {Sanchez}, {Saulder}, {Schaan}, {Schlafly}, {Schlegel}, {Schmoll}, {Scholte}, {Schubnell}, {Secroun}, {Seo}, {Serrano}, {Sharples}, {Sholl}, {Silber}, {Silva}, {Sirk}, {Siudek}, {Smith}, {Sprayberry}, {Staten}, {Stupak}, {Tan}, {Tarl{\'e}}, {Sien Tie}, {Tojeiro}, {Ure{\~n}a-L{\'o}pez}, {Valdes}, {Valenzuela}, {Valluri}, {Vargas-Maga{\~n}a}, {Verde}, {Walther}, {Wang}, {Wang}, {Weaver}, {Weaverdyck}, {Wechsler}, {Wilson},
  {Yang}, {Yu}, {Yuan}, {Y{\`e}che}, {Zhang}, {Zhang}, {Zhao}, {Zhou}, {Zhou}, {Zou}, {Zou}, {Zou}, \& {Zu}}]{desi-collaboration22a}
{DESI Collaboration}, {Abareshi}, B., {Aguilar}, J., {et~al.} 2022, arXiv e-prints, arXiv:2205.10939.
\newblock \doarXiv{2205.10939}

\bibitem[{{DESI Collaboration} {et~al.}(2024{\natexlab{a}}){DESI Collaboration}, {Adame}, {Aguilar}, {Ahlen}, {Alam}, {Aldering}, {Alexander}, {Alfarsy}, {Allende Prieto}, {Alvarez}, {Alves}, {Anand}, {Andrade-Oliveira}, {Armengaud}, {Asorey}, {Avila}, {Aviles}, {Bailey}, {Balaguera-Antol{\'\i}nez}, {Ballester}, {Baltay}, {Bault}, {Bautista}, {Behera}, {Beltran}, {BenZvi}, {Beraldo e Silva}, {Bermejo-Climent}, {Berti}, {Besuner}, {Beutler}, {Bianchi}, {Blake}, {Blum}, {Bolton}, {Brieden}, {Brodzeller}, {Brooks}, {Brown}, {Buckley-Geer}, {Burtin}, {Cabayol-Garcia}, {Cai}, {Canning}, {Cardiel-Sas}, {Carnero Rosell}, {Castander}, {Cervantes-Cota}, {Chabanier}, {Chaussidon}, {Chaves-Montero}, {Chen}, {Chen}, {Chuang}, {Claybaugh}, {Cole}, {Cooper}, {Cuceu}, {Davis}, {Dawson}, {de Belsunce}, {de la Cruz}, {de la Macorra}, {de Mattia}, {Demina}, {Demirbozan}, {DeRose}, {Dey}, {Dey}, {Dhungana}, {Ding}, {Ding}, {Doel}, {Doshi}, {Douglass}, {Edge}, {Eftekharzadeh}, {Eisenstein}, {Elliott}, {Escoffier}, {Fagrelius},
  {Fan}, {Fanning}, {Fawcett}, {Ferraro}, {Ereza}, {Flaugher}, {Font-Ribera}, {Forero-S{\'a}nchez}, {Forero-Romero}, {Frenk}, {G{\"a}nsicke}, {Garc{\'\i}a}, {Garc{\'\i}a-Bellido}, {Garcia-Quintero}, {Garrison}, {Gil-Mar{\'\i}n}, {Golden-Marx}, {Gontcho A Gontcho}, {Gonzalez-Morales}, {Gonzalez-Perez}, {Gordon}, {Graur}, {Green}, {Gruen}, {Guy}, {Hadzhiyska}, {Hahn}, {Han}, {Hanif}, {Herrera-Alcantar}, {Honscheid}, {Hou}, {Howlett}, {Huterer}, {Ir{\v{s}}i{\v{c}}}, {Ishak}, {Jana}, {Jiang}, {Jimenez}, {Jing}, {Joudaki}, {Jullo}, {Joyce}, {Juneau}, {Kizhuprakkat}, {Kara{\c{c}}ayl{\i}}, {Karim}, {Kehoe}, {Kent}, {Khederlarian}, {Kim}, {Kirkby}, {Kisner}, {Kitaura}, {Kneib}, {Koposov}, {Kov{\'a}cs}, {Kremin}, {Krolewski}, {L'Huillier}, {Lahav}, {Lambert}, {Lamman}, {Lan}, {Landriau}, {Lang}, {Lange}, {Lasker}, {Le Guillou}, {Leauthaud}, {Levi}, {Li}, {Linder}, {Lyons}, {Magneville}, {Manera}, {Manser}, {Margala}, {Martini}, {McDonald}, {Medina}, {Medina-Varela}, {Meisner}, {Mena-Fern{\'a}ndez}, {Meneses-Rizo},
  {Mezcua}, {Miquel}, {Montero-Camacho}, {Moon}, {Moore}, {Moustakas}, {Mueller}, {Mundet}, {Mu{\~n}oz-Guti{\'e}rrez}, {Myers}, {Nadathur}, {Napolitano}, {Neveux}, {Newman}, {Nie}, {Niz}, {Norberg}, {Noriega}, {Paillas}, {Palanque-Delabrouille}, {Palmese}, {Zhiwei}, {Parkinson}, {Penmetsa}, {Percival}, {P{\'e}rez-Fern{\'a}ndez}, {P{\'e}rez-R{\`a}fols}, {Pieri}, {Poppett}, {Porredon}, {Prada}, {Pucha}, {Raichoor}, {Ram{\'\i}rez-P{\'e}rez}, {Ramirez-Solano}, {Rashkovetskyi}, {Ravoux}, {Rocher}, {Rockosi}, {Ross}, {Rossi}, {Ruggeri}, {Ruhlmann-Kleider}, {Sabiu}, {Said}, {Saintonge}, {Samushia}, {Sanchez}, {Saulder}, {Schaan}, {Schlafly}, {Schlegel}, {Scholte}, {Schubnell}, {Seo}, {Shafieloo}, {Sharples}, {Sheu}, {Silber}, {Sinigaglia}, {Siudek}, {Slepian}, {Smith}, {Sprayberry}, {Stephey}, {Su{\'a}rez-P{\'e}rez}, {Sun}, {Tan}, {Tarl{\'e}}, {Tojeiro}, {Ure{\~n}a-L{\'o}pez}, {Vaisakh}, {Valcin}, {Valdes}, {Valluri}, {Vargas-Maga{\~n}a}, {Variu}, {Verde}, {Walther}, {Wang}, {Wang}, {Weaver}, {Weaverdyck},
  {Wechsler}, {White}, {Xie}, {Yang}, {Y{\`e}che}, {Yu}, {Yuan}, {Zhang}, {Zhang}, {Zhao}, {Zheng}, {Zhou}, {Zhou}, {Zou}, {Zou}, {Zu}, \& {DESI Collaboration}}]{DESI2023a.KP1.SV}
{DESI Collaboration}, {Adame}, A.~G., {Aguilar}, J., {et~al.} 2024{\natexlab{a}}, \aj, 167, 62, \dodoi{10.3847/1538-3881/ad0b08}

\bibitem[{{DESI Collaboration} {et~al.}(2024{\natexlab{b}}){DESI Collaboration}, {Adame}, {Aguilar}, {Ahlen}, {Alam}, {Aldering}, {Alexander}, {Alfarsy}, {Allende Prieto}, {Alvarez}, {Alves}, {Anand}, {Andrade-Oliveira}, {Armengaud}, {Asorey}, {Avila}, {Aviles}, {Bailey}, {Balaguera-Antol{\'\i}nez}, {Ballester}, {Baltay}, {Bault}, {Bautista}, {Behera}, {Beltran}, {BenZvi}, {Beraldo e Silva}, {Bermejo-Climent}, {Berti}, {Besuner}, {Beutler}, {Bianchi}, {Blake}, {Blum}, {Bolton}, {Brieden}, {Brodzeller}, {Brooks}, {Brown}, {Buckley-Geer}, {Burtin}, {Cabayol-Garcia}, {Cai}, {Canning}, {Cardiel-Sas}, {Carnero Rosell}, {Castander}, {Cervantes-Cota}, {Chabanier}, {Chaussidon}, {Chaves-Montero}, {Chen}, {Chen}, {Chuang}, {Claybaugh}, {Cole}, {Cooper}, {Cuceu}, {Davis}, {Dawson}, {de Belsunce}, {de la Cruz}, {de la Macorra}, {Della Costa}, {de Mattia}, {Demina}, {Demirbozan}, {DeRose}, {Dey}, {Dey}, {Dhungana}, {Ding}, {Ding}, {Doel}, {Doshi}, {Douglass}, {Edge}, {Eftekharzadeh}, {Eisenstein}, {Elliott}, {Ereza},
  {Escoffier}, {Fagrelius}, {Fan}, {Fanning}, {Fawcett}, {Ferraro}, {Flaugher}, {Font-Ribera}, {Forero-Romero}, {Forero-S{\'a}nchez}, {Frenk}, {G{\"a}nsicke}, {Garc{\'\i}a}, {Garc{\'\i}a-Bellido}, {Garcia-Quintero}, {Garrison}, {Gil-Mar{\'\i}n}, {Golden-Marx}, {Gontcho A Gontcho}, {Gonzalez-Morales}, {Gonzalez-Perez}, {Gordon}, {Graur}, {Green}, {Gruen}, {Guy}, {Hadzhiyska}, {Hahn}, {Han}, {Hanif}, {Herrera-Alcantar}, {Honscheid}, {Hou}, {Howlett}, {Huterer}, {Ir{\v{s}}i{\v{c}}}, {Ishak}, {Jacques}, {Jana}, {Jiang}, {Jimenez}, {Jing}, {Joudaki}, {Joyce}, {Jullo}, {Juneau}, {Kara{\c{c}}ayl{\i}}, {Karim}, {Kehoe}, {Kent}, {Khederlarian}, {Kim}, {Kirkby}, {Kisner}, {Kitaura}, {Kizhuprakkat}, {Kneib}, {Koposov}, {Kov{\'a}cs}, {Kremin}, {Krolewski}, {L'Huillier}, {Lahav}, {Lambert}, {Lamman}, {Lan}, {Landriau}, {Lang}, {Lange}, {Lasker}, {Leauthaud}, {Le Guillou}, {Levi}, {Li}, {Linder}, {Lyons}, {Magneville}, {Manera}, {Manser}, {Margala}, {Martini}, {McDonald}, {Medina}, {Medina-Varela}, {Meisner},
  {Mena-Fern{\'a}ndez}, {Meneses-Rizo}, {Mezcua}, {Miquel}, {Montero-Camacho}, {Moon}, {Moore}, {Moustakas}, {Mueller}, {Mundet}, {Mu{\~n}oz-Guti{\'e}rrez}, {Myers}, {Nadathur}, {Napolitano}, {Neveux}, {Newman}, {Nie}, {Nikutta}, {Niz}, {Norberg}, {Noriega}, {Paillas}, {Palanque-Delabrouille}, {Palmese}, {Pan}, {Parkinson}, {Penmetsa}, {Percival}, {P{\'e}rez-Fern{\'a}ndez}, {P{\'e}rez-R{\`a}fols}, {Pieri}, {Poppett}, {Porredon}, {Pothier}, {Prada}, {Pucha}, {Raichoor}, {Ram{\'\i}rez-P{\'e}rez}, {Ramirez-Solano}, {Rashkovetskyi}, {Ravoux}, {Rocher}, {Rockosi}, {Ross}, {Rossi}, {Ruggeri}, {Ruhlmann-Kleider}, {Sabiu}, {Said}, {Saintonge}, {Samushia}, {Sanchez}, {Saulder}, {Schaan}, {Schlafly}, {Schlegel}, {Scholte}, {Schubnell}, {Seo}, {Shafieloo}, {Sharples}, {Sheu}, {Silber}, {Sinigaglia}, {Siudek}, {Slepian}, {Smith}, {Soumagnac}, {Sprayberry}, {Stephey}, {Su{\'a}rez-P{\'e}rez}, {Sun}, {Tan}, {Tarl{\'e}}, {Tojeiro}, {Ure{\~n}a-L{\'o}pez}, {Vaisakh}, {Valcin}, {Valdes}, {Valluri}, {Vargas-Maga{\~n}a}, {Variu},
  {Verde}, {Walther}, {Wang}, {Wang}, {Weaver}, {Weaverdyck}, {Wechsler}, {White}, {Xie}, {Yang}, {Y{\`e}che}, {Yu}, {Yuan}, {Zhang}, {Zhang}, {Zhao}, {Zheng}, {Zhou}, {Zhou}, {Zou}, {Zou}, \& {Zu}}]{DESI2023b.KP1.EDR}
---. 2024{\natexlab{b}}, \aj, 168, 58, \dodoi{10.3847/1538-3881/ad3217}

\bibitem[{{DESI Collaboration} {et~al.}(2024{\natexlab{c}}){DESI Collaboration}, {Adame}, {Aguilar}, {Ahlen}, {Alam}, {Alexander}, {Alvarez}, {Alves}, {Anand}, {Andrade}, {Armengaud}, {Avila}, {Aviles}, {Awan}, {Bailey}, {Baltay}, {Bault}, {Behera}, {BenZvi}, {Beutler}, {Bianchi}, {Blake}, {Blum}, {Brieden}, {Brodzeller}, {Brooks}, {Brown}, {Buckley-Geer}, {Burtin}, {Calderon}, {Canning}, {Carnero Rosell}, {Cereskaite}, {Cervantes-Cota}, {Chabanier}, {Chaussidon}, {Chaves-Montero}, {Chen}, {Chen}, {Claybaugh}, {Cole}, {Cuceu}, {Davis}, {Dawson}, {de la Macorra}, {de Mattia}, {Deiosso}, {Demina}, {Dey}, {Dey}, {Ding}, {Doel}, {Edelstein}, {Eftekharzadeh}, {Eisenstein}, {Elliott}, {Fagrelius}, {Fanning}, {Ferraro}, {Ereza}, {Findlay}, {Flaugher}, {Font-Ribera}, {Forero-S{\'a}nchez}, {Forero-Romero}, {Frenk}, {Garcia-Quintero}, {Gazta{\~n}aga}, {Gil-Mar{\'\i}n}, {Gontcho}, {Gonzalez-Morales}, {Gonzalez-Perez}, {Gordon}, {Green}, {Gruen}, {Gsponer}, {Gutierrez}, {Guy}, {Hadzhiyska}, {Hahn}, {Hanif},
  {Herrera-Alcantar}, {Honscheid}, {Hou}, {Howlett}, {Huterer}, {Ir{\v{s}}i{\v{c}}}, {Ishak}, {Juneau}, {Kara{\c{c}}ayl{\i}}, {Kehoe}, {Kent}, {Kirkby}, {Kitaura}, {Kong}, {Kremin}, {Krolewski}, {Lai}, {Lan}, {Landriau}, {Lang}, {Lasker}, {Le Goff}, {Le Guillou}, {Leauthaud}, {Levi}, {Li}, {Lodha}, {Magneville}, {Manera}, {Margala}, {Martini}, {Maus}, {McDonald}, {Medina-Varela}, {Meisner}, {Mena-Fern{\'a}ndez}, {Miquel}, {Moon}, {Moore}, {Moustakas}, {Mudur}, {Mueller}, {Mu{\~n}oz-Guti{\'e}rrez}, {Myers}, {Nadathur}, {Napolitano}, {Neveux}, {Newman}, {Nguyen}, {Nie}, {Niz}, {Noriega}, {Padmanabhan}, {Paillas}, {Palanque-Delabrouille}, {Pan}, {Penmetsa}, {Percival}, {Pieri}, {Pinon}, {Poppett}, {Porredon}, {Prada}, {P{\'e}rez-Fern{\'a}ndez}, {P{\'e}rez-R{\`a}fols}, {Rabinowitz}, {Raichoor}, {Ram{\'\i}rez-P{\'e}rez}, {Ramirez-Solano}, {Rashkovetskyi}, {Ravoux}, {Rezaie}, {Rich}, {Rocher}, {Rockosi}, {Roe}, {Rosado-Marin}, {Ross}, {Rossi}, {Ruggeri}, {Ruhlmann-Kleider}, {Samushia}, {Sanchez}, {Saulder},
  {Schlafly}, {Schlegel}, {Scholte}, {Schubnell}, {Seo}, {Sharples}, {Silber}, {Slosar}, {Smith}, {Sprayberry}, {Tan}, {Tarl{\'e}}, {Trusov}, {Vaisakh}, {Valcin}, {Valdes}, {Vargas-Maga{\~n}a}, {Verde}, {Walther}, {Wang}, {Wang}, {Weaver}, {Weaverdyck}, {Wechsler}, {Weinberg}, {White}, {Wilson}, {Yu}, {Yu}, {Yuan}, {Y{\`e}che}, {Zaborowski}, {Zarrouk}, {Zhang}, {Zhao}, {Zhao}, {Zhou}, \& {Zou}}]{DESI2024.II.KP3}
---. 2024{\natexlab{c}}, arXiv e-prints, arXiv:2411.12020, \dodoi{10.48550/arXiv.2411.12020}

\bibitem[{{DESI Collaboration} {et~al.}(2024{\natexlab{d}}){DESI Collaboration}, {Adame}, {Aguilar}, {Ahlen}, {Alam}, {Alexander}, {Alvarez}, {Alves}, {Anand}, {Andrade}, {Armengaud}, {Avila}, {Aviles}, {Awan}, {Bailey}, {Baltay}, {Bault}, {Behera}, {BenZvi}, {Beutler}, {Bianchi}, {Blake}, {Blum}, {Brieden}, {Brodzeller}, {Brooks}, {Buckley-Geer}, {Burtin}, {Calderon}, {Canning}, {Carnero Rosell}, {Cereskaite}, {Cervantes-Cota}, {Chabanier}, {Chaussidon}, {Chaves-Montero}, {Chen}, {Chen}, {Claybaugh}, {Cole}, {Cuceu}, {Davis}, {Dawson}, {de la Macorra}, {de Mattia}, {Deiosso}, {Dey}, {Dey}, {Ding}, {Doel}, {Edelstein}, {Eftekharzadeh}, {Eisenstein}, {Elliott}, {Fagrelius}, {Fanning}, {Ferraro}, {Ereza}, {Findlay}, {Flaugher}, {Font-Ribera}, {Forero-S{\'a}nchez}, {Forero-Romero}, {Garcia-Quintero}, {Garrison}, {Gazta{\~n}aga}, {Gil-Mar{\'\i}n}, {Gontcho}, {Gonzalez-Morales}, {Gonzalez-Perez}, {Gordon}, {Green}, {Gruen}, {Gsponer}, {Gutierrez}, {Guy}, {Hadzhiyska}, {Hahn}, {Hanif}, {Herrera-Alcantar},
  {Honscheid}, {Howlett}, {Huterer}, {Ir{\v{s}}i{\v{c}}}, {Ishak}, {Juneau}, {Kara{\c{c}}ayl{\i}}, {Kehoe}, {Kent}, {Kirkby}, {Kong}, {Koposov}, {Kremin}, {Krolewski}, {Lai}, {Lan}, {Landriau}, {Lang}, {Lasker}, {Le Goff}, {Le Guillou}, {Leauthaud}, {Levi}, {Li}, {Lodha}, {Magneville}, {Manera}, {Margala}, {Martini}, {Maus}, {McDonald}, {Medina-Varela}, {Meisner}, {Mena-Fern{\'a}ndez}, {Miquel}, {Moon}, {Moore}, {Moustakas}, {Mueller}, {Mu{\~n}oz-Guti{\'e}rrez}, {Myers}, {Nadathur}, {Napolitano}, {Neveux}, {Newman}, {Nguyen}, {Nie}, {Niz}, {Noriega}, {Padmanabhan}, {Paillas}, {Palanque-Delabrouille}, {Pan}, {Penmetsa}, {Percival}, {Pieri}, {Pinon}, {Poppett}, {Porredon}, {Prada}, {P{\'e}rez-Fern{\'a}ndez}, {P{\'e}rez-R{\`a}fols}, {Rabinowitz}, {Raichoor}, {Ram{\'\i}rez-P{\'e}rez}, {Ramirez-Solano}, {Rashkovetskyi}, {Ravoux}, {Rezaie}, {Rich}, {Rocher}, {Rockosi}, {Rodr{\'\i}guez-Mart{\'\i}nez}, {Roe}, {Rosado-Marin}, {Ross}, {Rossi}, {Ruggeri}, {Ruhlmann-Kleider}, {Samushia}, {Sanchez}, {Saulder}, {Schlafly},
  {Schlegel}, {Schubnell}, {Seo}, {Sharples}, {Silber}, {Slosar}, {Smith}, {Sprayberry}, {Tan}, {Tarl{\'e}}, {Trusov}, {Vaisakh}, {Valcin}, {Valdes}, {Vargas-Maga{\~n}a}, {Verde}, {Walther}, {Wang}, {Wang}, {Weaver}, {Weaverdyck}, {Wechsler}, {Weinberg}, {White}, {Wilson}, {Yu}, {Yu}, {Yuan}, {Y{\`e}che}, {Zaborowski}, {Zarrouk}, {Zhang}, {Zhao}, {Zhao}, {Zhou}, \& {Zou}}]{DESI2024.V.KP5}
---. 2024{\natexlab{d}}, arXiv e-prints, arXiv:2411.12021, \dodoi{10.48550/arXiv.2411.12021}

\bibitem[{{DESI Collaboration} {et~al.}(2024{\natexlab{e}}){DESI Collaboration}, {Adame}, {Aguilar}, {Ahlen}, {Alam}, {Alexander}, {Alvarez}, {Alves}, {Anand}, {Andrade}, {Armengaud}, {Avila}, {Aviles}, {Awan}, {Bailey}, {Baltay}, {Bault}, {Behera}, {BenZvi}, {Beutler}, {Bianchi}, {Blake}, {Blum}, {Brieden}, {Brodzeller}, {Brooks}, {Buckley-Geer}, {Burtin}, {Calderon}, {Canning}, {Carnero Rosell}, {Cereskaite}, {Cervantes-Cota}, {Chabanier}, {Chaussidon}, {Chaves-Montero}, {Chen}, {Chen}, {Claybaugh}, {Cole}, {Cuceu}, {Davis}, {Dawson}, {de la Macorra}, {de Mattia}, {Deiosso}, {Dey}, {Dey}, {Ding}, {Doel}, {Edelstein}, {Eftekharzadeh}, {Eisenstein}, {Elliott}, {Fagrelius}, {Fanning}, {Ferraro}, {Ereza}, {Findlay}, {Flaugher}, {Font-Ribera}, {Forero-S{\'a}nchez}, {Forero-Romero}, {Garcia-Quintero}, {Gazta{\~n}aga}, {Gil-Mar{\'\i}n}, {Gontcho}, {Gonzalez-Morales}, {Gonzalez-Perez}, {Gordon}, {Green}, {Gruen}, {Gsponer}, {Gutierrez}, {Guy}, {Hadzhiyska}, {Hahn}, {Hanif}, {Herrera-Alcantar}, {Honscheid},
  {Howlett}, {Huterer}, {Ir{\v{s}}i{\v{c}}}, {Ishak}, {Juneau}, {Kara{\c{c}}ayl{\i}}, {Kehoe}, {Kent}, {Kirkby}, {Kremin}, {Krolewski}, {Lai}, {Lan}, {Landriau}, {Lang}, {Lasker}, {Le Goff}, {Le Guillou}, {Leauthaud}, {Levi}, {Li}, {Linder}, {Lodha}, {Magneville}, {Manera}, {Margala}, {Martini}, {Maus}, {McDonald}, {Medina-Varela}, {Meisner}, {Mena-Fern{\'a}ndez}, {Miquel}, {Moon}, {Moore}, {Moustakas}, {Mudur}, {Mueller}, {Mu{\~n}oz-Guti{\'e}rrez}, {Myers}, {Nadathur}, {Napolitano}, {Neveux}, {Newman}, {Nguyen}, {Nie}, {Niz}, {Noriega}, {Padmanabhan}, {Paillas}, {Palanque-Delabrouille}, {Pan}, {Penmetsa}, {Percival}, {Pieri}, {Pinon}, {Poppett}, {Porredon}, {Prada}, {P{\'e}rez-Fern{\'a}ndez}, {P{\'e}rez-R{\`a}fols}, {Rabinowitz}, {Raichoor}, {Ram{\'\i}rez-P{\'e}rez}, {Ramirez-Solano}, {Rashkovetskyi}, {Rezaie}, {Rich}, {Rocher}, {Rockosi}, {Roe}, {Rosado-Marin}, {Ross}, {Rossi}, {Ruggeri}, {Ruhlmann-Kleider}, {Samushia}, {Sanchez}, {Saulder}, {Schlafly}, {Schlegel}, {Schubnell}, {Seo}, {Sharples}, {Silber},
  {Slosar}, {Smith}, {Sprayberry}, {Swanson}, {Tan}, {Tarl{\'e}}, {Trusov}, {Vaisakh}, {Valcin}, {Valdes}, {Vargas-Maga{\~n}a}, {Verde}, {Walther}, {Wang}, {Wang}, {Weaver}, {Weaverdyck}, {Wechsler}, {Weinberg}, {White}, {Yu}, {Yu}, {Yuan}, {Y{\`e}che}, {Zaborowski}, {Zarrouk}, {Zhang}, {Zhao}, {Zhao}, {Zhou}, \& {Zou}}]{2024arXiv240403000D}
---. 2024{\natexlab{e}}, arXiv e-prints, arXiv:2404.03000, \dodoi{10.48550/arXiv.2404.03000}

\bibitem[{{DESI Collaboration} {et~al.}(2024{\natexlab{f}}){DESI Collaboration}, {Adame}, {Aguilar}, {Ahlen}, {Alam}, {Alexander}, {Alvarez}, {Alves}, {Anand}, {Andrade}, {Armengaud}, {Avila}, {Aviles}, {Awan}, {Bahr-Kalus}, {Bailey}, {Baltay}, {Bault}, {Behera}, {BenZvi}, {Bera}, {Beutler}, {Bianchi}, {Blake}, {Blum}, {Brieden}, {Brodzeller}, {Brooks}, {Buckley-Geer}, {Burtin}, {Calderon}, {Canning}, {Carnero Rosell}, {Cereskaite}, {Cervantes-Cota}, {Chabanier}, {Chaussidon}, {Chaves-Montero}, {Chen}, {Chen}, {Claybaugh}, {Cole}, {Cuceu}, {Davis}, {Dawson}, {de la Macorra}, {de Mattia}, {Deiosso}, {Dey}, {Dey}, {Ding}, {Doel}, {Edelstein}, {Eftekharzadeh}, {Eisenstein}, {Elliott}, {Fagrelius}, {Fanning}, {Ferraro}, {Ereza}, {Findlay}, {Flaugher}, {Font-Ribera}, {Forero-S{\'a}nchez}, {Forero-Romero}, {Frenk}, {Garcia-Quintero}, {Gazta{\~n}aga}, {Gil-Mar{\'\i}n}, {Gontcho}, {Gonzalez-Morales}, {Gonzalez-Perez}, {Gordon}, {Green}, {Gruen}, {Gsponer}, {Gutierrez}, {Guy}, {Hadzhiyska}, {Hahn}, {Hanif},
  {Herrera-Alcantar}, {Honscheid}, {Howlett}, {Huterer}, {Ir{\v{s}}i{\v{c}}}, {Ishak}, {Juneau}, {Kara{\c{c}}ayl{\i}}, {Kehoe}, {Kent}, {Kirkby}, {Kremin}, {Krolewski}, {Lai}, {Lan}, {Landriau}, {Lang}, {Lasker}, {Le Goff}, {Le Guillou}, {Leauthaud}, {Levi}, {Li}, {Linder}, {Lodha}, {Magneville}, {Manera}, {Margala}, {Martini}, {Maus}, {McDonald}, {Medina-Varela}, {Meisner}, {Mena-Fern{\'a}ndez}, {Miquel}, {Moon}, {Moore}, {Moustakas}, {Mudur}, {Mueller}, {Mu{\~n}oz-Guti{\'e}rrez}, {Myers}, {Nadathur}, {Napolitano}, {Neveux}, {Newman}, {Nguyen}, {Nie}, {Niz}, {Noriega}, {Padmanabhan}, {Paillas}, {Palanque-Delabrouille}, {Pan}, {Penmetsa}, {Percival}, {Pieri}, {Pinon}, {Poppett}, {Porredon}, {Prada}, {P{\'e}rez-Fern{\'a}ndez}, {P{\'e}rez-R{\`a}fols}, {Rabinowitz}, {Raichoor}, {Ram{\'\i}rez-P{\'e}rez}, {Ramirez-Solano}, {Ravoux}, {Rashkovetskyi}, {Rezaie}, {Rich}, {Rocher}, {Rockosi}, {Roe}, {Rosado-Marin}, {Ross}, {Rossi}, {Ruggeri}, {Ruhlmann-Kleider}, {Samushia}, {Sanchez}, {Saulder}, {Schlafly}, {Schlegel},
  {Schubnell}, {Seo}, {Shafieloo}, {Sharples}, {Silber}, {Slosar}, {Smith}, {Sprayberry}, {Tan}, {Tarl{\'e}}, {Taylor}, {Trusov}, {Ure{\~n}a-L{\'o}pez}, {Vaisakh}, {Valcin}, {Valdes}, {Vargas-Maga{\~n}a}, {Verde}, {Walther}, {Wang}, {Wang}, {Weaver}, {Weaverdyck}, {Wechsler}, {Weinberg}, {White}, {Yu}, {Yu}, {Yuan}, {Y{\`e}che}, {Zaborowski}, {Zarrouk}, {Zhang}, {Zhao}, {Zhao}, {Zhou}, {Zhuang}, \& {Zou}}]{2024arXiv240403002D}
---. 2024{\natexlab{f}}, arXiv e-prints, arXiv:2404.03002, \dodoi{10.48550/arXiv.2404.03002}

\bibitem[{{DESI Collaboration} {et~al.}(2024{\natexlab{g}}){DESI Collaboration}, {Adame}, {Aguilar}, {Ahlen}, {Alam}, {Alexander}, {Allende Prieto}, {Alvarez}, {Alves}, {Anand}, {Andrade}, {Armengaud}, {Avila}, {Aviles}, {Awan}, {Bahr-Kalus}, {Bailey}, {Baltay}, {Bault}, {Behera}, {BenZvi}, {Beutler}, {Bianchi}, {Blake}, {Blum}, {Bonici}, {Brieden}, {Brodzeller}, {Brooks}, {Buckley-Geer}, {Burtin}, {Calderon}, {Canning}, {Carnero Rosell}, {Cereskaite}, {Cervantes-Cota}, {Chabanier}, {Chaussidon}, {Chaves-Montero}, {Chebat}, {Chen}, {Chen}, {Claybaugh}, {Cole}, {Cuceu}, {Davis}, {Dawson}, {de la Macorra}, {de Mattia}, {Deiosso}, {Dey}, {Dey}, {Ding}, {Doel}, {Edelstein}, {Eftekharzadeh}, {Eisenstein}, {Elbers}, {Elliott}, {Fagrelius}, {Fanning}, {Ferraro}, {Ereza}, {Findlay}, {Flaugher}, {Font-Ribera}, {Forero-S{\'a}nchez}, {Forero-Romero}, {Frenk}, {Garcia-Quintero}, {Garrison}, {Gazta{\~n}aga}, {Gil-Mar{\'\i}n}, {Gontcho}, {Gonzalez-Morales}, {Gonzalez-Perez}, {Gordon}, {Green}, {Gruen}, {Gsponer},
  {Gutierrez}, {Guy}, {Hadzhiyska}, {Hahn}, {Hanif}, {Herrera-Alcantar}, {Honscheid}, {Howlett}, {Huterer}, {Ir{\v{s}}i{\v{c}}}, {Ishak}, {Joyce}, {Juneau}, {Kara{\c{c}}ayl{\i}}, {Kehoe}, {Kent}, {Kirkby}, {Kong}, {Koposov}, {Kremin}, {Krolewski}, {Lahav}, {Lai}, {Lan}, {Landriau}, {Lang}, {Lasker}, {Le Goff}, {Le Guillou}, {Leauthaud}, {Levi}, {Li}, {Lodha}, {Magneville}, {Manera}, {Margala}, {Martini}, {Matthewson}, {Maus}, {McDonald}, {Medina-Varela}, {Meisner}, {Mena-Fern{\'a}ndez}, {Miquel}, {Moon}, {Moore}, {Moustakas}, {Mudur}, {Mueller}, {Mu{\~n}oz-Guti{\'e}rrez}, {Myers}, {Nadathur}, {Napolitano}, {Neveux}, {Newman}, {Nguyen}, {Nie}, {Niz}, {Noriega}, {Padmanabhan}, {Paillas}, {Palanque-Delabrouille}, {Pan}, {Penmetsa}, {Percival}, {Pieri}, {Pinon}, {Poppett}, {Porredon}, {Prada}, {P{\'e}rez-Fern{\'a}ndez}, {P{\'e}rez-R{\`a}fols}, {Rabinowitz}, {Raichoor}, {Ram{\'\i}rez-P{\'e}rez}, {Ramirez-Solano}, {Rashkovetskyi}, {Ravoux}, {Rezaie}, {Rich}, {Rocher}, {Rockosi}, {Roe}, {Rosado-Marin}, {Ross},
  {Rossi}, {Ruggeri}, {Ruhlmann-Kleider}, {Samushia}, {Sanchez}, {Saulder}, {Schlafly}, {Schlegel}, {Schubnell}, {Seo}, {Shafieloo}, {Sharples}, {Silber}, {Slosar}, {Smith}, {Sprayberry}, {Tan}, {Tarl{\'e}}, {Taylor}, {Trusov}, {Vaisakh}, {Valcin}, {Valdes}, {Valogiannis}, {Vargas-Maga{\~n}a}, {Verde}, {Walther}, {Wang}, {Wang}, {Weaver}, {Weaverdyck}, {Wechsler}, {Weinberg}, {White}, {Wilson}, {Yi}, {Yu}, {Yu}, {Yuan}, {Y{\`e}che}, {Zaborowski}, {Zarrouk}, {Zhang}, {Zhao}, {Zhao}, {Zhou}, {Zhuang}, \& {Zou}}]{DESI2024.VII.KP7B}
---. 2024{\natexlab{g}}, arXiv e-prints, arXiv:2411.12022, \dodoi{10.48550/arXiv.2411.12022}

\bibitem[{{DESI Collaboration} {et~al.}(2024{\natexlab{h}}){DESI Collaboration}, {Adame}, {Aguilar}, {Ahlen}, {Alam}, {Alexander}, {Alvarez}, {Alves}, {Anand}, {Andrade}, {Armengaud}, {Avila}, {Aviles}, {Awan}, {Bailey}, {Baltay}, {Bault}, {Behera}, {BenZvi}, {Beutler}, {Bianchi}, {Blake}, {Blum}, {Brieden}, {Brodzeller}, {Brooks}, {Brown}, {Buckley-Geer}, {Burtin}, {Calderon}, {Canning}, {Carnero Rosell}, {Cereskaite}, {Cervantes-Cota}, {Chabanier}, {Chaussidon}, {Chaves-Montero}, {Chen}, {Chen}, {Claybaugh}, {Cole}, {Cuceu}, {Davis}, {Dawson}, {de la Macorra}, {de Mattia}, {Deiosso}, {Demina}, {Dey}, {Dey}, {Ding}, {Doel}, {Edelstein}, {Eftekharzadeh}, {Eisenstein}, {Elliott}, {Fagrelius}, {Fanning}, {Ferraro}, {Ereza}, {Findlay}, {Flaugher}, {Font-Ribera}, {Forero-S{\'a}nchez}, {Forero-Romero}, {Frenk}, {Garcia-Quintero}, {Gazta{\~n}aga}, {Gil-Mar{\'\i}n}, {Gontcho}, {Gonzalez-Morales}, {Gonzalez-Perez}, {Gordon}, {Green}, {Gruen}, {Gsponer}, {Gutierrez}, {Guy}, {Hadzhiyska}, {Hahn}, {Hanif},
  {Herrera-Alcantar}, {Honscheid}, {Hou}, {Howlett}, {Huterer}, {Ir{\v{s}}i{\v{c}}}, {Ishak}, {Juneau}, {Kara{\c{c}}ayl{\i}}, {Kehoe}, {Kent}, {Kirkby}, {Kitaura}, {Kong}, {Kremin}, {Krolewski}, {Lai}, {Lan}, {Landriau}, {Lang}, {Lasker}, {Le Goff}, {Le Guillou}, {Leauthaud}, {Levi}, {Li}, {Lodha}, {Magneville}, {Manera}, {Margala}, {Martini}, {Maus}, {McDonald}, {Medina-Varela}, {Meisner}, {Mena-Fern{\'a}ndez}, {Miquel}, {Moon}, {Moore}, {Moustakas}, {Mudur}, {Mueller}, {Mu{\~n}oz-Guti{\'e}rrez}, {Myers}, {Nadathur}, {Napolitano}, {Neveux}, {Newman}, {Nguyen}, {Nie}, {Niz}, {Noriega}, {Padmanabhan}, {Paillas}, {Palanque-Delabrouille}, {Pan}, {Penmetsa}, {Percival}, {Pieri}, {Pinon}, {Poppett}, {Porredon}, {Prada}, {P{\'e}rez-Fern{\'a}ndez}, {P{\'e}rez-R{\`a}fols}, {Rabinowitz}, {Raichoor}, {Ram{\'\i}rez-P{\'e}rez}, {Ramirez-Solano}, {Rashkovetskyi}, {Ravoux}, {Rezaie}, {Rich}, {Rocher}, {Rockosi}, {Roe}, {Rosado-Marin}, {Ross}, {Rossi}, {Ruggeri}, {Ruhlmann-Kleider}, {Samushia}, {Sanchez}, {Saulder},
  {Schlafly}, {Schlegel}, {Scholte}, {Schubnell}, {Seo}, {Sharples}, {Silber}, {Slosar}, {Smith}, {Sprayberry}, {Tan}, {Tarl{\'e}}, {Trusov}, {Vaisakh}, {Valcin}, {Valdes}, {Vargas-Maga{\~n}a}, {Verde}, {Walther}, {Wang}, {Wang}, {Weaver}, {Weaverdyck}, {Wechsler}, {Weinberg}, {White}, {Wilson}, {Yu}, {Yu}, {Yuan}, {Y{\`e}che}, {Zaborowski}, {Zarrouk}, {Zhang}, \& {Zhao}}]{2024arXiv241112020D}
---. 2024{\natexlab{h}}, arXiv e-prints, arXiv:2411.12020, \dodoi{10.48550/arXiv.2411.12020}

\bibitem[{{Dey} {et~al.}(2019){Dey}, {Schlegel}, {Lang}, {Blum}, {Burleigh}, {Fan}, {Findlay}, {Finkbeiner}, {Herrera}, {Juneau}, {Landriau}, {Levi}, {McGreer}, {Meisner}, {Myers}, {Moustakas}, {Nugent}, {Patej}, {Schlafly}, {Walker}, {Valdes}, {Weaver}, {Y{\`e}che}, {Zou}, {Zhou}, {Abareshi}, {Abbott}, {Abolfathi}, {Aguilera}, {Alam}, {Allen}, {Alvarez}, {Annis}, {Ansarinejad}, {Aubert}, {Beechert}, {Bell}, {BenZvi}, {Beutler}, {Bielby}, {Bolton}, {Brice{\~n}o}, {Buckley-Geer}, {Butler}, {Calamida}, {Carlberg}, {Carter}, {Casas}, {Castander}, {Choi}, {Comparat}, {Cukanovaite}, {Delubac}, {DeVries}, {Dey}, {Dhungana}, {Dickinson}, {Ding}, {Donaldson}, {Duan}, {Duckworth}, {Eftekharzadeh}, {Eisenstein}, {Etourneau}, {Fagrelius}, {Farihi}, {Fitzpatrick}, {Font-Ribera}, {Fulmer}, {G{\"a}nsicke}, {Gaztanaga}, {George}, {Gerdes}, {Gontcho}, {Gorgoni}, {Green}, {Guy}, {Harmer}, {Hernand ez}, {Honscheid}, {Huang}, {James}, {Jannuzi}, {Jiang}, {Joyce}, {Karcher}, {Karkar}, {Kehoe}, {Kneib}, {Kueter-Young}, {Lan},
  {Lauer}, {Le Guillou}, {Le Van Suu}, {Lee}, {Lesser}, {Perreault Levasseur}, {Li}, {Mann}, {Marshall}, {Mart{\'\i}nez-V{\'a}zquez}, {Martini}, {du Mas des Bourboux}, {McManus}, {Meier}, {M{\'e}nard}, {Metcalfe}, {Mu{\~n}oz-Guti{\'e}rrez}, {Najita}, {Napier}, {Narayan}, {Newman}, {Nie}, {Nord}, {Norman}, {Olsen}, {Paat}, {Palanque-Delabrouille}, {Peng}, {Poppett}, {Poremba}, {Prakash}, {Rabinowitz}, {Raichoor}, {Rezaie}, {Robertson}, {Roe}, {Ross}, {Ross}, {Rudnick}, {Safonova}, {Saha}, {S{\'a}nchez}, {Savary}, {Schweiker}, {Scott}, {Seo}, {Shan}, {Silva}, {Slepian}, {Soto}, {Sprayberry}, {Staten}, {Stillman}, {Stupak}, {Summers}, {Sien Tie}, {Tirado}, {Vargas-Maga{\~n}a}, {Vivas}, {Wechsler}, {Williams}, {Yang}, {Yang}, {Yapici}, {Zaritsky}, {Zenteno}, {Zhang}, {Zhang}, {Zhou}, \& {Zhou}}]{2019AJ....157..168D}
{Dey}, A., {Schlegel}, D.~J., {Lang}, D., {et~al.} 2019, \aj, 157, 168, \dodoi{10.3847/1538-3881/ab089d}

\bibitem[{{Driver} {et~al.}(1994){Driver}, {Phillipps}, {Davies}, {Morgan}, \& {Disney}}]{1994MNRAS.268..393D}
{Driver}, S.~P., {Phillipps}, S., {Davies}, J.~I., {Morgan}, I., \& {Disney}, M.~J. 1994, \mnras, 268, 393, \dodoi{10.1093/mnras/268.2.393}

\bibitem[{{Gu} {et~al.}(2024){Gu}, {Guo}, {Cautun}, {Shao}, {Pei}, {Wang}, {Gao}, \& {Wang}}]{2024NatAs...8..538G}
{Gu}, Q., {Guo}, Q., {Cautun}, M., {et~al.} 2024, Nature Astronomy, 8, 538, \dodoi{10.1038/s41550-023-02192-6}

\bibitem[{{Guo} {et~al.}(2011{\natexlab{a}}){Guo}, {Cole}, {Eke}, \& {Frenk}}]{2011MNRAS.417..370G}
{Guo}, Q., {Cole}, S., {Eke}, V., \& {Frenk}, C. 2011{\natexlab{a}}, \mnras, 417, 370, \dodoi{10.1111/j.1365-2966.2011.19270.x}

\bibitem[{{Guo} {et~al.}(2012){Guo}, {Cole}, {Eke}, \& {Frenk}}]{2012MNRAS.427..428G}
---. 2012, \mnras, 427, 428, \dodoi{10.1111/j.1365-2966.2012.21882.x}

\bibitem[{{Guo} {et~al.}(2013){Guo}, {Cole}, {Eke}, {Frenk}, \& {Helly}}]{2013MNRAS.434.1838G}
{Guo}, Q., {Cole}, S., {Eke}, V., {Frenk}, C., \& {Helly}, J. 2013, \mnras, 434, 1838, \dodoi{10.1093/mnras/stt903}

\bibitem[{{Guo} {et~al.}(2015){Guo}, {Tempel}, \& {Libeskind}}]{2015ApJ...800..112G}
{Guo}, Q., {Tempel}, E., \& {Libeskind}, N.~I. 2015, \apj, 800, 112, \dodoi{10.1088/0004-637X/800/2/112}

\bibitem[{{Guo} {et~al.}(2010){Guo}, {White}, {Li}, \& {Boylan-Kolchin}}]{2010MNRAS.404.1111G}
{Guo}, Q., {White}, S., {Li}, C., \& {Boylan-Kolchin}, M. 2010, \mnras, 404, 1111, \dodoi{10.1111/j.1365-2966.2010.16341.x}

\bibitem[{{Guo} {et~al.}(2011{\natexlab{b}}){Guo}, {White}, {Boylan-Kolchin}, {De Lucia}, {Kauffmann}, {Lemson}, {Li}, {Springel}, \& {Weinmann}}]{2011MNRAS.413..101G}
{Guo}, Q., {White}, S., {Boylan-Kolchin}, M., {et~al.} 2011{\natexlab{b}}, \mnras, 413, 101, \dodoi{10.1111/j.1365-2966.2010.18114.x}

\bibitem[{{Guy} {et~al.}(2023){Guy}, {Bailey}, {Kremin}, {Alam}, {Alexander}, {Allende Prieto}, {BenZvi}, {Bolton}, {Brooks}, {Chaussidon}, {Cooper}, {Dawson}, {de la Macorra}, {Dey}, {Dey}, {Dhungana}, {Eisenstein}, {Font-Ribera}, {Forero-Romero}, {Gazta{\~n}aga}, {Gontcho A Gontcho}, {Green}, {Honscheid}, {Ishak}, {Kehoe}, {Kirkby}, {Kisner}, {Koposov}, {Lan}, {Landriau}, {Le Guillou}, {Levi}, {Magneville}, {Manser}, {Martini}, {Meisner}, {Miquel}, {Moustakas}, {Myers}, {Newman}, {Nie}, {Palanque-Delabrouille}, {Percival}, {Poppett}, {Prada}, {Raichoor}, {Ravoux}, {Ross}, {Schlafly}, {Schlegel}, {Schubnell}, {Sharples}, {Tarl{\'e}}, {Weaver}, {Y{\'e}che}, {Zhou}, {Zhou}, \& {Zou}}]{Spectro.Pipeline.Guy.2023}
{Guy}, J., {Bailey}, S., {Kremin}, A., {et~al.} 2023, \aj, 165, 144, \dodoi{10.3847/1538-3881/acb212}

\bibitem[{{Hahn} \& {DESI Team}(2022)}]{2022APS..APRH13003H}
{Hahn}, C., \& {DESI Team}. 2022, in APS Meeting Abstracts, Vol. 2022, APS April Meeting Abstracts, H13.003

\bibitem[{{Hahn} {et~al.}(2023){Hahn}, {Wilson}, {Ruiz-Macias}, {Cole}, {Weinberg}, {Moustakas}, {Kremin}, {Tinker}, {Smith}, {Wechsler}, {Ahlen}, {Alam}, {Bailey}, {Brooks}, {Cooper}, {Davis}, {Dawson}, {Dey}, {Dey}, {Eftekharzadeh}, {Eisenstein}, {Fanning}, {Forero-Romero}, {Frenk}, {Gazta{\~n}aga}, {Gontcho A Gontcho}, {Guy}, {Honscheid}, {Ishak}, {Juneau}, {Kehoe}, {Kisner}, {Lan}, {Landriau}, {Le Guillou}, {Levi}, {Magneville}, {Martini}, {Meisner}, {Myers}, {Nie}, {Norberg}, {Palanque-Delabrouille}, {Percival}, {Poppett}, {Prada}, {Raichoor}, {Ross}, {Safonova}, {Saulder}, {Schlafly}, {Schlegel}, {Sierra-Porta}, {Tarle}, {Weaver}, {Y{\`e}che}, {Zarrouk}, {Zhou}, {Zhou}, \& {Zou}}]{2023AJ....165..253H}
{Hahn}, C., {Wilson}, M.~J., {Ruiz-Macias}, O., {et~al.} 2023, \aj, 165, 253, \dodoi{10.3847/1538-3881/accff8}

\bibitem[{{Harsono} \& {De Propris}(2009)}]{2009AJ....137.3091H}
{Harsono}, D., \& {De Propris}, R. 2009, \aj, 137, 3091, \dodoi{10.1088/0004-6256/137/2/3091}

\bibitem[{{Jenkins} {et~al.}(2007){Jenkins}, {Hornschemeier}, {Mobasher}, {Alexander}, \& {Bauer}}]{2007ApJ...666..846J}
{Jenkins}, L.~P., {Hornschemeier}, A.~E., {Mobasher}, B., {Alexander}, D.~M., \& {Bauer}, F.~E. 2007, \apj, 666, 846, \dodoi{10.1086/520035}

\bibitem[{{Jiang} {et~al.}(2012){Jiang}, {Jing}, \& {Li}}]{2012ApJ...760...16J}
{Jiang}, C.~Y., {Jing}, Y.~P., \& {Li}, C. 2012, \apj, 760, 16, \dodoi{10.1088/0004-637X/760/1/16}

\bibitem[{{Juneau} {et~al.}(2025){Juneau}, {Canning}, {Alexander}, {Pucha}, {Fawcett}, {Myers}, {Moustakas}, {Ruiz-Macias}, {Cole}, {Pan}, {Aguilar}, {Ahlen}, {Alam}, {Bailey}, {Brooks}, {Chaussidon}, {Circosta}, {Claybaugh}, {Davis}, {Dawson}, {de la Macorra}, {Dey}, {Doel}, {Fanning}, {Forero-Romero}, {Gazta{\~n}aga}, {Gontcho A Gontcho}, {Gutierrez}, {Hahn}, {Honscheid}, {Kehoe}, {Kisner}, {Kremin}, {Lambert}, {Landriau}, {Le Guillou}, {Manera}, {Martini}, {Meisner}, {Miquel}, {Mu{\~n}oz-Guti{\'e}rrez}, {Nie}, {Palanque-Delabrouille}, {Percival}, {Poppett}, {Prada}, {Ravoux}, {Rezaie}, {Rossi}, {Sanchez}, {Schlafly}, {Schlegel}, {Schubnell}, {Seo}, {Silber}, {Siudek}, {Sprayberry}, {Tan}, {Tarl{\'e}}, {Y{\`e}che}, {Zhou}, \& {Zou}}]{2025AJ....169..157J}
{Juneau}, S., {Canning}, R., {Alexander}, D.~M., {et~al.} 2025, \aj, 169, 157, \dodoi{10.3847/1538-3881/adabc9}

\bibitem[{{Kawinwanichakij} {et~al.}(2014){Kawinwanichakij}, {Papovich}, {Quadri}, {Tran}, {Spitler}, {Kacprzak}, {Labb{\'e}}, {Straatman}, {Glazebrook}, {Allen}, {Cowley}, {Dav{\'e}}, {Dekel}, {Ferguson}, {Hartley}, {Koekemoer}, {Koo}, {Lu}, {Mehrtens}, {Nanayakkara}, {Persson}, {Rees}, {Salmon}, {Tilvi}, {Tomczak}, \& {van Dokkum}}]{2014ApJ...792..103K}
{Kawinwanichakij}, L., {Papovich}, C., {Quadri}, R.~F., {et~al.} 2014, \apj, 792, 103, \dodoi{10.1088/0004-637X/792/2/103}

\bibitem[{{Klypin} {et~al.}(1999){Klypin}, {Kravtsov}, {Valenzuela}, \& {Prada}}]{1999ApJ...522...82K}
{Klypin}, A., {Kravtsov}, A.~V., {Valenzuela}, O., \& {Prada}, F. 1999, \apj, 522, 82, \dodoi{10.1086/307643}

\bibitem[{{Koposov} {et~al.}(2024){Koposov}, {Allende Prieto}, {Cooper}, {Li}, {Beraldo e Silva}, {Kim}, {Carrillo}, {Dey}, {Manser}, {Nikakhtar}, {Riley}, {Rockosi}, {Valluri}, {Aguilar}, {Ahlen}, {Bailey}, {Blum}, {Brooks}, {Claybaugh}, {Cole}, {de la Macorra}, {Dey}, {Forero-Romero}, {Gazta{\~n}aga}, {Guy}, {Kremin}, {Le Guillou}, {Levi}, {Manera}, {Meisner}, {Miquel}, {Moustakas}, {Nie}, {Palanque-Delabrouille}, {Percival}, {Rezaie}, {Rossi}, {Sanchez}, {Schlafly}, {Schubnell}, {Tarl{\'e}}, {Weaver}, \& {Zhou}}]{2024MNRAS.533.1012K}
{Koposov}, S.~E., {Allende Prieto}, C., {Cooper}, A.~P., {et~al.} 2024, \mnras, 533, 1012, \dodoi{10.1093/mnras/stae1842}

\bibitem[{{Lan} {et~al.}(2016){Lan}, {M{\'e}nard}, \& {Mo}}]{2016MNRAS.459.3998L}
{Lan}, T.-W., {M{\'e}nard}, B., \& {Mo}, H. 2016, \mnras, 459, 3998, \dodoi{10.1093/mnras/stw898}

\bibitem[{{Lang} {et~al.}(2016){Lang}, {Hogg}, \& {Mykytyn}}]{2016ascl.soft04008L}
{Lang}, D., {Hogg}, D.~W., \& {Mykytyn}, D. 2016, {The Tractor: Probabilistic astronomical source detection and measurement}, Astrophysics Source Code Library, record ascl:1604.008

\bibitem[{{Lares} {et~al.}(2011){Lares}, {Lambas}, \& {Dom{\'\i}nguez}}]{2011AJ....142...13L}
{Lares}, M., {Lambas}, D.~G., \& {Dom{\'\i}nguez}, M.~J. 2011, \aj, 142, 13, \dodoi{10.1088/0004-6256/142/1/13}

\bibitem[{{Lasker} {et~al.}(2025){Lasker}, {Carnero Rosell}, {Myers}, {Ross}, {Bianchi}, {Hanif}, {Kehoe}, {de Mattia}, {Napolitano}, {Percival}, {Staten}, {Aguilar}, {Ahlen}, {Bigwood}, {Brooks}, {Claybaugh}, {Cole}, {de la Macorra}, {Ding}, {Doel}, {Fanning}, {Forero-Romero}, {Gazta{\~n}aga}, {Gontcho A Gontcho}, {Gutierrez}, {Honscheid}, {Howlett}, {Juneau}, {Kremin}, {Landriau}, {Le Guillou}, {Levi}, {Manera}, {Meisner}, {Miquel}, {Moustakas}, {Mueller}, {Nie}, {Niz}, {Oh}, {Palanque-Delabrouille}, {Poppett}, {Prada}, {Rezaie}, {Rossi}, {Sanchez}, {Schlegel}, {Schubnell}, {Seo}, {Sprayberry}, {Tarl{\'e}}, {Vargas-Maga{\~n}a}, {Weaver}, {Wilson}, {Zheng}, \& {DESI Collaboration}}]{2025JCAP...01..127L}
{Lasker}, J., {Carnero Rosell}, A., {Myers}, A.~D., {et~al.} 2025, \jcap, 2025, 127, \dodoi{10.1088/1475-7516/2025/01/127}

\bibitem[{{Levi} {et~al.}(2013){Levi}, {Bebek}, {Beers}, {Blum}, {Cahn}, {Eisenstein}, {Flaugher}, {Honscheid}, {Kron}, {Lahav}, {McDonald}, {Roe}, {Schlegel}, \& {representing the DESI collaboration}}]{Snowmass2013.Levi}
{Levi}, M., {Bebek}, C., {Beers}, T., {et~al.} 2013, arXiv e-prints, arXiv:1308.0847.
\newblock \doarXiv{1308.0847}

\bibitem[{{Li} {et~al.}(2022){Li}, {Wang}, {Mo}, {Huang}, {Katz}, {Luo}, {Cui}, {Li}, {Yang}, {Jiang}, \& {Zhang}}]{2022ApJ...936...11L}
{Li}, R., {Wang}, H., {Mo}, H.~J., {et~al.} 2022, \apj, 936, 11, \dodoi{10.3847/1538-4357/ac8359}

\bibitem[{{Lorrimer} {et~al.}(1994){Lorrimer}, {Frenk}, {Smith}, {White}, \& {Zaritsky}}]{1994MNRAS.269..696L}
{Lorrimer}, S.~J., {Frenk}, C.~S., {Smith}, R.~M., {White}, S.~D.~M., \& {Zaritsky}, D. 1994, \mnras, 269, 696, \dodoi{10.1093/mnras/269.3.696}

\bibitem[{{Loveday} {et~al.}(2012){Loveday}, {Norberg}, {Baldry}, {Driver}, {Hopkins}, {Peacock}, {Bamford}, {Liske}, {Bland-Hawthorn}, {Brough}, {Brown}, {Cameron}, {Conselice}, {Croom}, {Frenk}, {Gunawardhana}, {Hill}, {Jones}, {Kelvin}, {Kuijken}, {Nichol}, {Parkinson}, {Phillipps}, {Pimbblet}, {Popescu}, {Prescott}, {Robotham}, {Sharp}, {Sutherland}, {Taylor}, {Thomas}, {Tuffs}, {van Kampen}, \& {Wijesinghe}}]{2012MNRAS.420.1239L}
{Loveday}, J., {Norberg}, P., {Baldry}, I.~K., {et~al.} 2012, \mnras, 420, 1239, \dodoi{10.1111/j.1365-2966.2011.20111.x}

\bibitem[{{Lovell} {et~al.}(2014){Lovell}, {Frenk}, {Eke}, {Jenkins}, {Gao}, \& {Theuns}}]{2014MNRAS.439..300L}
{Lovell}, M.~R., {Frenk}, C.~S., {Eke}, V.~R., {et~al.} 2014, \mnras, 439, 300, \dodoi{10.1093/mnras/stt2431}

\bibitem[{{Lovell} {et~al.}(2012){Lovell}, {Eke}, {Frenk}, {Gao}, {Jenkins}, {Theuns}, {Wang}, {White}, {Boyarsky}, \& {Ruchayskiy}}]{2012MNRAS.420.2318L}
{Lovell}, M.~R., {Eke}, V., {Frenk}, C.~S., {et~al.} 2012, \mnras, 420, 2318, \dodoi{10.1111/j.1365-2966.2011.20200.x}

\bibitem[{{Luo} {et~al.}(2024){Luo}, {Wang}, {Cui}, {Mo}, {Li}, {Jing}, {Katz}, {Dav{\'e}}, {Yang}, {Chen}, {Li}, \& {Huang}}]{2024ApJ...966..236L}
{Luo}, X., {Wang}, H., {Cui}, W., {et~al.} 2024, \apj, 966, 236, \dodoi{10.3847/1538-4357/ad392e}

\bibitem[{{Miller} {et~al.}(2024){Miller}, {Doel}, {Gutierrez}, {Besuner}, {Brooks}, {Gallo}, {Heetderks}, {Jelinsky}, {Kent}, {Lampton}, {Levi}, {Liang}, {Meisner}, {Sholl}, {Silber}, {Sprayberry}, {Aguilar}, {de la Macorra}, {Eisenstein}, {Fanning}, {Font-Ribera}, {Gazta{\~n}aga}, {Gontcho A Gontcho}, {Honscheid}, {Jimenez}, {Joyce}, {Kehoe}, {Kisner}, {Kremin}, {Landriau}, {Le Guillou}, {Magneville}, {Martini}, {Miquel}, {Moustakas}, {Nie}, {Percival}, {Poppett}, {Prada}, {Rossi}, {Schlegel}, {Schubnell}, {Seo}, {Sharples}, {Tarl{\'e}}, {Vargas-Maga{\~n}a}, {Zhou}, \& {the DESI Collaboration}}]{Corrector.Miller.2023}
{Miller}, T.~N., {Doel}, P., {Gutierrez}, G., {et~al.} 2024, \aj, 168, 95, \dodoi{10.3847/1538-3881/ad45fe}

\bibitem[{{Milne} {et~al.}(2007){Milne}, {Pritchet}, {Poole}, {Gwyn}, {Kavelaars}, {Harris}, \& {Hanes}}]{2007AJ....133..177M}
{Milne}, M.~L., {Pritchet}, C.~J., {Poole}, G.~B., {et~al.} 2007, \aj, 133, 177, \dodoi{10.1086/509733}

\bibitem[{{Moore} {et~al.}(1999){Moore}, {Ghigna}, {Governato}, {Lake}, {Quinn}, {Stadel}, \& {Tozzi}}]{1999ApJ...524L..19M}
{Moore}, B., {Ghigna}, S., {Governato}, F., {et~al.} 1999, \apjl, 524, L19, \dodoi{10.1086/312287}

\bibitem[{{Moretti} {et~al.}(2015){Moretti}, {Bettoni}, {Poggianti}, {Fasano}, {Varela}, {D'Onofrio}, {Vulcani}, {Cava}, {Fritz}, {Couch}, {Moles}, \& {Kj{\ae}rgaard}}]{2015A&A...581A..11M}
{Moretti}, A., {Bettoni}, D., {Poggianti}, B.~M., {et~al.} 2015, \aap, 581, A11, \dodoi{10.1051/0004-6361/201526080}

\bibitem[{{Moustakas} {et~al.}(2023){Moustakas}, {Scholte}, {Dey}, \& {Khederlarian}}]{2023ascl.soft08005M}
{Moustakas}, J., {Scholte}, D., {Dey}, B., \& {Khederlarian}, A. 2023, {FastSpecFit: Fast spectral synthesis and emission-line fitting of DESI spectra}, Astrophysics Source Code Library, record ascl:2308.005.
\newblock \doeprint{2308.005}

\bibitem[{{M{\"u}ller} \& {Jerjen}(2020)}]{2020A&A...644A..91M}
{M{\"u}ller}, O., \& {Jerjen}, H. 2020, \aap, 644, A91, \dodoi{10.1051/0004-6361/202038862}

\bibitem[{{Myers} {et~al.}(2023){Myers}, {Moustakas}, {Bailey}, {Weaver}, {Cooper}, {Forero-Romero}, {Abolfathi}, {Alexander}, {Brooks}, {Chaussidon}, {Chuang}, {Dawson}, {Dey}, {Dey}, {Dhungana}, {Doel}, {Fanning}, {Gazta{\~n}aga}, {Gontcho A Gontcho}, {Gonzalez-Morales}, {Hahn}, {Herrera-Alcantar}, {Honscheid}, {Ishak}, {Karim}, {Kirkby}, {Kisner}, {Koposov}, {Kremin}, {Lan}, {Landriau}, {Lang}, {Levi}, {Magneville}, {Napolitano}, {Martini}, {Meisner}, {Newman}, {Palanque-Delabrouille}, {Percival}, {Poppett}, {Prada}, {Raichoor}, {Ross}, {Schlafly}, {Schlegel}, {Schubnell}, {Tan}, {Tarle}, {Wilson}, {Y{\`e}che}, {Zhou}, {Zhou}, \& {Zou}}]{2023AJ....165...50M}
{Myers}, A.~D., {Moustakas}, J., {Bailey}, S., {et~al.} 2023, \aj, 165, 50, \dodoi{10.3847/1538-3881/aca5f9}

\bibitem[{{Nierenberg} {et~al.}(2013){Nierenberg}, {Treu}, {Menci}, {Lu}, \& {Wang}}]{2013ApJ...772..146N}
{Nierenberg}, A.~M., {Treu}, T., {Menci}, N., {Lu}, Y., \& {Wang}, W. 2013, \apj, 772, 146, \dodoi{10.1088/0004-637X/772/2/146}

\bibitem[{{Peebles} \& {Nusser}(2010)}]{2010Natur.465..565P}
{Peebles}, P.~J.~E., \& {Nusser}, A. 2010, \nat, 465, 565, \dodoi{10.1038/nature09101}

\bibitem[{{Planck Collaboration} {et~al.}(2020){Planck Collaboration}, {Aghanim}, {Akrami}, {Ashdown}, {Aumont}, {Baccigalupi}, {Ballardini}, {Banday}, {Barreiro}, {Bartolo}, {Basak}, {Battye}, {Benabed}, {Bernard}, {Bersanelli}, {Bielewicz}, {Bock}, {Bond}, {Borrill}, {Bouchet}, {Boulanger}, {Bucher}, {Burigana}, {Butler}, {Calabrese}, {Cardoso}, {Carron}, {Challinor}, {Chiang}, {Chluba}, {Colombo}, {Combet}, {Contreras}, {Crill}, {Cuttaia}, {de Bernardis}, {de Zotti}, {Delabrouille}, {Delouis}, {Di Valentino}, {Diego}, {Dor{\'e}}, {Douspis}, {Ducout}, {Dupac}, {Dusini}, {Efstathiou}, {Elsner}, {En{\ss}lin}, {Eriksen}, {Fantaye}, {Farhang}, {Fergusson}, {Fernandez-Cobos}, {Finelli}, {Forastieri}, {Frailis}, {Fraisse}, {Franceschi}, {Frolov}, {Galeotta}, {Galli}, {Ganga}, {G{\'e}nova-Santos}, {Gerbino}, {Ghosh}, {Gonz{\'a}lez-Nuevo}, {G{\'o}rski}, {Gratton}, {Gruppuso}, {Gudmundsson}, {Hamann}, {Handley}, {Hansen}, {Herranz}, {Hildebrandt}, {Hivon}, {Huang}, {Jaffe}, {Jones}, {Karakci}, {Keih{\"a}nen},
  {Keskitalo}, {Kiiveri}, {Kim}, {Kisner}, {Knox}, {Krachmalnicoff}, {Kunz}, {Kurki-Suonio}, {Lagache}, {Lamarre}, {Lasenby}, {Lattanzi}, {Lawrence}, {Le Jeune}, {Lemos}, {Lesgourgues}, {Levrier}, {Lewis}, {Liguori}, {Lilje}, {Lilley}, {Lindholm}, {L{\'o}pez-Caniego}, {Lubin}, {Ma}, {Mac{\'\i}as-P{\'e}rez}, {Maggio}, {Maino}, {Mandolesi}, {Mangilli}, {Marcos-Caballero}, {Maris}, {Martin}, {Martinelli}, {Mart{\'\i}nez-Gonz{\'a}lez}, {Matarrese}, {Mauri}, {McEwen}, {Meinhold}, {Melchiorri}, {Mennella}, {Migliaccio}, {Millea}, {Mitra}, {Miville-Desch{\^e}nes}, {Molinari}, {Montier}, {Morgante}, {Moss}, {Natoli}, {N{\o}rgaard-Nielsen}, {Pagano}, {Paoletti}, {Partridge}, {Patanchon}, {Peiris}, {Perrotta}, {Pettorino}, {Piacentini}, {Polastri}, {Polenta}, {Puget}, {Rachen}, {Reinecke}, {Remazeilles}, {Renzi}, {Rocha}, {Rosset}, {Roudier}, {Rubi{\~n}o-Mart{\'\i}n}, {Ruiz-Granados}, {Salvati}, {Sandri}, {Savelainen}, {Scott}, {Shellard}, {Sirignano}, {Sirri}, {Spencer}, {Sunyaev}, {Suur-Uski}, {Tauber}, {Tavagnacco},
  {Tenti}, {Toffolatti}, {Tomasi}, {Trombetti}, {Valenziano}, {Valiviita}, {Van Tent}, {Vibert}, {Vielva}, {Villa}, {Vittorio}, {Wandelt}, {Wehus}, {White}, {White}, {Zacchei}, \& {Zonca}}]{2020A&A...641A...6P}
{Planck Collaboration}, {Aghanim}, N., {Akrami}, Y., {et~al.} 2020, \aap, 641, A6, \dodoi{10.1051/0004-6361/201833910}

\bibitem[{{Popesso} {et~al.}(2005{\natexlab{a}}){Popesso}, {Biviano}, {B{\"o}hringer}, \& {Romaniello}}]{2005nfcd.conf..346P}
{Popesso}, P., {Biviano}, A., {B{\"o}hringer}, H., \& {Romaniello}, M. 2005{\natexlab{a}}, in IAU Colloq. 198: Near-fields cosmology with dwarf elliptical galaxies, ed. H.~{Jerjen} \& B.~{Binggeli}, 346--350, \dodoi{10.1017/S1743921305004035}

\bibitem[{{Popesso} {et~al.}(2005{\natexlab{b}}){Popesso}, {B{\"o}hringer}, {Romaniello}, \& {Voges}}]{2005A&A...433..415P}
{Popesso}, P., {B{\"o}hringer}, H., {Romaniello}, M., \& {Voges}, W. 2005{\natexlab{b}}, \aap, 433, 415, \dodoi{10.1051/0004-6361:20041870}

\bibitem[{{Poppett} {et~al.}(2024){Poppett}, {Tyas}, {Aguilar}, {Bebek}, {Bramall}, {Claybaugh}, {Edelstein}, {Fagrelius}, {Heetderks}, {Jelinsky}, {Jelinsky}, {Lafever}, {Lambert}, {Lampton}, {Levi}, {Martini}, {Rockosi}, {Schmoll}, {Sharples}, {Sirk}, {Wishnow}, {Yu}, {Ahlen}, {Bault}, {BenZvi}, {Brooks}, {Cole}, {de la Macorra}, {Dey}, {Doel}, {Fanning}, {Font-Ribera}, {Forero-Romero}, {Gazta{\~n}aga}, {Gontcho A Gontcho}, {Gonzalez-Morales}, {Hahn}, {Honscheid}, {Jimenez}, {Juneau}, {Kirkby}, {Kremin}, {Landriau}, {Le Guillou}, {Manera}, {Meisner}, {Miquel}, {Moustakas}, {Mueller}, {Mu{\~n}oz-Guti{\'e}rrez}, {Myers}, {Nie}, {Niz}, {Palanque-Delabrouille}, {Percival}, {Prada}, {Rabinowitz}, {Rezaie}, {Rossi}, {Sanchez}, {Schlafly}, {Schlegel}, {Schubnell}, {Seo}, {Sprayberry}, {Tarl{\'e}}, {Vargas-Maga{\~n}a}, {Weaver}, \& {Zhou}}]{FiberSystem.Poppett.2024}
{Poppett}, C., {Tyas}, L., {Aguilar}, J., {et~al.} 2024, \aj, 168, 245, \dodoi{10.3847/1538-3881/ad76a4}

\bibitem[{{Raichoor} {et~al.}(2023){Raichoor}, {Moustakas}, {Newman}, {Karim}, {Ahlen}, {Alam}, {Bailey}, {Brooks}, {Dawson}, {de la Macorra}, {de Mattia}, {Dey}, {Dey}, {Dhungana}, {Eftekharzadeh}, {Eisenstein}, {Fanning}, {Font-Ribera}, {Garc{\'\i}a-Bellido}, {Gazta{\~n}aga}, {A Gontcho}, {Guy}, {Honscheid}, {Ishak}, {Kehoe}, {Kisner}, {Kremin}, {Lan}, {Landriau}, {Le Guillou}, {Levi}, {Magneville}, {Manera}, {Martini}, {Meisner}, {Myers}, {Nie}, {Palanque-Delabrouille}, {Percival}, {Poppett}, {Prada}, {Ross}, {Ruhlmann-Kleider}, {Sabiu}, {Schlafly}, {Schlegel}, {Tarl{\'e}}, {Weaver}, {Y{\`e}che}, {Zhou}, {Zhou}, \& {Zou}}]{2023AJ....165..126R}
{Raichoor}, A., {Moustakas}, J., {Newman}, J.~A., {et~al.} 2023, \aj, 165, 126, \dodoi{10.3847/1538-3881/acb213}

\bibitem[{{Rines} \& {Geller}(2008)}]{2008AJ....135.1837R}
{Rines}, K., \& {Geller}, M.~J. 2008, \aj, 135, 1837, \dodoi{10.1088/0004-6256/135/5/1837}

\bibitem[{{Roberts} {et~al.}(2021){Roberts}, {Nierenberg}, \& {Peter}}]{2021MNRAS.502.1205R}
{Roberts}, D.~M., {Nierenberg}, A.~M., \& {Peter}, A. H.~G. 2021, \mnras, 502, 1205, \dodoi{10.1093/mnras/stab069}

\bibitem[{{Ross} {et~al.}(2024){Ross}, {Aguilar}, {Ahlen}, {Alam}, {Anand}, {Bailey}, {Bianchi}, {Brieden}, {Brooks}, {Burtin}, {Carnero Rosell}, {Chaussidon}, {Claybaugh}, {Cole}, {Dawson}, {de la Macorra}, {de Mattia}, {Dey}, {Dey}, {Doel}, {Fanning}, {Ferraro}, {Ereza}, {Font-Ribera}, {Forero-Romero}, {Gazta{\~n}aga}, {Gil-Mar{\'\i}n}, {Gontcho}, {Gonzalez-Morales}, {Guy}, {Hahn}, {Heydenreich}, {Honscheid}, {Howlett}, {Ishak}, {Karim}, {Kirkby}, {Kisner}, {Kong}, {Kremin}, {Krolewski}, {Lambert}, {Landriau}, {Lasker}, {Le Guillou}, {Levi}, {Manera}, {Martini}, {McDonald}, {Meisner}, {Miquel}, {Moon}, {Moustakas}, {Mu{\~n}oz-Guti{\'e}rrez}, {Myers}, {Nadathur}, {Napolitano}, {Newman}, {Nie}, {Niz}, {Palanque-Delabrouille}, {Percival}, {Poppett}, {Prada}, {Raichoor}, {Ravoux}, {Rezaie}, {Rosado-Marin}, {Rossi}, {Samushia}, {Sanchez}, {Schlafly}, {Schlegel}, {Seo}, {Smith}, {Sprayberry}, {Tarl{\'e}}, {Valcin}, {Vargas-Maga{\~n}a}, {Weaver}, {Wilson}, {Yu}, {Zarrouk}, {Zhao}, {Zhou}, \&
  {Zou}}]{2024arXiv240516593R}
{Ross}, A.~J., {Aguilar}, J., {Ahlen}, S., {et~al.} 2024, arXiv e-prints, arXiv:2405.16593, \dodoi{10.48550/arXiv.2405.16593}

\bibitem[{{Sales} {et~al.}(2013){Sales}, {Wang}, {White}, \& {Navarro}}]{2013MNRAS.428..573S}
{Sales}, L.~V., {Wang}, W., {White}, S. D.~M., \& {Navarro}, J.~F. 2013, \mnras, 428, 573, \dodoi{10.1093/mnras/sts054}

\bibitem[{{Schlafly} {et~al.}(2023){Schlafly}, {Kirkby}, {Schlegel}, {Myers}, {Raichoor}, {Dawson}, {Aguilar}, {Allende Prieto}, {Bailey}, {BenZvi}, {Bermejo-Climent}, {Brooks}, {de la Macorra}, {Dey}, {Doel}, {Fanning}, {Font-Ribera}, {Forero-Romero}, {Garc{\'\i}a-Bellido}, {Gontcho A Gontcho}, {Guy}, {Hahn}, {Honscheid}, {Ishak}, {Juneau}, {Kehoe}, {Kisner}, {Kremin}, {Landriau}, {Lang}, {Lasker}, {Levi}, {Magneville}, {Manser}, {Martini}, {Meisner}, {Miquel}, {Moustakas}, {Newman}, {Nie}, {Palanque-Delabrouille}, {Percival}, {Poppett}, {Rockosi}, {Ross}, {Rossi}, {Tarl{\'e}}, {Weaver}, {Y{\`e}che}, {Zhou}, \& {DESI Collaboration}}]{SurveyOps.Schlafly.2023}
{Schlafly}, E.~F., {Kirkby}, D., {Schlegel}, D.~J., {et~al.} 2023, \aj, 166, 259, \dodoi{10.3847/1538-3881/ad0832}

\bibitem[{{Shi} {et~al.}(2024){Shi}, {Wang}, {Li}, {Zhu}, {Smith}, {Cole}, {Gao}, {Chen}, {Li}, \& {Han}}]{2024ApJ...973...82S}
{Shi}, R., {Wang}, W., {Li}, Z., {et~al.} 2024, \apj, 973, 82, \dodoi{10.3847/1538-4357/ad64cf}

\bibitem[{{Silber} {et~al.}(2022){Silber}, {Fagrelius}, {Fanning}, {Schubnell}, {Aguilar}, {Ahlen}, {Ameel}, {Ballester}, {Baltay}, {Bebek}, {Beard}, {Besuner}, {Cardiel-Sas}, {Casas}, {Castander}, {Claybaugh}, {Dobson}, {Duan}, {Dunlop}, {Edelstein}, {Emmet}, {Elliott}, {Evatt}, {Gershkovich}, {Guy}, {Harris}, {Heetderks}, {Heetderks}, {Honscheid}, {Illa}, {Jelinsky}, {Jelinsky}, {Jimenez}, {Karcher}, {Kent}, {Kirkby}, {Kneib}, {Lambert}, {Lampton}, {Leitner}, {Levi}, {McCauley}, {Meisner}, {Miller}, {Miquel}, {Mundet}, {Poppett}, {Rabinowitz}, {Reil}, {Roman}, {Schlegel}, {Serrano}, {Van Shourt}, {Sprayberry}, {Tarl{\'e}}, {Sien Tie}, {Weaverdyck}, {Zhang}, {Azzaro}, {Bailey}, {Becerril}, {Blackwell}, {Bouri}, {Brooks}, {Buckley-Geer}, {Pe{\~n}ate Castro}, {Derwent}, {Dey}, {Dhungana}, {Doel}, {Eisenstein}, {Fahim}, {Garcia-Bellido}, {Gazta{\~n}aga}, {Gontcho}, {Gutierrez}, {H{\"o}rler}, {Kehoe}, {Kisner}, {Kremin}, {Kronig}, {Landriau}, {Le Guillou}, {Martini}, {Moustakas}, {Palanque-Delabrouille}, {Peng},
  {Percival}, {Prada}, {Allende Prieto}, {Gonzalez de Rivera}, {Sanchez}, {Sanchez}, {Sharples}, {Soares-Santos}, {Schlafly}, {Weaver}, {Zhou}, {Zhu}, \& {Zou}}]{silber22a}
{Silber}, J.~H., {Fagrelius}, P., {Fanning}, K., {et~al.} 2022, arXiv e-prints, arXiv:2205.09014.
\newblock \doarXiv{2205.09014}

\bibitem[{{Siudek} {et~al.}(2024){Siudek}, {Pucha}, {Mezcua}, {Juneau}, {Aguilar}, {Ahlen}, {Brooks}, {Circosta}, {Claybaugh}, {Cole}, {Dawson}, {de la Macorra}, {Dey}, {Dey}, {Doel}, {Font-Ribera}, {Forero-Romero}, {Gazta{\~n}aga}, {Gontcho A Gontcho}, {Gutierrez}, {Honscheid}, {Howlett}, {Ishak}, {Kehoe}, {Kirkby}, {Kisner}, {Kremin}, {Lambert}, {Landriau}, {Le Guillou}, {Manera}, {Martini}, {Meisner}, {Miquel}, {Moustakas}, {Newman}, {Niz}, {Pan}, {Percival}, {Poppett}, {Prada}, {Rossi}, {Saintonge}, {Sanchez}, {Schlegel}, {Scholte}, {Schubnell}, {Seo}, {Speranza}, {Sprayberry}, {Tarl{\'e}}, {Weaver}, \& {Zou}}]{2024A&A...691A.308S}
{Siudek}, M., {Pucha}, R., {Mezcua}, M., {et~al.} 2024, \aap, 691, A308, \dodoi{10.1051/0004-6361/202451761}

\bibitem[{{Smith} {et~al.}(2019){Smith}, {He}, {Cole}, {Stothert}, {Norberg}, {Baugh}, {Bianchi}, {Wilson}, {Brooks}, {Forero-Romero}, {Moustakas}, {Percival}, {Tarle}, \& {Wechsler}}]{2019MNRAS.484.1285S}
{Smith}, A., {He}, J.-h., {Cole}, S., {et~al.} 2019, \mnras, 484, 1285, \dodoi{10.1093/mnras/stz059}

\bibitem[{{Somerville}(2002)}]{2002ApJ...572L..23S}
{Somerville}, R.~S. 2002, \apjl, 572, L23, \dodoi{10.1086/341444}

\bibitem[{{Tinker} {et~al.}(2021){Tinker}, {Cao}, {Alpaslan}, {DeRose}, {Mao}, \& {Wechsler}}]{2021MNRAS.505.5370T}
{Tinker}, J.~L., {Cao}, J., {Alpaslan}, M., {et~al.} 2021, \mnras, 505, 5370, \dodoi{10.1093/mnras/stab1576}

\bibitem[{{Tully}(1988)}]{1988ngc..book.....T}
{Tully}, R.~B. 1988, {Nearby galaxies catalog}

\bibitem[{{Wang} {et~al.}(2014{\natexlab{a}}){Wang}, {Mo}, {Yang}, {Jing}, \& {Lin}}]{2014ApJ...794...94W}
{Wang}, H., {Mo}, H.~J., {Yang}, X., {Jing}, Y.~P., \& {Lin}, W.~P. 2014{\natexlab{a}}, \apj, 794, 94, \dodoi{10.1088/0004-637X/794/1/94}

\bibitem[{{Wang} {et~al.}(2016){Wang}, {Mo}, {Yang}, {Zhang}, {Shi}, {Jing}, {Liu}, {Li}, {Kang}, \& {Gao}}]{2016ApJ...831..164W}
{Wang}, H., {Mo}, H.~J., {Yang}, X., {et~al.} 2016, \apj, 831, 164, \dodoi{10.3847/0004-637X/831/2/164}

\bibitem[{{Wang} {et~al.}(2011){Wang}, {Jing}, {Li}, {Okumura}, \& {Han}}]{2011ApJ...734...88W}
{Wang}, W., {Jing}, Y.~P., {Li}, C., {Okumura}, T., \& {Han}, J. 2011, \apj, 734, 88, \dodoi{10.1088/0004-637X/734/2/88}

\bibitem[{{Wang} {et~al.}(2014{\natexlab{b}}){Wang}, {Sales}, {Henriques}, \& {White}}]{2014MNRAS.442.1363W}
{Wang}, W., {Sales}, L.~V., {Henriques}, B.~M.~B., \& {White}, S.~D.~M. 2014{\natexlab{b}}, \mnras, 442, 1363, \dodoi{10.1093/mnras/stu988}

\bibitem[{{Wang} \& {White}(2012)}]{2012MNRAS.424.2574W}
{Wang}, W., \& {White}, S.~D.~M. 2012, \mnras, 424, 2574, \dodoi{10.1111/j.1365-2966.2012.21256.x}

\bibitem[{{Wang} {et~al.}(2019){Wang}, {Han}, {Sonnenfeld}, {Yasuda}, {Li}, {Jing}, {More}, {Price}, {Lupton}, {Rykoff}, {Stark}, {Lan}, {Takada}, {Huang}, {Luo}, {Bahcall}, \& {Komiyama}}]{2019MNRAS.487.1580W}
{Wang}, W., {Han}, J., {Sonnenfeld}, A., {et~al.} 2019, \mnras, 487, 1580, \dodoi{10.1093/mnras/stz1339}

\bibitem[{{Wang} {et~al.}(2021{\natexlab{a}}){Wang}, {Takada}, {Li}, {Carlsten}, {Lan}, {Shi}, {Miyatake}, {More}, {Beaton}, {Lupton}, {Lin}, {Qiu}, \& {Luo}}]{2021MNRAS.500.3776W}
{Wang}, W., {Takada}, M., {Li}, X., {et~al.} 2021{\natexlab{a}}, \mnras, 500, 3776, \dodoi{10.1093/mnras/staa3495}

\bibitem[{{Wang} {et~al.}(2021{\natexlab{b}}){Wang}, {Li}, {Shi}, {Han}, {Yasuda}, {Jing}, {More}, {Takada}, {Miyatake}, \& {Nishizawa}}]{2021ApJ...919...25W}
{Wang}, W., {Li}, X., {Shi}, J., {et~al.} 2021{\natexlab{b}}, \apj, 919, 25, \dodoi{10.3847/1538-4357/ac0e38}

\bibitem[{{Wang} {et~al.}(2024){Wang}, {Yang}, {Gu}, {Xu}, {Xu}, {Wang}, {Katsianis}, {Han}, {He}, {Zheng}, {Li}, {Wang}, {Hong}, {Wang}, {Tan}, {Zou}, {Lange}, {Hahn}, {Behroozi}, {Aguilar}, {Ahlen}, {Brooks}, {Claybaugh}, {Cole}, {de la Macorra}, {Dey}, {Doel}, {Forero-Romero}, {Honscheid}, {Kehoe}, {Kisner}, {Lambert}, {Manera}, {Meisner}, {Miquel}, {Moustakas}, {Nie}, {Poppett}, {Rezaie}, {Rossi}, {Sanchez}, {Schubnell}, {Tarl{\'e}}, {Weaver}, \& {Zhou}}]{2024ApJ...971..119W}
{Wang}, Y., {Yang}, X., {Gu}, Y., {et~al.} 2024, \apj, 971, 119, \dodoi{10.3847/1538-4357/ad5294}

\bibitem[{{Wegner}(2011)}]{2011MNRAS.413.1333W}
{Wegner}, G.~A. 2011, \mnras, 413, 1333, \dodoi{10.1111/j.1365-2966.2011.18218.x}

\bibitem[{{Westra} {et~al.}(2010){Westra}, {Geller}, {Kurtz}, {Fabricant}, \& {Dell'Antonio}}]{2010PASP..122.1258W}
{Westra}, E., {Geller}, M.~J., {Kurtz}, M.~J., {Fabricant}, D.~G., \& {Dell'Antonio}, I. 2010, \pasp, 122, 1258, \dodoi{10.1086/657452}

\bibitem[{{Xie} {et~al.}(2014){Xie}, {Gao}, \& {Guo}}]{2014MNRAS.441..933X}
{Xie}, L., {Gao}, L., \& {Guo}, Q. 2014, \mnras, 441, 933, \dodoi{10.1093/mnras/stu513}

\bibitem[{{Xu} \& {Jing}(2022)}]{2022ApJ...926..130X}
{Xu}, K., \& {Jing}, Y. 2022, \apj, 926, 130, \dodoi{10.3847/1538-4357/ac4707}

\bibitem[{{Xu} {et~al.}(2022{\natexlab{a}}){Xu}, {Jing}, \& {Gao}}]{2022ApJ...939..104X}
{Xu}, K., {Jing}, Y.~P., \& {Gao}, H. 2022{\natexlab{a}}, \apj, 939, 104, \dodoi{10.3847/1538-4357/ac8f47}

\bibitem[{{Xu} {et~al.}(2023){Xu}, {Jing}, {Zheng}, \& {Gao}}]{2023ApJ...944..200X}
{Xu}, K., {Jing}, Y.~P., {Zheng}, Y., \& {Gao}, H. 2023, \apj, 944, 200, \dodoi{10.3847/1538-4357/acb13e}

\bibitem[{{Xu} {et~al.}(2022{\natexlab{b}}){Xu}, {Zheng}, \& {Jing}}]{2022ApJ...925...31X}
{Xu}, K., {Zheng}, Y., \& {Jing}, Y. 2022{\natexlab{b}}, \apj, 925, 31, \dodoi{10.3847/1538-4357/ac38a2}

\bibitem[{{Xu} {et~al.}(2025){Xu}, {Jing}, {Cole}, {Frenk}, {Bose}, {Elbers}, {Wang}, {Wang}, {Moore}, {Aguilar}, {Ahlen}, {Bianchi}, {Brooks}, {Claybaugh}, {de la Macorra}, {Dey}, {Forero-Romero}, {Gazta{\~n}aga}, {Gontcho}, {Gutierrez}, {Honscheid}, {Ishak}, {Kisner}, {Koposov}, {Landriau}, {Le Guillou}, {Miquel}, {Moustakas}, {Poppett}, {Prada}, {P{\'e}rez-R{\`a}fols}, {Rossi}, {Sanchez}, {Sprayberry}, {Tarl{\'e}}, {Weaver}, \& {Zou}}]{Xu2025}
{Xu}, K., {Jing}, Y.~P., {Cole}, S., {et~al.} 2025, arXiv e-prints, arXiv:2503.01948, \dodoi{10.48550/arXiv.2503.01948}

\bibitem[{{Yang} {et~al.}(2018){Yang}, {Zhang}, {Wang}, {Liu}, {Lu}, {Li}, {Shi}, {Jing}, {Mo}, {van den Bosch}, {Kang}, {Cui}, {Guo}, {Li}, {Lim}, {Lu}, {Luo}, {Wei}, \& {Yang}}]{2018ApJ...860...30Y}
{Yang}, X., {Zhang}, Y., {Wang}, H., {et~al.} 2018, \apj, 860, 30, \dodoi{10.3847/1538-4357/aac2ce}

\bibitem[{{Yang} {et~al.}(2021){Yang}, {Xu}, {He}, {Gu}, {Katsianis}, {Meng}, {Shi}, {Zou}, {Zhang}, {Liu}, {Wang}, {Dong}, {Lu}, {Li}, {Chen}, {Wang}, {Mo}, {Fu}, {Guo}, {Leauthaud}, {Luo}, {Zhang}, \& {Zu}}]{2021ApJ...909..143Y}
{Yang}, X., {Xu}, H., {He}, M., {et~al.} 2021, \apj, 909, 143, \dodoi{10.3847/1538-4357/abddb2}

\bibitem[{{Zhou} {et~al.}(2023){Zhou}, {Dey}, {Newman}, {Eisenstein}, {Dawson}, {Bailey}, {Berti}, {Guy}, {Lan}, {Zou}, {Aguilar}, {Ahlen}, {Alam}, {Brooks}, {de la Macorra}, {Dey}, {Dhungana}, {Fanning}, {Font-Ribera}, {Gontcho}, {Honscheid}, {Ishak}, {Kisner}, {Kov{\'a}cs}, {Kremin}, {Landriau}, {Levi}, {Magneville}, {Manera}, {Martini}, {Meisner}, {Miquel}, {Moustakas}, {Myers}, {Nie}, {Palanque-Delabrouille}, {Percival}, {Poppett}, {Prada}, {Raichoor}, {Ross}, {Schlafly}, {Schlegel}, {Schubnell}, {Tarl{\'e}}, {Weaver}, {Wechsler}, {Y{\'e}che}, \& {Zhou}}]{2023AJ....165...58Z}
{Zhou}, R., {Dey}, B., {Newman}, J.~A., {et~al.} 2023, \aj, 165, 58, \dodoi{10.3847/1538-3881/aca5fb}

\bibitem[{{Zou} {et~al.}(2017){Zou}, {Zhou}, {Fan}, {Zhang}, {Zhou}, {Nie}, {Peng}, {McGreer}, {Jiang}, {Dey}, {Fan}, {He}, {Jiang}, {Lang}, {Lesser}, {Ma}, {Mao}, {Schlegel}, \& {Wang}}]{2017PASP..129f4101Z}
{Zou}, H., {Zhou}, X., {Fan}, X., {et~al.} 2017, \pasp, 129, 064101, \dodoi{10.1088/1538-3873/aa65ba}

\end{thebibliography}
\bibliographystyle{aasjournal}

\end{document}